\documentclass[trackchanges, twocolumn]{aastex701}

\usepackage{hyperref, courier}
\usepackage{amsmath, amssymb}
\usepackage{float}
\usepackage{xcolor}

\newcommand{\DEL}[1]{}
\newcommand{\no}{\nodata}

\newcommand\eg{{e.g.},}
\defcitealias{Ortiz2024}{O24}

\shortauthors{Bowling et al.}
\shorttitle{Distinct Star-Forming Properties of NIRCam AGN}

\begin{document}
\title{{PEARLS: Distinct Star-Forming Properties of the Host Galaxies of NIRCam-Selected AGN}}

\author[orcid=0009-0007-0782-0721]{Gibson B.\ Bowling}
\email{gbbowlin@asu.edu}
\affiliation{School of Earth and Space Exploration, Arizona State University,
Tempe, AZ 85287-1404, USA}

\author[0000-0002-6150-833X]{Rafael {Ortiz~III}} 
\email{rortizii@asu.edu}
\affiliation{School of Earth and Space Exploration, Arizona State University,
Tempe, AZ 85287-1404, USA}

\author[0000-0002-9895-5758]{S.\ P.\ Willner}
\email{swillner@cfa.harvard.edu}
\affiliation{Center for Astrophysics \textbar\ Harvard \& Smithsonian, 
60 Garden Street, Cambridge, MA, 02138, USA}

\author[0000-0003-3329-1337]{Seth H.\ Cohen}
\email{seth.cohen@asu.edu}
\affiliation{School of Earth and Space Exploration, Arizona State University,
Tempe, AZ 85287-1404, USA}

\author[0000-0001-6650-2853]{Timothy Carleton}
\email{tmcarlet@asu.edu}
\affiliation{School of Earth and Space Exploration, Arizona State University,
Tempe, AZ 85287-1404, USA}

\author[0000-0001-8156-6281]{Rogier A.\ Windhorst}
\email{Rogier.Windhorst@gmail.com}
\affiliation{School of Earth and Space Exploration, Arizona State University,
Tempe, AZ 85287-1404, USA}

\author[0000-0003-1268-5230]{Rolf A.\ Jansen}
\email{rolfjansen.work@gmail.com}
\affiliation{School of Earth and Space Exploration, Arizona State University,
Tempe, AZ 85287-1404, USA}

\author[0000-0002-6917-0214]{Kevin~S.~Croker}
\email{kevin.croker@asu.edu}
\affiliation{School of Earth and Space Exploration, Arizona State University,
Tempe, AZ 85287-1404, USA}
\affiliation{Department of Physics and Astronomy, University of Hawai`i at M\=anoa,
2505 Correa Rd., Honolulu, HI 96822 USA}

\author[0000-0001-9262-9997]{Christopher N.\ A.\ Willmer}
\email{cnawillmer@gmail.com}
\affiliation{Steward Observatory, University of Arizona,
933 N Cherry Ave, Tucson, AZ, 85721-0009, USA}

\author[0000-0002-2203-7889]{W. Peter Maksym}
\email{walter.p.maksym@nasa.gov}
\affiliation{NASA Marshall Space Flight Center, Huntsville, AL 35812, USA}

\author[0000-0001-6434-7845]{Madeline A.\ Marshall}
\email{madeline_marshall@outlook.com}
\affiliation{Los Alamos National Laboratory, Los Alamos, NM 87545, USA}

\author[0000-0003-3351-0878]{Rosalia O'Brien}
\email{robrien5@asu.edu}
\affiliation{University of Maryland/CRESST2, NASA Goddard Space Flight Center, Greenbelt, MD 20771, USA}
\affiliation{School of Earth and Space Exploration, Arizona State University, Tempe, AZ 85287-1404, USA}

\author[0000-0002-5319-6620]{Payaswini Saikia}
\email{payaswini.ssc@gmail.com}
\affiliation{Department of Astronomy, Yale University, PO Box 208101, New Haven, CT 06520-8101, USA}

\author[0000-0003-4223-7324]{Massimo Ricotti}
\email{ricotti@umd.edu}
\affiliation{Dept. of Astronomy, University of Maryland, College Park, MD 20742, USA}

\author[0000-0002-9816-1931]{Jordan C.\ J.\ D'Silva}
\email{jordan.dsilva@research.uwa.edu.au}
\affiliation{International Centre for Radio Astronomy Research (ICRAR) and the
International Space Centre (ISC), The University of Western Australia, M468,
35 Stirling Highway, Crawley, WA 6009, Australia}
\affiliation{ARC Centre of Excellence for All Sky Astrophysics in 3 Dimensions
(ASTRO 3D), Australia}

\author[0000-0001-7410-7669]{Dan Coe}
\email{dcoe@stsci.edu}
\affiliation{Space Telescope Science Institute, 3700 San Martin Drive, Baltimore, MD 21218, USA}
\affiliation{Association of Universities for Research in Astronomy (AURA) for the European Space Agency (ESA), STScI, Baltimore, MD 21218, USA}
\affiliation{Center for Astrophysical Sciences, Department of Physics and Astronomy, The Johns Hopkins University, 3400 N Charles St. Baltimore, MD 21218, USA}

\author[0000-0003-1949-7638]{Christopher J.\ Conselice}
\email{conselice@gmail.com}
\affiliation{Jodrell Bank Centre for Astrophysics, Alan Turing Building,
University of Manchester, Oxford Road, Manchester M13 9PL, UK}

\author[0000-0001-9065-3926]{Jose M. Diego}
\email{chemadiegor@gmail.com}
\affiliation{Instituto de F\'isica de Cantabria (CSIC-UC). Avenida. Los Castros
s/n. 39005 Santander, Spain}

\author[0000-0001-9491-7327]{Simon P.\ Driver}
\email{Simon.Driver@icrar.org}
\affiliation{International Centre for Radio Astronomy Research (ICRAR) and the
International Space Centre (ISC), The University of Western Australia, M468,
35 Stirling Highway, Crawley, WA 6009, Australia}

\author[0000-0003-1625-8009]{Brenda L.\ Frye}
\email{brendafrye@gmail.com}
\affiliation{Department of Astronomy/Steward Observatory, University of Arizona, 933 N Cherry Ave,
Tucson, AZ, 85721-0009, USA}

\author[0000-0001-9440-8872]{Norman A.\ Grogin}
\email{nagrogin@stsci.edu}
\affiliation{Space Telescope Science Institute,
3700 San Martin Drive, Baltimore, MD 21218, USA}

\author[0000-0002-9984-4937]{Rachel Honor}
\email{rchonor@asu.edu}
\affiliation{School of Earth and Space Exploration, Arizona State University, Tempe, AZ 85287-1404, USA}

\author[0000-0002-7265-7920]{Jake Summers}
\email{jsummers@caltech.edu}
\affiliation{TAPIR, California Institute of Technology, Pasadena, CA 91125, USA}
\affiliation{LIGO Laboratory, California Institute of Technology, Pasadena, CA 91125, USA}

\author[0000-0003-3382-5941]{Nor Pirzkal}
\email{npirzkal@stsci.edu}
\affiliation{Space Telescope Science Institute,
3700 San Martin Drive, Baltimore, MD 21218, USA}

\author[0000-0003-0429-3579]{Aaron Robotham}
\email{aaron.robotham@uwa.edu.au}
\affiliation{International Centre for Radio Astronomy Research (ICRAR) and the
International Space Centre (ISC), The University of Western Australia, M468,
35 Stirling Highway, Crawley, WA 6009, Australia}

\author[0000-0003-0894-1588]{Russell E.\ Ryan, Jr.}
\email{rryan@stsci.edu}
\affiliation{Space Telescope Science Institute,
3700 San Martin Drive, Baltimore, MD 21218, USA}

\author[0000-0002-0648-1699]{Brent M.~Smith}
\email{bsmith18@asu.edu}
\affiliation{School of Earth and Space Exploration, Arizona State University, Tempe, AZ 85287-1404, USA}

\author[0000-0001-7592-7714]{Haojing Yan}
\email{yanhaojing@gmail.com}
\affiliation{Department of Physics and Astronomy, University of Missouri,
Columbia, MO 65211, USA}

\author[orcid=0000-0003-0202-0534]{Cheng Cheng}
\email{chengcheng@bao.ac.cn}
\affiliation{Chinese Academy of Sciences South America Center for Astronomy, National Astronomical Observatories, CAS, Beijing 100101, China} 
\affiliation{Key Laboratory of Optical Astronomy, NAOC, 20A Datun Road, Chaoyang District, Beijing 100101, China}

\author[0000-0001-5769-0821]{Liam Nolan}
\email{liamjn2@illinois.edu}
\affiliation{Department of Astronomy, University of Illinois at Urbana-Champaign, Urbana, IL 61801, USA}

\author[0000-0001-8751-3463]{Heidi B.~Hammel}
\email{hbhammel@aura-astronomy.org}
\affiliation{Association of Universities for Research in Astronomy, 1331 Pennsylvania 
Avenue NW, Suite 1475, Washington, DC 20005, USA}

\author[0000-0001-7694-4129]{Stefanie N.~Milam}
\email{stefanie.n.milam@nasa.gov}
\affiliation{NASA Goddard Space Flight Center, Greenbelt, MD\,20771, USA}

\begin{abstract}
We present the results of AGN--host-galaxy decomposition using JWST/NIRCam, HST/ACS, and HST/WFC3 imaging of the North Ecliptic Pole Time Domain Field (NEP-TDF). The light profiles of 36 NIRCam-selected AGN candidates are modeled for measurement of their point sources, and point source-subtracted host-galaxy emission is used in SED modeling to estimate star-formation rate (SFR)\null. { The star formation rates, stellar masses, and fractional AGN contributions inferred for the NIRCam-selected sample are statistically distinct from those of comparable X-ray- and radio-selected AGN samples. Gaussian mixture modeling of the NIRCam sample in $f_{\rm AGN}$--$ \log_{10}\mathrm{sSFR}$ space favors two partially overlapping components with robust posterior membership assignments, which largely separate galaxies possessing X-ray and/or radio counterparts and bright point-source cores from those without. Based on their offsets from the canonical star-forming main sequence (SFMS), galaxies with bright point-source cores are likely AGN and also show evidence of ongoing or very recent star formation, possibly influenced by AGN feedback. The remaining objects likely either host low-luminosity or obscured AGN that are not prominent in UV--visible--NIR wavelengths or are galaxies with bright, compact cores unrelated to AGN activity. Neither component shows a preference for higher or lower stellar mass or redshift, suggesting that their observed differences correlate more strongly with AGN activity than with stellar mass or redshift.}
\end{abstract}

\correspondingauthor{Gibson B. Bowling}
\email{gbbowlin@asu.edu}

\section{Introduction}
The centers of most galaxies harbor supermassive black holes \citep[SMBHs; \eg][]{Kormendy1995, Magorrian1998, Richstone1998}, and their distinct signatures give rise to observations unique to the nuclear regions of galaxies \citep[\eg][]{Seyfert1943, Burbidge1970, Khachikian1974, Heckman1980, Tran1995a, Tran1995b, Tran1995c}. Accreting SMBHs are the central engines of active galactic nuclei (AGN) that radiate across the entirety of the electromagnetic spectrum \citep[\eg][]{Seyfert1943, FanaroffRiley1974, Elvis1978, Stern2005}. One observational signature of many AGN is a nuclear point source \citep[\eg][]{Schmidt1963, Hazard1963, Kotilainen1992a, Kotilainen1992b, Ramirez2014, Giacconi2001, Dunlop2003, Alexander2003, Lacy2004, Zakamska2006, Gezari2013, Ortiz2024, Payaswini2025} that appears as the telescope aperture's point spread function (PSF) superimposed on the host-galaxy light profile. 

The interactions of AGN and their host galaxies are complex and represent a critical phase in the evolution of massive galaxies \citep[\eg][]{Kauffmann2003, Kewley2006}. AGN are known both to suppress \citep[quench; \eg][]{Silk1998, Fabian1999, Sturm2011, Fabian2012, Laha2021} and enhance \citep[\eg][]{vanBreugel2004, Zinn2013, Nayakshin2014, Kirkpatrick2020, Zhuang2021, Joseph2022, Duncan2023} the star-formation rates (SFRs) of their host galaxies in negative or positive feedback processes, respectively. 

The star formation in galaxies is strongly dependent on mass and redshift \citep[\eg][]{MadauDickinson2014, Bisigello2018}, and so it is useful to contextualize  SFRs with respect to a canonical star-forming main sequence \citep[SFMS;][]{Noeske2007} relating the SFR to stellar mass and redshift calibrated by observations of many star-forming galaxies \citep[\eg][]{Speagle2014, Popesso2023}. The result is a 4-tiered classification scheme for galaxies that includes (1) nominal star formation consistent with the SFMS, (2) starburst (enhanced star formation significantly above the SFMS), (3) transitional (star formation that is below the SFMS and that may be ``ramping down''), and (4) quiescence, which is characterized by star formation significantly below the SFMS or not detectable. In the context of AGN-host studies, quiescence may be taken to mean quenched, though they are not always synonymous. Observationally calibrated models predict that starburst galaxies smoothly become post-starburst galaxies with rapidly decreasing SFR \citep[\eg][]{Whitaker2012, Patel2012, Wang2019}, eventually becoming quiescent. Galaxies are not confined to a single phase during their lifetimes, and this is in part due to AGN.

Historically, photometric classification of IR-bright galaxies has been subject to degeneracy due to the detection bias at IR wavelengths: stars and dust-obscured AGN both shine bright in the NIR \citep[\eg][]{Sajina2005}, and correlation between IR luminosity and starburst activity is common \citep[\eg][]{Farrah2003, Blank2012}. Corrections to AGN selection schemes \citep[\eg][]{Donley2012} and spectral-energy distribution (SED)-fitting routines \citep[\eg][]{Stalevski2012, Stalevski2016} have been developed to address this, but they have been forced by limited angular resolution to use the combined AGN+host emission to derive an AGN fraction and galaxy properties rather than deriving host-galaxy properties from galaxy emission alone.  This makes the fits subject to systematic biases, especially in the NIR and at high redshifts, where the spatial and spectroscopic resolutions have been most limited \citep[\eg][]{Roberts-Borsani2021}, even with flagship observatories \citep[\eg][]{Fazio2004}.

Since becoming operational in 2022, {the James Webb Space Telescope's (JWST) Near-Infrared Camera (NIRCam)} has brought next-generation sensitivity and resolution to the NIR \citep[][]{Gardner2023, Rieke2023}. NIRCam is uniquely suited to resolve the host galaxies of IR-bright AGN and probe their star-forming properties. A recent { JWST} study {\citep[][]{Ortiz2024}  selected AGN on the basis of a central point source and validated that selection via the selected candidates' colors \citep[\eg][]{Stern2012, Assef2013} and by the AGN fractions determined from their SEDs. Visual selection of apparent point-source cores on high-resolution NIRCam images thus selects strong AGN candidates.} Additional confirmation has come from 3~GHz VLA observations \citep{Willner2023,Willner2026} and from VLBA detections \citep{Payaswini2025}. There is no single way to detect all AGN without extensive multi-wavelength coverage, and the infrared offers a critical vantage point in the search for less-luminous AGN that are undetected in the UV--visible \citep[\eg][]{Kocevski2023, Yang2023, Lyu2024, Latif2024, Rieke2025, Bonaventura2026}. 

This study examines the host galaxies of 36 NIRCam-selected AGN candidates { by deriving AGN-subtracted host-galaxy photometry, fitting the hosts' SEDs,} and comparing { the resulting measurements of the hosts'} star-forming properties to those of X-ray- and radio-selected AGN\null.

The paper is organized as follows. Section~\ref{sec:data+cat} discusses the data used in this study and the construction of a sample of NIRCam-selected AGN candidates. Section~\ref{sec:ps-measurement} describes the light-profile fitting of the sample and measurement of point sources. Section~\ref{sec:results+discussion} describes the SED fitting and its results and discusses the star-forming properties of the sample with respect to X-ray- and radio-selected samples. A summary of the results and discussion of future prospects is in Section~\ref{sec:summary}.
All magnitudes are presented in AB units \citep{Oke1983}, and, where necessary, we assume a flat $\Lambda$CDM cosmology with $H_0 = 68$ km s$^{-1}$ Mpc$^{-1}$, $\Omega_\text{matter} = 0.32$, and $T_\text{CMB} = 2.725$ \citep[][]{Planck2016, Planck2020}.

\section{Data \& Sample Selection}
\label{sec:data+cat}
\subsection{Observations}
\label{sec:data}
In a previous study, \citet[][\citetalias{Ortiz2024} henceforth]{Ortiz2024} identified AGN candidates in the JWST North Ecliptic Pole Time Domain Field (NEP-TDF: RA, Decl.\ = 17:22:47.9, +65:49:22 J2000,  $\sim$15\arcmin\ diameter; \citealt{Jansen2018}.) JWST/NIRCam images of the NEP-TDF in the F090W, F115W, F150W, F200W, F277W, F356W, F410M, and F444W filters from the Prime Extragalactic Areas for Reionization and Lensing Science (PEARLS) GTO program \citep[P.I.: R. A. Windhorst, PID 2738][]{Windhorst2023} were used. Images were first processed using the STScI \texttt{CALWEBB} pipeline for basic data reduction{{\footnote{\url{https://github.com/spacetelescope/jwst}}}} {{\citep[][]{Bushouse2023}}}. A customized sky-subtraction and image combination routine (described by \citealt{Windhorst2023}) was then used to generate final mosaics with pixel scales of $0\farcs030~\text{pixel}^{-1}$ and alignment to a GAIA DR3\footnote{\url{https://www.cosmos.esa.int/web/gaia/dr3}} absolute astrometric grid{, following the
techniques described in detail by \citet{Koekemoer2011}}. The $5\sigma$ point-source depth extends down to $\sim$29.5~{mag} \citep{Windhorst2023}.  \citetalias{Ortiz2024} gave further details of observations and reduction. 

Images of the NEP-TDF using the Hubble Space Telescope (HST) Wide Field Camera 3 UVIS channel (WFC3/UVIS) F275W filter and the Advanced Camera for Surveys Wide Field Camera (ACS/WFC) F435W and F606W filters provide additional rest-frame UV--visible constraints on the sample galaxies' star formation. The images were constructed with $0\farcs030~\text{pixel}^{-1}$ pixel scales and are the data products of HST GO 15278 (P.I.: R.\,A. Jansen) and the HST TREASUREHUNT program (GO 16252+16793; PIs: R.\,A. Jansen \& N.\,A. Grogin) with $2\sigma$ limiting magnitudes down to $\sim$29.5~{mag}. \citet{Obrien2024} and R. A. Jansen et al.\ (in prep.)\ give further details regarding observations and reduction.

For measurements of galaxy-integrated flux in JWST and HST bands, the photometry provided by the \citetalias{Ortiz2024} catalog was used. Where available,  spectroscopic redshifts from the MMT Observatory's Binospec and Hectospec instruments  \citep[][and C. N. A. Willmer et al., in prep.]{Zhao2024,Silver2026}  were used. For objects without spectroscopic redshifts, the photometric redshifts from \citetalias{Ortiz2024} were used instead. 

\subsection{AGN- \& QSO-like Object Selection}
\begin{figure*}
    \centering
    \includegraphics[width=\linewidth]{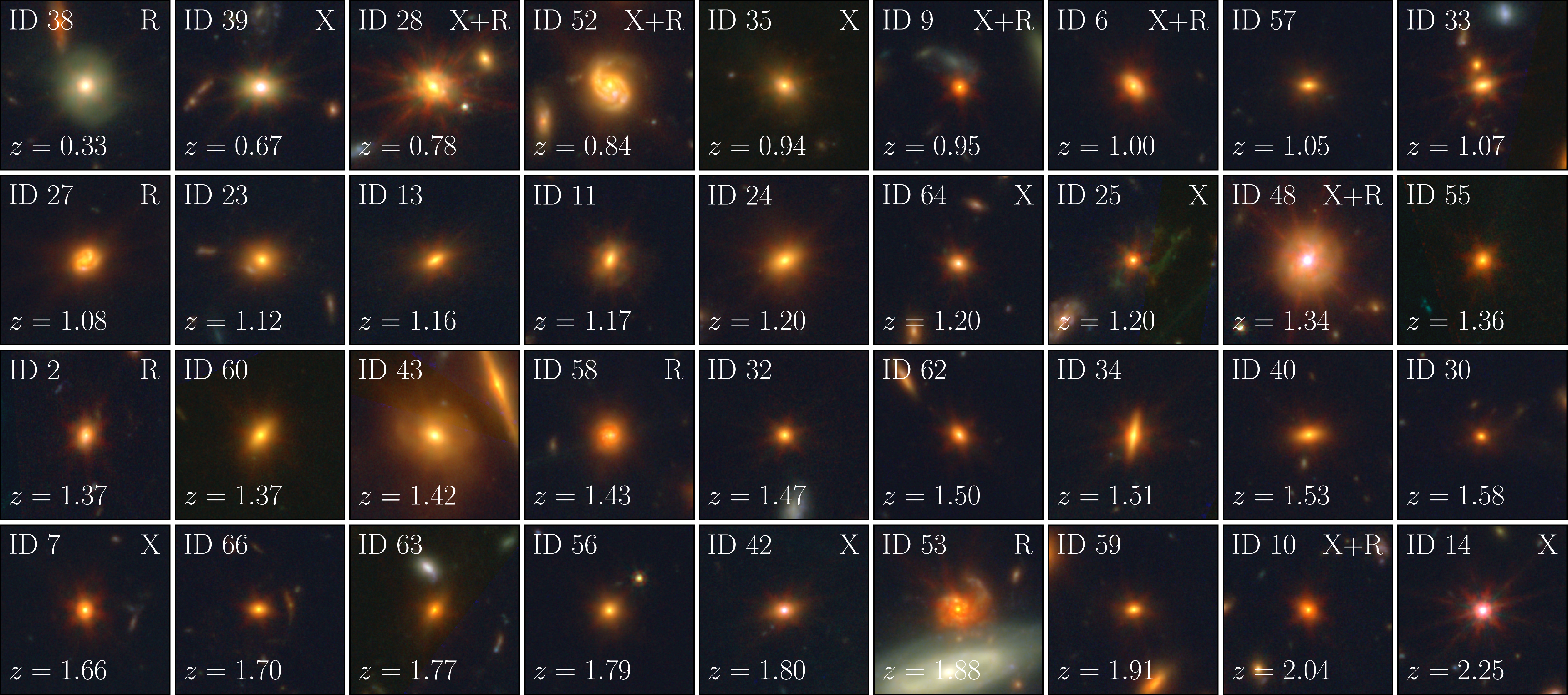}
    \caption{RGB image cutouts of the sample galaxies. Stamps are shown in order of increasing redshift and are $\sim$6\arcsec\ on a side. Each object's redshift and \citetalias{Ortiz2024} ID are shown in the lower and upper left corners of their stamps, respectively. Objects with X-ray counterparts have an X in the top right corner, objects with radio counterparts have an R, and objects with both have X+R. The image RGB colors use the \textsc{Trilogy}\footnote{\url{https://github.com/dancoe/Trilogy}} \citep{Coe2012} prescription from the 11 filters of HST+JWST coverage. The color channels are R = F444W+F410M+F356W, G = F277W+F200W+F150W, and B = F115W+F090W+F606W+F435W+F275W.}
    \label{fig:stamps}
\end{figure*}

\label{sec:sample}
{  \citetalias{Ortiz2024} constructed their sample of central point-source galaxies (CPGs) by visually selecting apparent point-source cores. \citetalias{Ortiz2024} then modeled the objects' light profiles  to  determine whether the core is an unresolved point-source or a compact bulge-like component and used NIRCam photometry to derive colors equivalent to W1$-$W2 \citep{Assef2013}. SEDs were then fit to estimate AGN contributions, stellar masses, and star-formation rates. The resulting catalog contains 66 objects with tabulated photometry, AGN contribution to the SED $f_{\rm AGN}$, W1$-$W2-like colors, and NIR morphological parameters.}

{ For this study,} objects were selected from the \citetalias{Ortiz2024} CPG catalog to have point-source cores or  $f_\text{AGN} \geq 0.3$. {{This was done to produce the highest-fidelity sample of morphologically-selected AGN while including candidates that were potentially misclassified morphologically but have significant AGN contribution to their SEDs}}.
A point-source core was not used as the sole selection criterion because of the uncertainty in the \citetalias{Ortiz2024} morphological classification (e.g., ID 48, considered for this study, was classified as { having} a bulge but has X-ray and radio counterparts { and therefore is a likely AGN})\null. Central point-source features indicate a highly luminous and compact nuclear region suggestive of an AGN, while $f_\text{AGN} \geq 0.3$ additionally suggests a substantial AGN contribution to the SED, which is useful especially in cases where mid-IR or sub-millimeter photometry are not available \citep[\eg][]{Ciesla2015}. Near-infrared (NIR) colors were not considered in the selection because some objects in the \citetalias{Ortiz2024} catalog are not classified as AGN by their colors despite having point-source cores and/or $f_\text{AGN} \geq 0.3$ (e.g., ID 63, which is considered in this study). { In addition, AGN selection by W1$-$W2 color is successful only over a limited redshift range and is sensitive to galaxies' $A_{V}$ and star-formation histories \citepalias[\eg][their Figure 9 for CPGs and CPG-like galaxies specifically]{Ortiz2024}, so not color-selecting  avoids bias that might otherwise exclude some objects.} These criteria select 36/66 CPGs from the \citetalias{Ortiz2024} catalog.\footnote{The F444W fit failed for \citetalias{Ortiz2024} ID~1{, the most obvious point-source core candidate in the NIRCam imaging of the NEP-TDF,} because the image core was saturated in this filter\null. A similar fit in F277W \citep{Willner2026} succeeded in finding a point-source core. This object is nevertheless excluded from our sample because saturation in the three longest-wavelength filters prevents reliable host-galaxy photometry, necessary for the analysis in this work.}

The sample is morphologically diverse, as shown in Figure~\ref{fig:stamps}. It contains both early- (e.g., IDs 39 and 60) and late-type (e.g., IDs 27 and 53) galaxies. Several objects have resolved structure in their hosts (e.g., off-center UV--visible clumps in ID 52 or apparent flocculent and anemic spiral structure in IDs 58 and 48, respectively), though the structures of most of the hosts are outshone by the central point sources. 

\subsection{Comparative X-Ray and Radio Samples}
\label{sec:other-catalogs}
{{AGN are most reliably detected by their X-ray and/or radio emission, and therefore it's useful to compare the stellar masses and star-formation rates of the NIRCam-selected sample to those of X-ray- and radio-selected AGN host galaxies.}} {{For the X-ray part of this comparison,}} we used the {{\citet[][]{Ortiz2026}}}\ catalog of SED parameters of X-ray sources as seen by \citet{Zhao2024} and \citet{Silver2026}. This catalog is the result of extensive multi-wavelength analysis combining NuSTAR and XMM-Newton X-ray fluxes with GALEX \citep{Martin2005}, SDSS \citep{York2000}, DECaLS \citep{Dey2019}, MMT/MMIRS \citep{Willmer2023}, Subaru/HSC \citep{Miyazaki2018}, WISE \citep{Schlafly2019}, HST, and JWST photometry. Objects in this catalog are required to have 4 broadband detections. To prevent comparison with ill-constrained SED results, only objects whose SED had $\chi^2_\nu \leq 10$ are considered in this study. Within this quality cut, we considered only objects with $f_\text{AGN} \geq 0.3$ to ensure that the X-ray-selected sample was comparable to the NIRCam-selected sample.

For similar comparisons to radio-selected objects, we used the \citet{Willner2026} catalog of SED parameters of NIRCam counterparts to radio detections in the field. {{Stellar masses, star formation rates, and SED AGN fractions}} were {{calculated from the}} UV--NIR photometry from HST and JWST {{described in \S\ref{sec:data}}} in combination with VLA 3~GHz fluxes. Only objects with $f_\text{AGN} \geq 0.3$ were selected from this catalog so that the radio-selected sample was comparable to the NIRCam-selected sample.

{{Neither the X-ray nor the radio sample was matched to the NIRCam sample using derived properties other than $f_{\rm AGN}$, but they are appropriate for comparison for two reasons. First, the physical parameters for both were estimated using the same SED-fitting code as the one used in this study (\textsc{CIGALE}, see \S\ref{sec:host-gals-seds}), so systematic differences between SED-fitting prescriptions are not present, and the stellar masses and star-formation rates are directly comparable. In addition, these samples were constructed from deep NIRCam surveys in the NEP-TDF, avoiding bias towards the brightest objects in the field.}}

{{We also consider X-ray and radio detections of galaxies in this paper's sample.}} To identify X-ray counterparts, we used the X-ray catalog from W.\ P.\ Maksym et al.\ (in prep.), which was compiled from Chandra X-ray Observatory (CXO) ACIS-I and ACIS-S observations of the entire field over a net $1.8~\text{Ms}$ exposure time.  { X-ray counterparts to the NIRCam-selected sample are the nearest X-ray detections to NIRCam/F444W detections within $1''$ and were identified by W. P. Maksym et al. (in prep.).} 
To identify radio counterparts, we used the \citet{Hyun2023} 3~GHz catalog  with $1\sigma$ noise of $\sim$1~$\mu\text{Jy}~\text{beam}^{-1}$ based on observations with the NSF's Karl G.\ Jansky Very Large Array (VLA).\footnote{The National Radio Astronomy Observatory is a facility of the National Science Foundation operated under cooperative agreement by Associated Universities, Inc.}
Radio counterparts were identified by \citetalias{Ortiz2024} using the methodology outlined by \citet{Willner2023}: for each 3~GHz source, the nearest NIRCam/F444W source was identified, after which all matches were visually inspected and  corrected where necessary in cases of, e.g., radio lobes, blended galaxies, or other complex morphologies.  { \citet{Willner2026} give a full list of NIRCam counterparts to NEP-TDF radio sources.}

After cross-matching X-ray and radio detections, 13/36 objects have X-ray counterparts, and 11/36 objects have radio counterparts with 6/36 having both. The sample is tabulated in Table~\ref{tab:objects}, where the core type and $f_\text{AGN}$ from \citetalias{Ortiz2024} are listed for reference alongside the X-ray and/or radio counterpart ID(s) if applicable.

\section{Measurement of Nuclear point sources}
\label{sec:ps-measurement}
\subsection{Light-Profile Modeling}
\label{sec:galfit}
A crucial element of morphological AGN selection is  distinguishing nuclear point sources from bright, compact bulges that may appear with point-source features. For galaxies in the NEP-TDF, \citetalias{Ortiz2024} fit model light profiles { with \textsc{Galfit} \citep[][]{Peng2002,Peng2010}} using only the JWST/F444W images rather than the full suite of available HST+JWST filters, and point-source presence was determined only on the basis of best-fit { light-profile} $\chi^2_\nu$ comparison between two models {{including or not including a central point-source component}}. For the purposes of \citetalias{Ortiz2024}, the $\chi^2_\nu$ treatment was sufficient, but using all available broadband imaging best treats the wavelength dependence of the AGN emission within the host galaxy { and is necessary for panchromatic measurement of point-source cores}.

\begin{figure*}
    \centering
    \includegraphics[scale=.56]{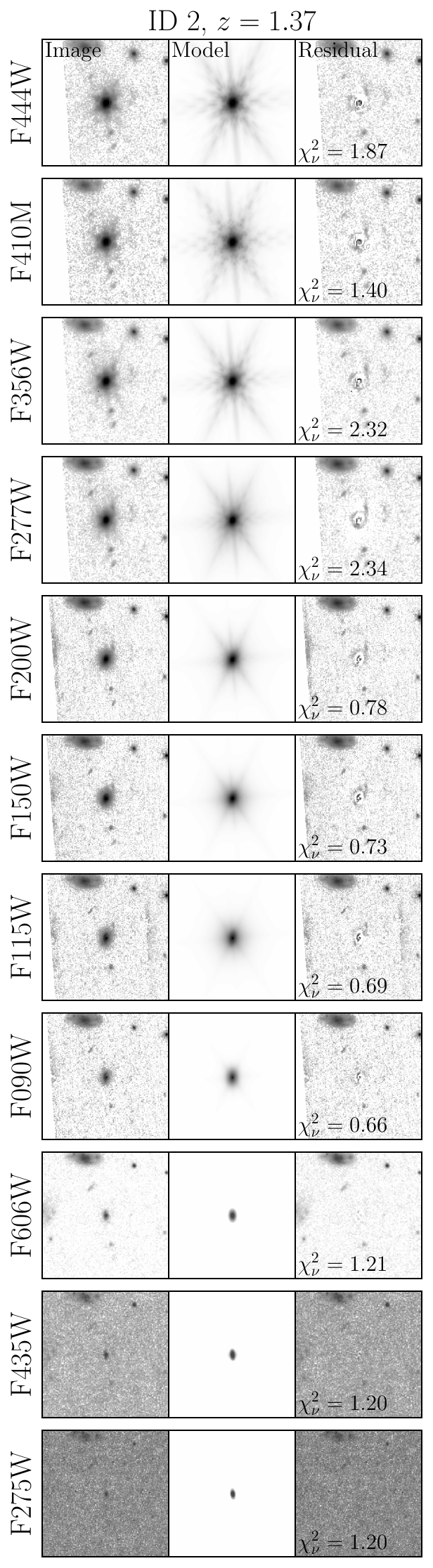}
    \includegraphics[scale=.56]{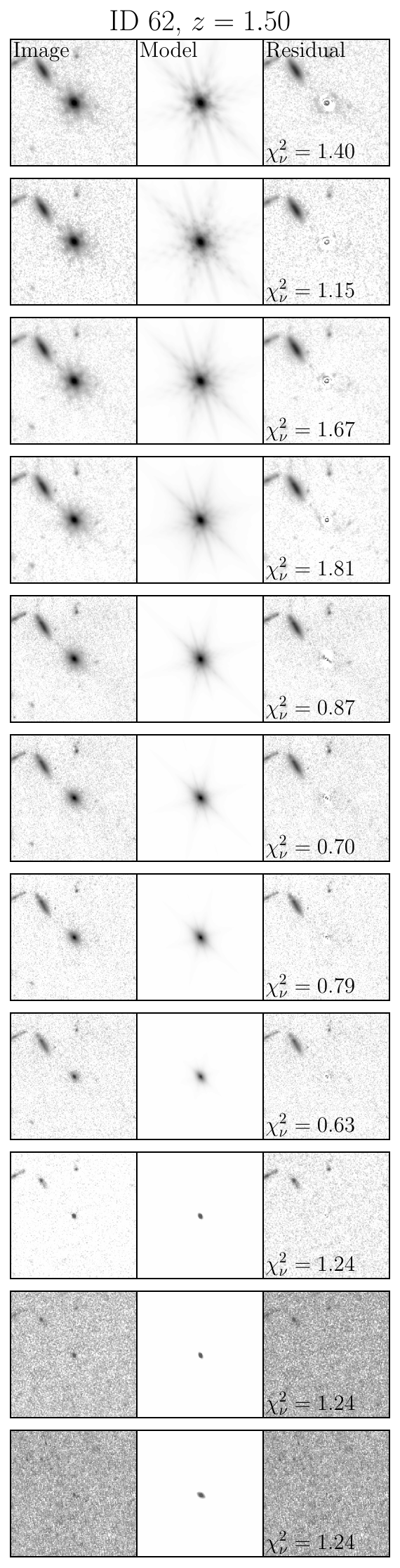}
    \includegraphics[scale=.56]{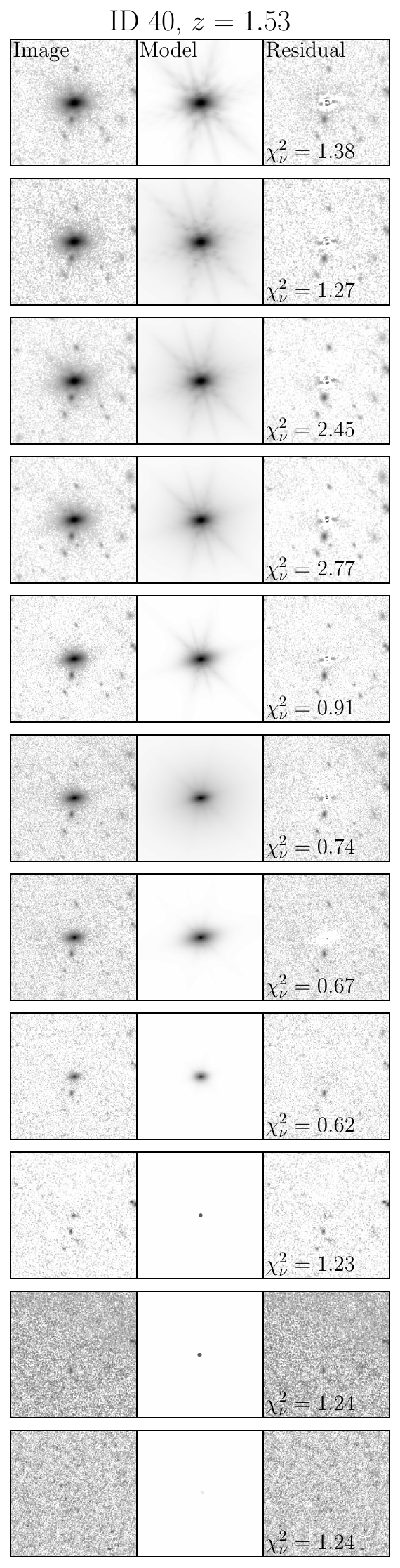}
    \caption{Images of the modeling done for IDs 2 (left), 62 (middle), and 40 (right). Each set of 3 images corresponds to the indicated filter; the leftmost image is the galaxy image, the middle image is the best-fit model light profile found by \textsc{Galfit}, and the right image is the residual. The goodness-of-fit statistic $\chi^2_\nu$ of each is shown in the bottom right of the residual panel of each. All images are $\sim4''$ on a side. Compared to detections in JWST {{filters}}, detections in HST {{filters}} were less frequent.}
    \label{fig:example-fits}
\end{figure*}

For this work, three-component model light profiles were fit to the sample in all filters with detections in the \citetalias[][]{Ortiz2024} catalog. This was done using \textsc{Galfit} version 3.0.5. The model components included one point-source and two S\'ersic \citep{Sersic1963} profiles, conceptually representing a compact bulge and a more extended disk but with a wide range of free parameters. The intention was not a complete representation of the galaxy's structural parameters but only a measurement of the galaxy's central flux to allow an accurate measurement of the central point-source magnitude. { The PSF models used in the fitting were those simulated using \textsc{STPSF} \citep[formerly \textsc{WebbPSF},][]{Perrin2014}, and} the technical \textsc{Galfit} prescription is {{included}} in Appendices~\ref{app:galfit-configuration} and~\ref{app:psfs}\null. The \textsc{Galfit} parameter space was left free enough to iterate towards a physically sensible best-fit solution, and a point-source magnitude was always recovered, even when its flux was negligible compared to the host galaxy. Results of the 11-filter fitting procedure for three objects are shown in Figure \ref{fig:example-fits}, and the best-fit parameters for the JWST filters are listed in Table~\ref{tab:galfit-best-fit-params}.

Figure~\ref{fig:example-fits} shows a smooth increase in the point-source brightness between filters, going from nearly invisible in the HST and short-wavelength NIRCam filters to dominant in the redder NIRCam filters. Using a three-component model for all galaxy images means that differences between types of models cannot produce spurious results concerning the \textit{presence} of a point source. This is especially important for ``model continuity'' purposes, where, e.g., if a point source is present in a JWST/F277W image, one would expect it also to be present in the corresponding JWST/F356W and redder images. This was achieved in the fitting results. 

Occasionally, ``leftover'' flux can be seen in the JWST filter residuals. In many cases, the residual structures appear to be from non-axisymmetric features of the host galaxies (e.g., clumps, spiral structure, dust lanes, etc.) that cannot be accounted for using a strictly S\'ersic-based model. Direct measurement of these features from the residuals with \textsc{SourceExtractor} using the same configuration as \citetalias{Ortiz2024}\footnote{Specifically, both \texttt{DETECT\_THRESH} and \texttt{ANALYSIS\_THRESH} were set to $1.5\sigma$ above the local background. Higher thresholds were tested but resulted in non-detection of any residual structure from the host galaxies.} showed that their mean magnitudes were in the range of $\sim$27--29~mag in F444W, which is magnitudes fainter than the disk or bulge components found by \textsc{Galfit} (see Table \ref{tab:galfit-best-fit-params}) and is therefore not problematic \textit{for measuring point-source magnitudes}. 

\subsection{Uncertainty in the Point-Source Magnitude}
\label{sec:PS-mag-uncertainties}
The purpose of modeling a galaxy's light profile is to gain some insight into what its ``true'' light profile may be.
{{Per \S2.2 of \cite{Peng2002}, \textsc{Galfit}'s model-parameter uncertainties may not be completely accurate. \citet{Haussler2007} stress-tested \textsc{Galfit} and showed that model-parameter uncertainties must be measured explicitly in cases where the parameter space is nonlinear or may have local minima, as is likely here.}}

For this work, model-parameter uncertainties were determined by a method adapted from \S5.2 of \citet[][]{vanderWel2012}. \textsc{Galfit} was used to simulate images of each sample host galaxy in every filter using the best-fit parameters of their S\'ersic components. A point-source was added at $\pm0.5~\text{mag}$ from its best-fit magnitude in steps of $0.1~\text{mag}$, as \citet[][]{Gabor2009} showed that point-source magnitudes are recoverable in this fashion within $\pm0.5~\text{mag}$. Noise from empty parts of the science mosaics was added to each simulated image, resulting in 11 noisy simulated images, each with a known point-source magnitude. \textsc{Galfit} was run on all simulated images using the same configuration as described in Appendix \ref{app:galfit-configuration}, and
\begin{equation}
    \label{eqn:deltaM}
    \Delta m \equiv m'-m
\end{equation}
was measured for every fitting result using the ``recovered'' and ``true'' point-source magnitudes $m'$ and $m$. Then, for every filter, the mean of $\Delta m$ was computed as
\begin{equation}
    \label{eq:mean}
    \langle\Delta m\rangle = \frac{1}{N}\sum_{k=1}^{N} \Delta m_k
\end{equation}
for the number of convergent simulated image fitting runs $N$\null. For most objects, this was all 11 runs in each filter with a detection. For the entire sample, the mean and median of ${\Delta m}$ were $-0.08~\text{mag}$ and $-0.06~\text{mag}$, respectively, indicating that  \S\ref{sec:galfit}'s procedure for measuring point-source magnitudes  is robust.

The point-source magnitude uncertainty in each filter for each object was computed as the population standard deviation of the distribution of $\Delta m_k$,
\begin{equation}
    \label{eqn:ps-mag-err}
    \delta m_\text{PS} \equiv \left[\frac{1}{N}\sum_{k=1}^N \left( \Delta m_k - \langle\Delta m\rangle \right)^2\right]^{1/2}.
\end{equation}

\begin{figure}[t]
    \centering
    \includegraphics[width=\linewidth]{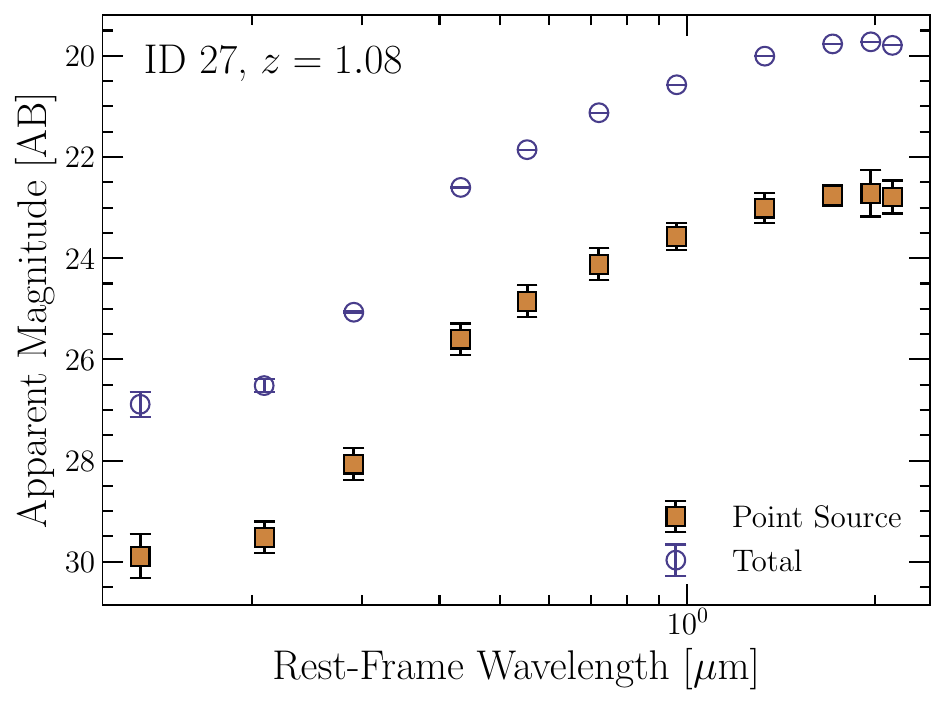}
    \caption{The UV--visible--NIR point source magnitude as a function of rest-frame wavelength for ID 27. Point-source magnitudes are drawn in gold with error bars from the procedure of \S\ref{sec:PS-mag-uncertainties}, and the total magnitudes and error bars from \citetalias{Ortiz2024} are drawn in blue. Their uncertainties are very small in the NIRCam filters, generally significant only  for faint objects in HST filters. The fitting procedure outlined in~\S\ref{sec:galfit} and Appendix~\ref{app:galfit-configuration} is capable of determining point-source magnitudes down to $\sim$30~mag, where they are unimportant compared to host-galaxy emission.}
    \label{fig:psf-mags}
\end{figure}

An example of the point-source measurement results is shown in Figure \ref{fig:psf-mags}. For most objects, the shape of the point-source magnitudes are consistent with the SED shapes of obscured AGN over the wavelength where power-law-like emission ``ramps up'' \citep[see, \eg][and the references therein]{Treister2004, Hatziminaoglou2009, Pozzi2010, Stalevski2012, Stalevski2016, Yang2023}. Because many objects in the \citetalias{Ortiz2024} catalog are probably obscured AGN, this is expected. 

Photometry of the host galaxies was computed using the \citetalias{Ortiz2024} catalog photometry and the point-source measurements from \S\ref{sec:ps-measurement}. Host-galaxy flux densities {{$S_{\rm host}$}} in mJy were computed using
\begin{equation}
    \label{eqn:host-gal-flux}
    S_\text{host} = \zeta\times\left( 10^{-0.4 m_\text{cat}} - 10^{-0.4 m_\text{PS}} \right),
\end{equation}
where $\zeta$ is the constant $ 3.631 \times 10^6~\text{mJy}$, $m_\text{cat}$ is the galaxy-integrated \textsc{SourceExtractor} magnitude from \citetalias{Ortiz2024} and $m_{\text{PS}}$ is the point-source magnitude from \S\ref{sec:galfit}{{, both in AB mag}}. Flux density uncertainties were computed as
\begin{equation}
    \label{eqn:host-gal-flux-err}
    \delta S_\text{host} = \zeta^\prime \times  \left(10^{-0.8 m_{\rm cat}}\delta m_{\rm cat}^2 + 10^{-0.8 m_\text{PS}}\delta m_\text{PS}^2\right)^{1/2}
\end{equation}
for $\delta m_{\rm cat}$ from \citetalias{Ortiz2024} photometry,\footnote{Photometric uncertainties in the \citetalias{Ortiz2024} photometry are not published, but they are the \textsc{SourceExtractor} \texttt{MAGERR\_AUTO} uncertainties on the published measurements.} $\delta m_\text{PS}$ from \S\ref{sec:PS-mag-uncertainties}, and the constant $\zeta^\prime = \ln 10^{0.4\zeta} = 3.344\times 10^6~\text{mJy}$. {{The point sources in HST filters' models were often so faint ($m_{\rm PS,HST} \lesssim 28~$AB) that the host-galaxy flux changed only by small amounts by point-source subtraction.}} 
The catalog flux densities capture irregular, non-axisymmetric features of the galaxies, and the host-galaxy flux densities depend on models only insofar as the models determine the galaxies' central magnitudes and thus their point-source magnitudes. 

\begin{figure*}
    \centering
    \includegraphics[width=\linewidth]{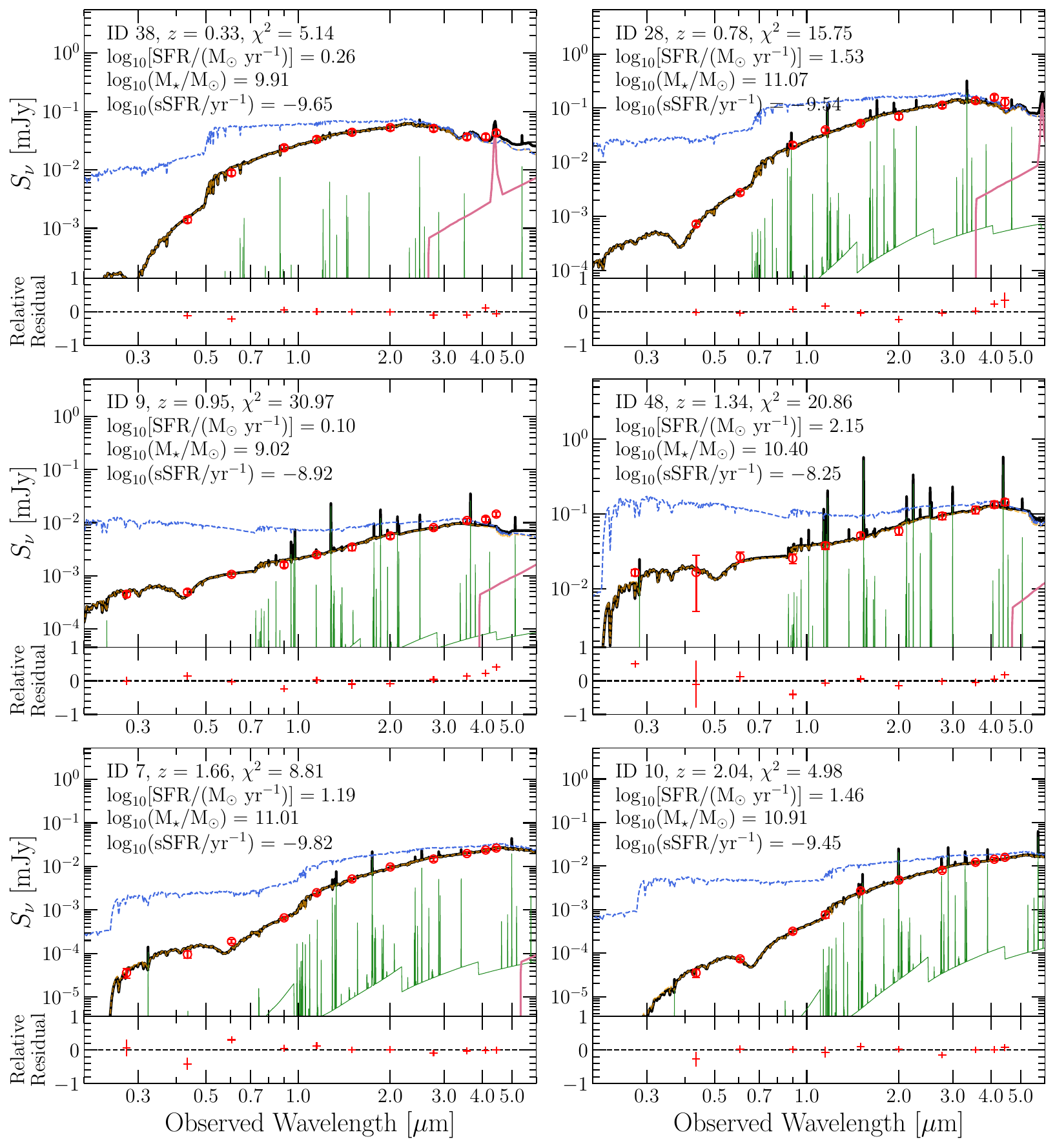}
    \caption{Best-fit host-galaxy SEDs from \textsc{CIGALE}\null. The redshift, $\chi^2$ of the SED fit, and logarithms of the SFR, stellar mass, and $\text{sSFR} \equiv \text{SFR}/\text{M}_\star$ are shown in the upper left of each panel. Host-galaxy flux densities are plotted in red with uncertainties computed via Equation~\eqref{eqn:host-gal-flux-err}, and model SEDs are drawn in black. Nebular emission is drawn in green, stellar attenuated and unattenuated are drawn in orange and blue, respectively, and dust emission, if present, is drawn in pink. The objects shown are representative of the redshift range of the sample, $z\sim0.3$--2 and are ordered by increasing redshift (down and right is higher). All those shown had detections in all 11 filters except for IDs 10, 28, and 38, which are non-detections in HST/F275W.
    }
    \label{fig:seds}
\end{figure*}

\section{Results \& Discussion}
\label{sec:results+discussion}
\subsection{Host Galaxy SEDs \& Parameter Estimation}
\label{sec:host-gals-seds}
We used the energy balance SED-fitting code \textsc{CIGALE}\footnote{\url{https://cigale.lam.fr/}} \citep{Boquien2019, Yang2020, Yang2022, Burgarella2025} version 2025.1 to fit SEDs {{to the point-source-subtracted photometry}} and estimate the {{stellar masses and star-formation rates of the host galaxies}}.
%
Table~\ref{tab:cigale-parameters} describes the \textsc{CIGALE} configuration. Five uniquely-seeded stochastic star-formation histories (SFHs)\footnote{The \citet{CarvajalBohorquez2025} SFHs are based on JWST observations of galaxies with $6 \lesssim z \lesssim 12$. However, stochastic star formation is evident at lower redshifts \citep[\eg][]{Papovich2004,Kim2019}.} from \citet[][]{CarvajalBohorquez2025} were considered for each object, with a delayed-$\tau$ SFH
as a baseline (\texttt{sfhstochastic\_carvajal2025} module). A \citet[][]{Bruzual2003} simple stellar population with solar metallicity with a \citet[][]{Chabrier2003} initial mass function (IMF) was chosen (\texttt{bc03} module). Nebular lines were computed according to the \textsc{CLOUDY}-based \citet[][]{Inoue2011} templates, and nebular continua including free-free, free-bound, and two-photon processes were modeled via the \texttt{nebular} module. Dust attenuation was modeled using a \citet[][]{Calzetti2000} law, modified to account for potential starburst or quenching activity (\texttt{dustatt\_modified\_starburst} module). Because the photometry extends into the NIR, dust emission was accounted for by the \citet[][]{Dale2014} templates (\texttt{dale2014} module) with AGN fraction set to zero. Redshift was not left as a free parameter in fitting; where available, spectroscopically-confirmed redshifts were given to \textsc{CIGALE}. Otherwise, the photometric redshifts from the \citetalias{Ortiz2024} catalog were used. This was done in the \texttt{redshifting} module, which also includes intergalactic medium (IGM) transmission from \citet[][]{Meiksin2006}.

\textsc{CIGALE} was configured to add an additional 10\% uncertainty to all input flux densities to account for any uncertainties that were not considered or propagated in \S\ref{sec:PS-mag-uncertainties}. The results found by \textsc{CIGALE}'s \texttt{bayes} routine were used to infer physical parameters as they are less model-dependent than the best-fit (by SED $\chi^2_\nu$) results. {{The \texttt{bayes} results are likelihood-weighted means from parameter posterior probability density functions (PDFs), and the corresponding uncertainties are the 1$\sigma$ spread in the PDFs.}}

\begin{figure}
	\centering
	\includegraphics[width=\linewidth]{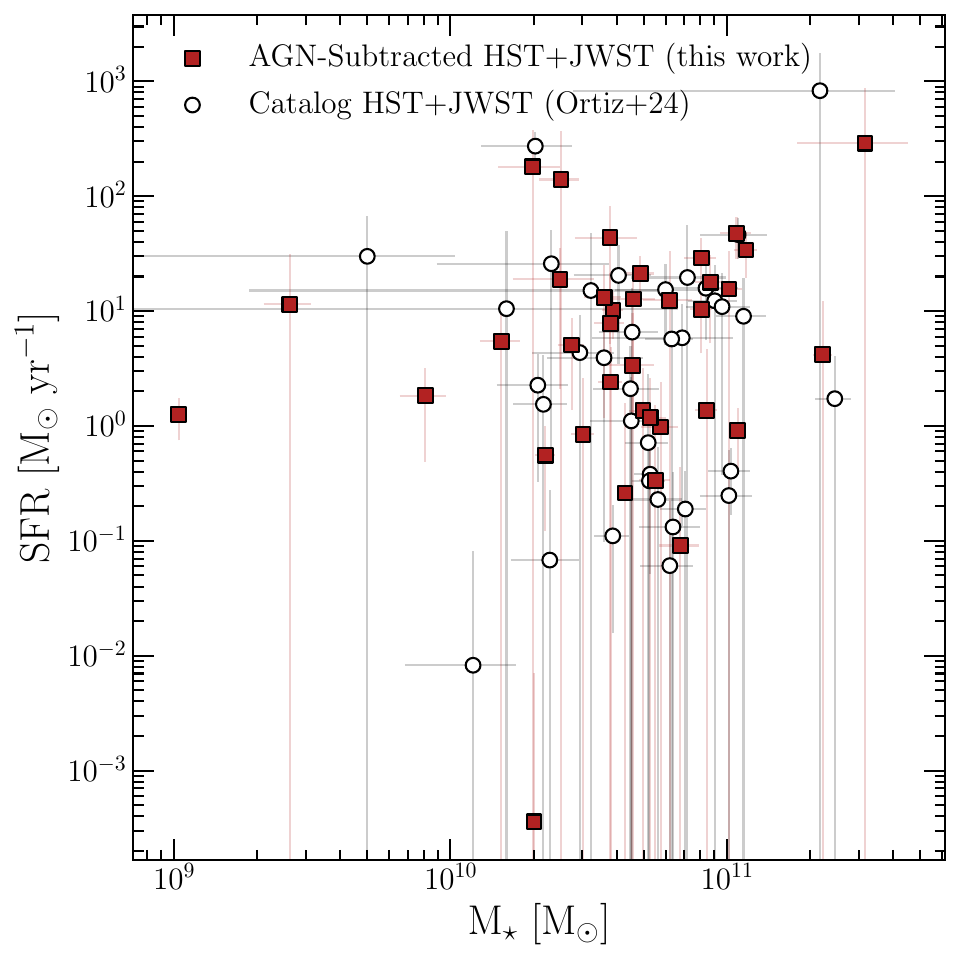}
	\caption{{{Star-formation rates and stellar masses for the NIRCam-selected sample. Red squares show values derived in this study, and white circles show those from  \citetalias{Ortiz2024}. Uncertainties shown are those from \textsc{CIGALE}'s \texttt{bayes} routine. 
    }}}
	\label{fig:SFR-compare}
\end{figure}

Even though the SEDs are informed by UV--visible measurements from HST, the 11-band photometric coverage is simply not enough to constrain SFH in detail. However, as delayed-$\tau$ SFHs are generally flexible \citep[see, \eg][and references therein]{Carnall2019, Leja2019}, and  stochastic feedback mechanisms may be at play in AGN--host systems, a stochastic SFH constructed around a baseline delayed-$\tau$ trend is sufficient for the purposes of this study (i.e., it does not restrict the possibility of starburst or quenching behaviors while still following the expected delayed-$\tau$ behavior).
Even though the {{specific shape of the}} SFH cannot be inferred from the 11-filter photometry alone,
a stochastic SFH with a delayed-$\tau$ baseline allows appropriately for the star formation preferred by the data. {{The SFR is sensitive to dust, however, and without far-IR (FIR) data to constrain dust luminosity, the uncertainties in SFR are large, as shown as Figures~\ref{fig:SFR-compare} and \ref{fig:sfms-offsets}. Fortunately, ongoing star formation characteristically emits in the UV, which, while sensitive to dust, is directly constrained by the HST filters for many objects. With the available input data, \textsc{CIGALE}'s calculated uncertainties should therefore encompass the allowed SFR range.}}

The SED fitting results' $\chi^2_\nu$ distribution has median, mean, and standard deviation of 0.90, 1.66, and 2.33, respectively. Example SEDs of objects whose redshifts trace the sample's redshift range are shown in Figure \ref{fig:seds}. Their shapes are typical of galaxies whose emission is dominated by stellar continuum \citep[\eg][and references therein]{Iyer2026}, as expected for galaxies whose AGN emission has been removed.

{{Because star-formation rates are a particular focus of this study,  Figure~\ref{fig:SFR-compare} shows how our SFRs compare to the \citetalias{Ortiz2024} SFRs derived from the host+AGN photometry.\footnote{These are unpublished, but they are the \texttt{bayes} results from the \citetalias{Ortiz2024} \textsc{CIGALE} output.}  
Our derived stellar masses are in close agreement with those of  \citetalias{Ortiz2024}, exhibiting a median offset of only $-0.05$~dex and MAD of $0.06$~dex. In contrast, the inferred star formation rates are systematically higher than \citetalias{Ortiz2024}'s by a median of 0.34~dex, indicating that the AGN subtraction primarily affects the recovered host-galaxy star-formation rates while leaving stellar mass estimates largely unchanged. Because both sets of SFR measurements are directly constrained in the UV--visible---a wavelength range over which emission from active star formation and AGN emission are degenerate in SED modeling---and ours are  derived from AGN-subtracted photometry, the SFRs in this work may be more reliable than those of \citetalias{Ortiz2024}. Because the SFR uncertainties in both cases are comparably large, offsets in central values need not indicate vastly different SFRs.}}

{{An independent measure of SFR \textit{could} in principle be derived from emission line fluxes \citep[\eg][]{BPT1981, Kennicutt1998}. However, aside from the problem of subtracting the AGN contribution, the spectroscopic data are from visible--NIR spectrographs (\S\ref{sec:data}) that do not contain the necessary lines in their wavelength ranges given the sample galaxies' redshifts---most notably H$\alpha$---to be useful for this, and the rest-frame blue or UV lines are subject to unknown extinction. The SFRs from \citet[][]{Willner2026}, however, offer a first-order consistency check on those in this study because they were derived from photometry including 3~GHz radio fluxes. For the 11/36 galaxies with counterparts in this band, the SFRs from AGN-subtracted photometry are broadly consistent with those of \citet[][]{Willner2026}, although they are systematically lower by $\sim$$0.20$~dex. The general agreement between the two SFR estimates suggests that the AGN-subtracted \textsc{CIGALE} fits recover plausible host-galaxy SFRs, though the small sample of 3~GHz counterparts limits the power of this comparison. That the 3~GHz SFRs are slightly higher than the ones derived in this study is consistent with additional AGN-related radio emission contributing to the observed 3~GHz flux, thus potentially biasing the SFR inferred from the 3~GHz fluxes.
All in all, the comparisons with the \citetalias{Ortiz2024} and \citet{Willner2026} SFRs suggest that the SFRs derived in this study are reasonable within their large uncertainties.}}

\subsection{Host Galaxy Star Formation}
\label{sec:canonical-SF}
Star formation in the Universe reached its peak at $z\sim2$ and has been on the decline since \citep[][]{MadauDickinson2014}, but  galaxies exhibit a variety of star-forming behaviors at all redshifts. Observations have parameterized  dependence on stellar mass and redshift with a canonical star-forming main sequence \citep[SFMS,][]{Noeske2007}, and galaxies' SFRs are compared by computing offsets,
\begin{equation}
    \label{eqn:sfms-offset-def}
    \Delta\rm SFMS \equiv \log_{10}\left(\rm SFR / SFR_{MS}\right)
\end{equation}
for main-sequence star-formation rate $\rm SFR_{MS}$.

\begin{figure*}
    \centering
    \includegraphics[scale=0.53]{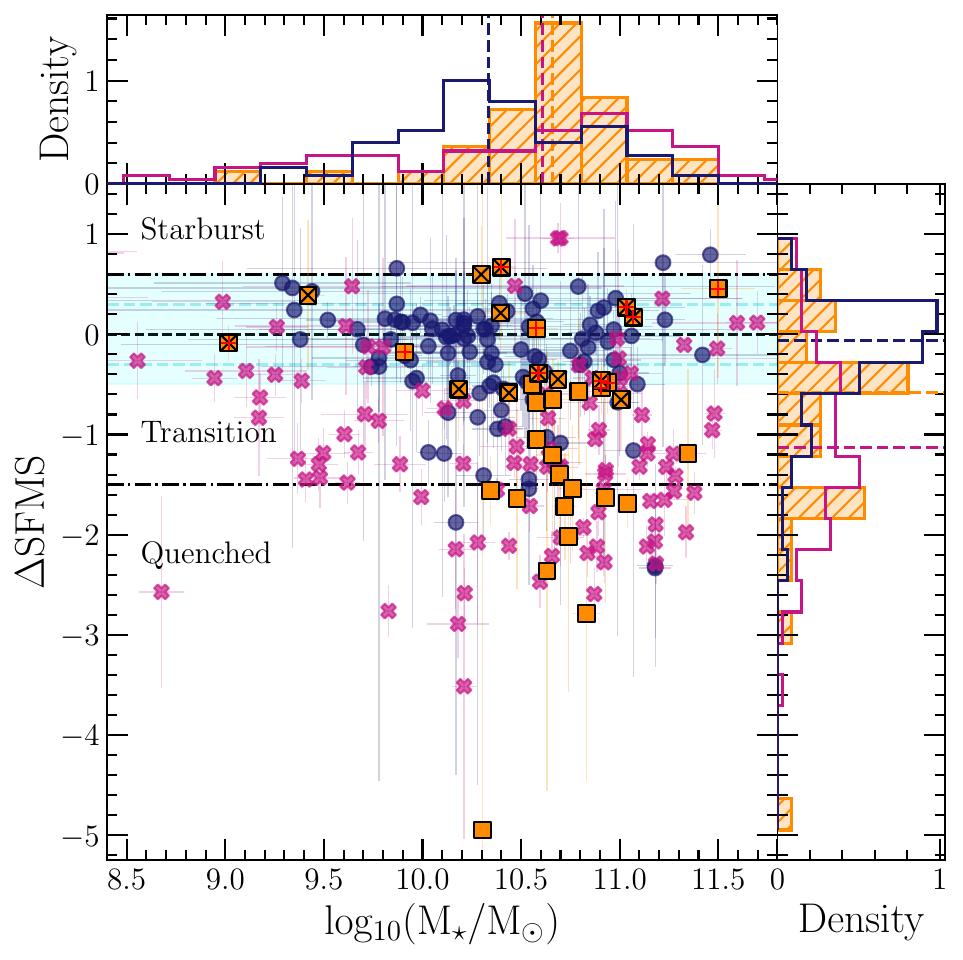}
    \includegraphics[scale=0.53]{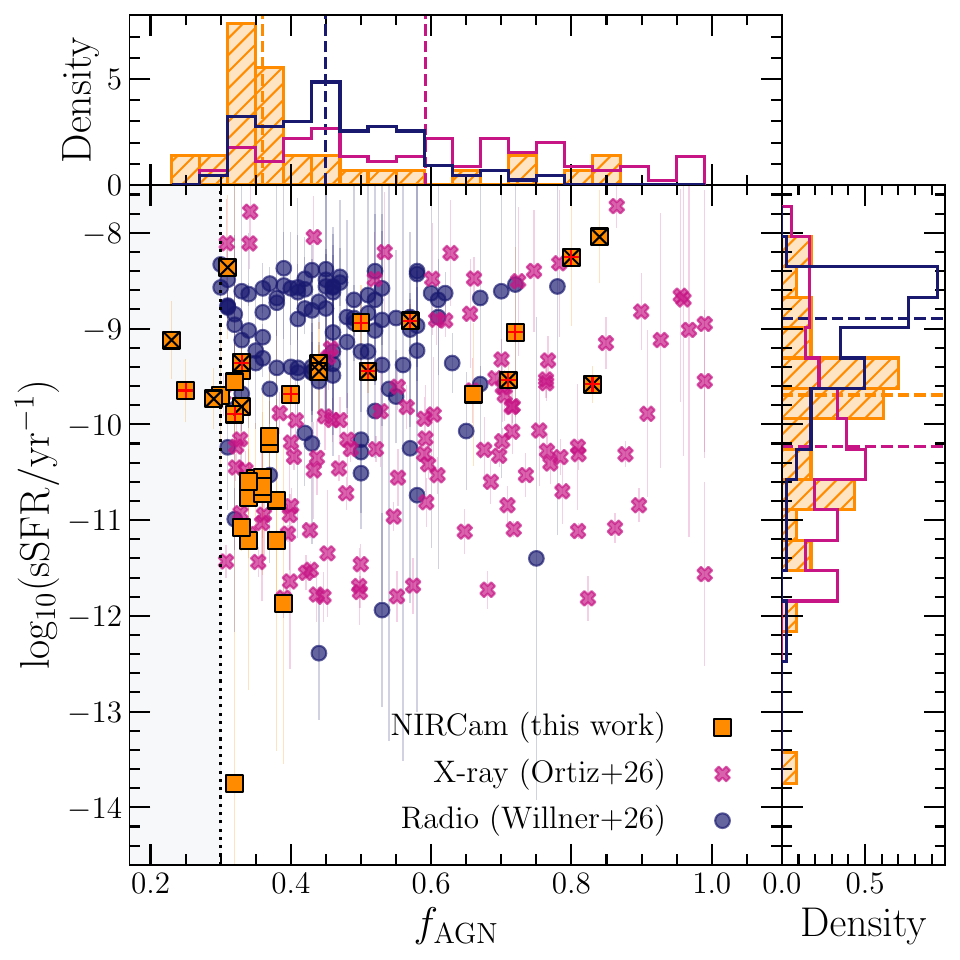}
\caption{Star-forming properties of the sample. This study's NIRCam-selected objects are plotted as orange squares, and the X-ray- {{\citep[][]{Ortiz2026}}}\  and radio-selected \citep[][]{Willner2026} objects  are plotted as pink Xs and blue circles, respectively. {{Uncertainties  shown are those calculated by \textsc{CIGALE}'s \texttt{bayes} routine. The mean  $f_{\rm AGN}$ uncertainties for radio and X-ray samples are 0.26 and 0.11, respectively, extending up to $\sim$0.35.  Individual uncertainties are not shown to avoid cluttering the plot.}} NIRCam-selected objects with an X-ray counterpart (W. P. Maksym et al.\ in prep.)\ are plotted with a black $\times$; those with a radio counterpart \citep[][]{Hyun2023} are plotted with a red $+$. Objects with both X-ray and radio counterparts are plotted with both symbols. The central plots are flanked by two normalized histograms of the quantities that their axes are parallel to. The distribution of each quantity is drawn in its sample's respective color, and{{, to guide the eye,}} the median values {{of the distributions for each sample}} are marked with colored dashed lines. \textit{Left panel:} Star-forming main sequence offsets using SFR$_\text{MS}$ from \citet[][]{Popesso2023} as a function of logarithmic stellar mass. The cyan-shaded region marks the extent of the SFMS with the upper boundary separating it from a starburst region \citep[\eg][]{Rodighiero2011} and the lower boundary separating it from the transition region between main-sequence and quenched \citep[][]{Renzini2015}. The $\pm0.3~\text{dex}$ region is bounded by darker dashed cyan lines in the cyan-shaded region. \textit{Right panel:} Logarithmic {{$\rm sSFR \equiv SFR / M_\star$}} as a function of $f_\text{AGN}$ (from \citetalias{Ortiz2024} for the NIRCam sample). A vertical dotted line marks $f_\text{AGN}= 0.3$. Three NIRCam-selected objects do not satisfy this cut but  have point-source cores  \citepalias[][\S\ref{sec:sample}]{Ortiz2024}. The outlying NIRCam-selected object well below the others is ID~30 at $z=1.58$, a non-detection in HST/F275W {{with a corresponding large uncertainty in its SFR}}.
}
    \label{fig:sfms-offsets}
\end{figure*}

The left panel of Figure \ref{fig:sfms-offsets} shows  $\Delta\text{SFMS}$ from the \citet{Popesso2023} SFMS for the NIRCam-selected objects in this study and for X-ray- and radio-selected objects from {{\citet[][]{Ortiz2026}}}\ and \citet[][]{Willner2026}, respectively.
{{The \citet{Popesso2023} SFMS was calibrated using both spectroscopic (e.g., H$\alpha$) and photometric (e.g., SED-derived) SFR estimates for galaxies at $0\lesssim z \lesssim 6$ with corrections applied to normalize them to each other. Many of the calibrators used SFR estimates from galaxy-integrated photometry, and the NIRCam sample's galaxies are within \citeauthor{Popesso2023}'s redshift range, making their SFMS an appropriate comparison for the SFRs derived in \S\ref{sec:host-gals-seds}.}}

The medians of the radio-selected, NIRCam-selected, and X-ray-selected objects fall at SFMS offsets of $-0.06$~dex, $-0.58$~dex, and $-1.13$~dex, respectively. { 
Pairwise Anderson--Darling tests \citep[][]{AD1952, AD1954} were performed on the unbinned $\rm \Delta SFMS$ distributions of all three samples and showed that the NIRCam-selected sample exhibits a statistically significant difference relative to both the X-ray- $(p=0.035)$ and radio-selected $(p=0.001)$ samples.
The X-ray- and radio-selected samples also differ $(p=0.001)$. Similar tests on the unbinned stellar mass give $p=0.078$ for NIRCam--X-ray and $p=0.001$ for NIRCam--radio and X-ray--radio. AGN fraction gives $p=0.001$ for all, and $\rm \log_{10} sSFR$ gives $p=0.042$ for NIRCam--X-ray and $p=0.001$ for NIRCam--radio and X-ray--radio distributions. In summary, all cases except for the NIRCam--X-ray stellar masses have statistically distinct distributions. Given that many of both the NIRCam- and X-ray-selected objects lie in the transition region of the left panel of Figure~\ref{fig:sfms-offsets}, the similarity between their stellar mass distributions is not surprising.
}

{ While the Anderson--Darling tests quantify differences in the one-dimensional distributions, they do not account for joint dependence on stellar mass. We performed 
pairwise energy-distance tests \citep[\eg][\S10.3]{StatCompR} on the joint $(\log_{10}M_\star, \rm \Delta SFMS)$ distributions. 
These evaluated the statistical distance between distributions by comparing between-sample and within-sample separations using an $L_1$ metric instead of the typical $L_2$ Euclidean one.
The joint distributions differ significantly for all three pairwise comparisons ($p=0.027$, $p=1.5\times10^{-5}$, and $p<10^{-6}$ for NIRCam--X-ray, NIRCam--radio, and X-ray--radio, respectively). For the joint $(f_{\rm AGN}, \log_{10}\mathrm{sSFR})$ distributions in the right panel of Figure~\ref{fig:sfms-offsets}, similar tests show statistically significant differences between all three samples ($p=0.013$ for NIRCam--X-ray and $p \lesssim 10^{-6}$ for NIRCam--radio and X-ray--radio). Thus the star-forming properties of the NIRCam-selected objects are distinct from those of the X-ray- and radio-selected objects.}

Simulations predict that inflows of gas from the pristine reservoirs of the circumgalactic medium (CGM) can trigger nuclear starbursts and rapid growth of the central SMBH \citep[\eg][]{DiMatteo2005, Hopkins2006, Dekel2014, Zolotov2015}. The AGN then grows in a heavily-obscured, IR-bright phase powered by CGM inflows \citep[\eg][]{Hopkins2009}, which is consistent with the IR-luminous galaxy-quasar evolutionary picture of \citet[][]{Sanders1998}. Negative AGN feedback can subsequently quench host star formation, giving rise to a population of galaxies with bright nuclear regions and on their way to quiescence. In this framework, it is possible that dusty NIRCam-selected AGN trace a transitional phase of AGN growth between short-timescale X-ray AGN activity \citep[\eg][]{Schawinski2015} and longer-lived radio AGN activity \citep[\eg][]{Bird2008}. 

The NIRCam-selected objects with X-ray and/or radio counterparts don't fall into the same parts of the parameter space as the X-ray- or radio-selected objects { that lack NIRCam point-source morphology}. The former are { mostly} contained within $\pm0.3$--0.6~{dex} of $\Delta\text{SFMS}=0$; there are 10 { of these objects within this range}, and only 6 within $\pm0.3$~{dex} of $\Delta\text{SFMS}=0$. Most ordinary galaxies have $|\Delta\rm SFMS|\le0.3$  \citep[\eg][]{Daddi2007, Pannella2009}, but many of these sources do not. 

The only starburst object in the NIRCam sample is ID 23 at $z=1.12$, which has both radio and X-ray counterparts. Its apparent morphology is, from Figure \ref{fig:stamps},  early-type,  consistent with its S\'ersic indices from \textsc{Galfit}\null. It is possible that ID 23 is a compact radio galaxy, similar to objects found by \citet[\eg][]{Windhorst1991, Windhorst1992}, but the detection of X-ray flux suggests current  SMBH accretion activity \citep[\eg][]{Alexander2012, Alexander2025}, which points to simultaneous AGN activity and star formation. \citetalias{Ortiz2024} (their Figure 7) found that 10/66 CPGs were starburst, 3 of which were shown to have point-source cores and were thus selected for this study; this fraction is much higher than the  single starburst object found in the NIRCam sample (Figure~\ref{fig:sfms-offsets} left).
The difference does not arise from the different SFMS prescriptions (\citealt{Speagle2014} by \citetalias{Ortiz2024}, \citealt{Popesso2023} here) because they produce similar results over the redshift range of our sample. Because Figure~\ref{fig:sfms-offsets} shows host-galaxy properties with minimal AGN contamination,  the \citetalias{Ortiz2024} SEDs appear to have been biased by rest UV--visible AGN emission. The likely scenario is that a fraction of the total emission in these wavelengths came from an AGN, but \textsc{CIGALE} interpreted it as an excess of flux from young stars, and thus the SFRs were biased upward\footnote{{While \S\ref{sec:host-gals-seds} reported that the {\it median} SFRs of all 36 objects increased relative to \citetalias{Ortiz2024},  for such small sample sizes the median is not necessarily representative of bulk behavior. Moreover, the median  is  insensitive to outlier behavior, which starburst objects are.}} \citep[as expected from, e.g.,][]{ErrozFerrer2013}.

Given that galaxies' stellar masses and SFRs are  strongly correlated \citep[\eg][]{Bisigello2018}, it's useful to consider the sSFR. This is done in the right panel of Figure \ref{fig:sfms-offsets} with respect to the $f_\text{AGN}$ for the samples. The three NIRCam-selected objects with $f_\text{AGN} \leq 0.3$,   selected on the basis of their point-source cores, are either X-ray or radio-detected, suggestive of NIRCam's ability to find \textit{bona fide} AGN by their nuclear point sources in the reddest filters. The median $f_\text{AGN}$ for the NIRCam-selected objects is $0.35$, which is almost certainly an underestimate. Many objects in the \citetalias{Ortiz2024} catalog have lower SED AGN fractions, indicating that the sample is biased to have a lower $f_{\rm AGN}$. 

The sSFRs have similar trends as the SFMS offsets: the median of the NIRCam-selected objects' sSFR is nearly equidistant from those of the X-ray and radio samples, with the X-ray median being lower and the radio median being higher. The NIRCam-selected sources with X-ray and/or radio counterparts are clustered around the radio sample's median sSFR. That the two panels' star-forming patterns are similar suggests that { the NIRCam-selected sample's shape in this parameter space} is a feature of the data, not just an artifact of the SFMS or the assumptions it makes.

\subsection{{Star-Forming Properties of the NIRCam-Selected Sample}}
\label{sec:substructure}
{ The joint $(f_{\rm AGN},\log_{10}\mathrm{sSFR})$ distribution shown in the right panel of Figure~\ref{fig:sfms-offsets} suggests the presence of internal structure within the NIRCam-selected sample, with one concentration characterized by higher sSFRs and another by lower AGN fractions. To determine whether this apparent structure emerges from multimodality in either quantity, we performed Hartigan--Hartigan dip tests \citep{HartiganDip} on the unbinned $f_{\rm AGN}$ and $\log_{10} \rm sSFR$ distributions and found that neither shows evidence of multimodality  ($p=0.936$ for $f_{\rm AGN}$ and $p=0.288$ for $\log_{10}\rm sSFR$). Similar tests conducted on the $\log_{10} M_\star$ and $\rm \Delta SFMS$ distributions of the NIRCam-selected sample gave the same conclusions ($p=0.952$ for $\log_{10} M_\star$ and $p=0.407$ for $\Delta \rm SFMS$), and thus there is no evidence for multiple subgroups in the one-dimensional distributions alone.}

\begin{figure}
    \centering
    \includegraphics[width=\linewidth]{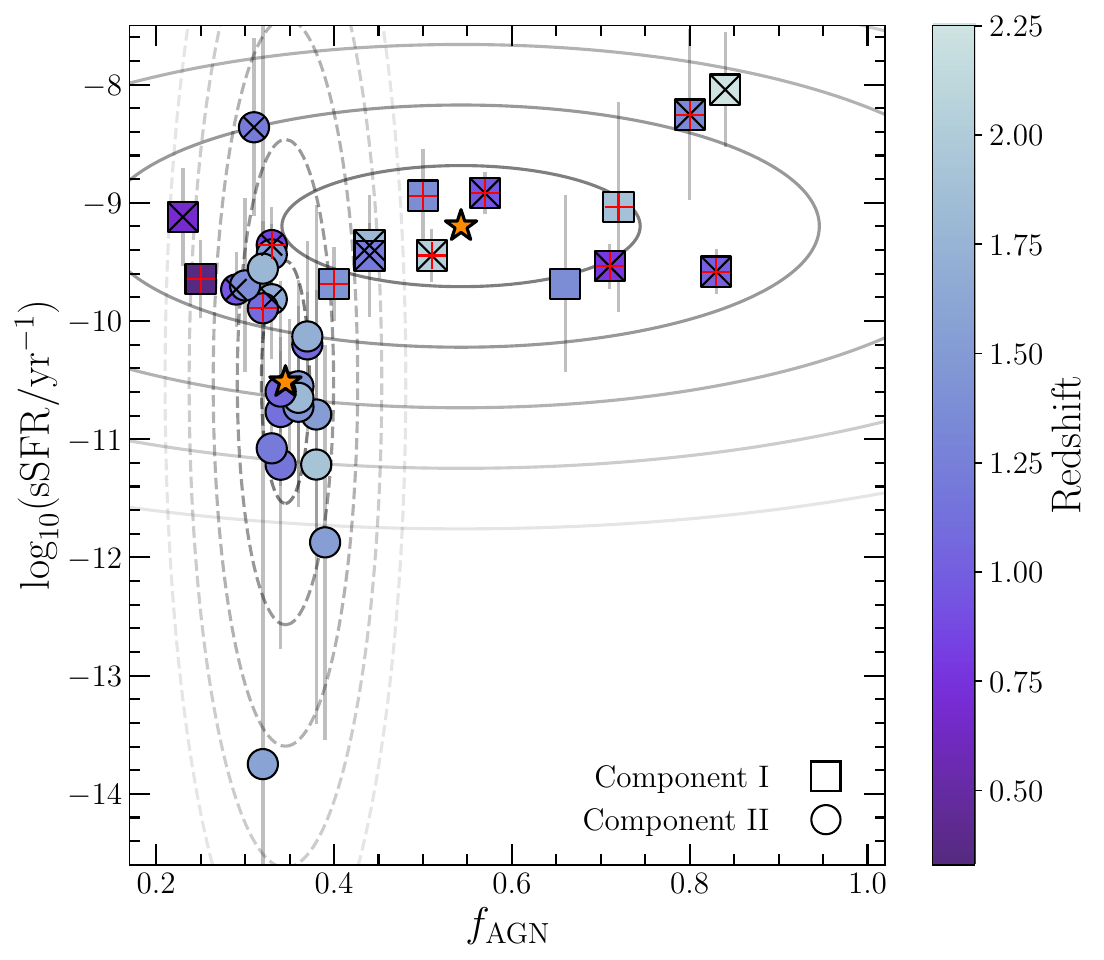}
    \caption{{{Separation of Components I and II\null. The parameter space is the same as the right panel of Figure~\ref{fig:sfms-offsets}, but only  the NIRCam-selected sample is shown.}} Hosts with X-ray and radio detections are marked with $\times$ and $+$, respectively, and marker colors correspond to redshift as shown in the {{color}} bar. {{The two gold stars mark the centers of the two components, and the 1--5$\sigma$ ellipses for each component are drawn around them. 
    }}
    }
    \label{fig:dSFMS-fAGN-z}
\end{figure}

\begin{figure*}
    \centering
    \includegraphics[width=0.45\linewidth]{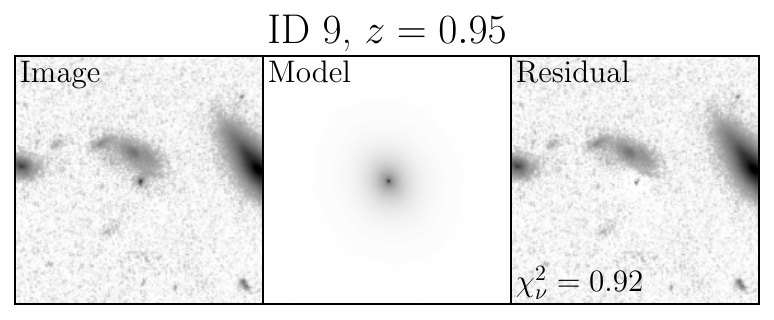}
    \includegraphics[width=0.45\linewidth]{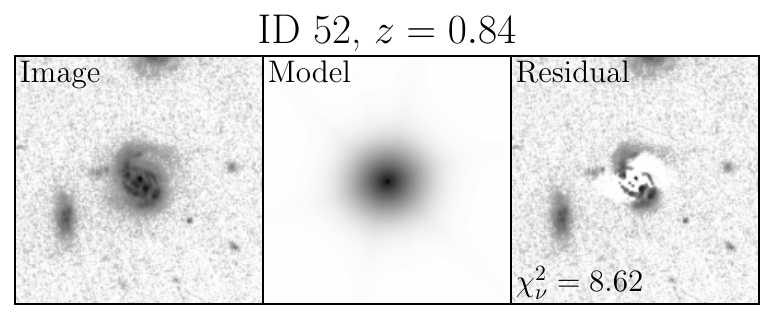}
    \includegraphics[width=0.45\linewidth]{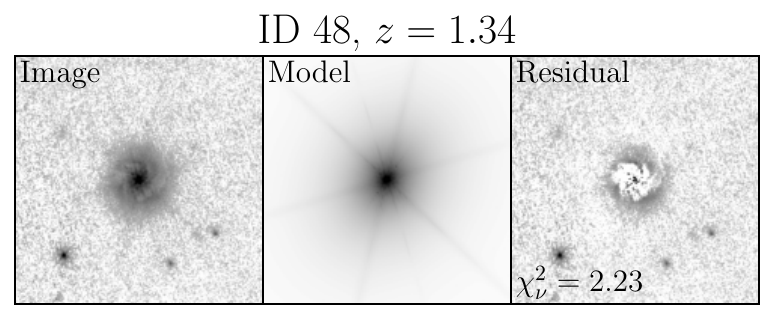}
    \includegraphics[width=0.45\linewidth]{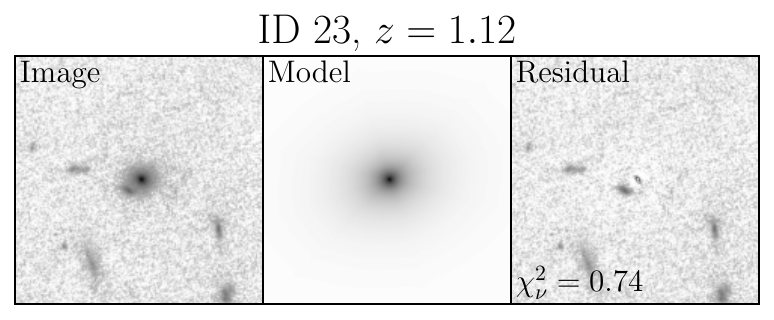}
    \includegraphics[width=0.45\linewidth]{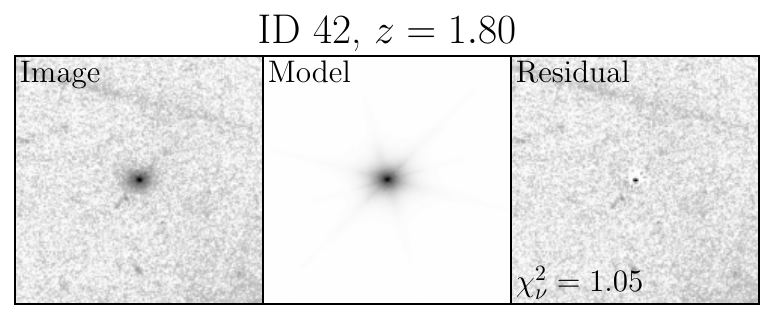}
    \includegraphics[width=0.45\linewidth]{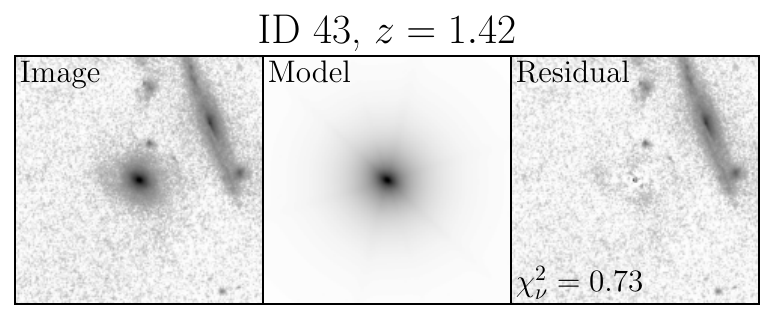}
    \caption{JWST/F090W light-profile models for six galaxies. The left column shows { Component I} galaxies and the right column { Component II} galaxies with IDs and redshifts labeled. For each galaxy, the left panel shows the observed image, the middle panel shows the model, and the right panel shows the residual { with the fit $\chi^2_\nu$ inset}. 
    }
    \label{fig:galfit-spirals}
\end{figure*}

{ The absence of one-dimensional multimodality, however, does not preclude the presence of subgroups in the joint $(f_{\rm AGN},\log_{10}\mathrm{sSFR})$ parameter space. To determine whether there is a statistical basis for the visually identified clusters, we performed Gaussian mixture modeling (GMM) on the two-dimensional $(f_{\rm AGN},\log_{10}\mathrm{sSFR})$ distribution.\footnote{The GMM treatment was chosen over the so-called ``two-dimensional Kolmogorov-Smirnov test'' \citep[\eg][]{2DKs1, 2DKs2} because it directly quantifies underlying structure instead of relying on hypothesis tests dependent on a choice of significance level.}
Models containing between one and ten Gaussian components were fit under four covariance parameterizations (full, tied, diagonal, and spherical), and the ``best''  model was chosen by virtue of its minimum Bayesian Information Criterion (BIC)\null. As expected, the global BIC minimum corresponded to a ten-component spherical model when no additional constraints were imposed; this fit, however, partitioned the sample into clusters containing as few as one or two galaxies and  therefore is an overfit rather than a physically meaningful decomposition. To obtain an interpretable model, we required each component to contain at least ten galaxies.\footnote{Varying this minimum in either direction did not change the resulting best-fit solution. Such stability means that the best-fit solution is strongly preferred and independent of the specific minimum so long as a reasonable one is chosen.} Under this constraint, the best-fit model consists of two distinct but partially overlapping components comprising 14 and 22 galaxies, respectively (diagonal and full covariance are about equally good and produce this same result). 
}
{ The results of the GMM fitting are shown in Figure~\ref{fig:dSFMS-fAGN-z}}.

{ The two components recovered by the GMM fitting correspond to the visually identified ones in the right panel of Figure~\ref{fig:sfms-offsets}. They are centered at $(f_{\rm AGN},\log_{10}\mathrm{sSFR})=(0.56,-9.19)$ (``Component~I'') and $(f_{\rm AGN},\log_{10}\mathrm{sSFR}) = (0.34,-10.45)$ (``Component~II''). Galaxies were assigned to the component for which they have the higher posterior probability of belonging. Most galaxies strongly prefer a single component: the median posterior probability is $P=0.997$, and 27/36 galaxies (75\%) are assigned to one component with $P>0.95$.
The preference of most galaxies for one component over the other suggests that the GMM was successful and that the fitted Gaussian components provide an adequate description of the observed distribution.\footnote{A formal residual analysis of the type described by, e.g., \citet{deSouza2017}, based on comparing observed and model densities, was not pursued. For a sample of only 36 galaxies, such diagnostics are very sensitive to kernel-density smoothing and sampling variance instead of model performance, making BIC model selection and posterior membership probabilities more informative indicators of model adequacy.}}

{ The 14 Component I galaxies and 22 Component II galaxies are identified in Table~\ref{tab:objects}.} There is no clear { redshift preference by either component,} though  the highest-redshift object in the sample---ID~14, an apparent quasar with detections in all 11 filters, spectroscopically confirmed to be at $z=2.2508$---is a { Component I} object. From the morphologies visible in Figure \ref{fig:stamps} and visual inspection of \textsc{Galfit} residuals, { Component I} galaxies commonly have spiral structure { or have their brightness dominated by point-source features (especially in F444W and the long-wavelength channel filters but even at shorter wavelengths, see Table~\ref{tab:galfit-best-fit-params}).} { Some Component II galaxies show spiral structure, but their light profiles are largely not dominated by a point source. Thus morphological interpretation of these groups is limited}. This is shown in Figure \ref{fig:galfit-spirals} for a selection of objects with these characteristics. { Given the prominence of point-sources in the Component I galaxies' light profiles, it is likely that they host more luminous AGN than Component II galaxies do, which is consistent with the characteristically lower $f_{\rm AGN} \sim 0.34$ of Component~II.}

The existence of two groups paints the picture of host galaxies that { have notably different star-forming properties} likely { due to} AGN activity. Those in { Component II} have lower $f_\text{AGN}$ but a range of specific star formation rates. They do not { often} have X-ray or radio detections and are { likely} heading towards a quenched state if they have not arrived there already. As { so few} of the { Component II} galaxies are detected in the X-ray or radio, there are two possibilities. One is that { they} are not AGN but are instead galaxies with  bright, compact central regions that appear pointlike when convolved with the JWST/NIRCam PSF{, possibly due to bright nuclear star clusters}. The other is that they host low-luminosity or highly obscured AGN\null. { This is consistent with  only 6/22 Component II galaxies meeting the \citet[][]{Assef2013} AGN color  criterion \citepalias{Ortiz2024}.} The available data are insufficient to distinguish these possibilities, but if the galaxies do not host AGN,  morphological AGN selection with NIRCam is subject to contamination from galaxies that are bright because { their compact cores are intrinsically bright}.

In contrast, { Component I galaxies} have the brightest point sources in the sample and tend to have X-ray and radio counterparts. Therefore these galaxies likely do host AGN\null. { This is supported  all but one (ID 39 at $z=0.6726$) of the Component I galaxies being identified by the \citet[][]{Assef2013} criterion as AGN \citepalias{Ortiz2024}.} These galaxies may be currently or very recently experiencing bursts or quenching of star formation in response to AGN feedback, as they are mostly still within $\pm$0.6~{dex} of the SFMS but further away than the expected $\pm$0.3~{dex} for genuine main-sequence galaxies. Longer-wavelength observations of NIRCam-selected objects will { offer crucial IR constraints on SFR and} make the presence or absence of AGN clearer, especially if the AGN are obscured, and can help fully quantify NIRCam's ability to sample a complete AGN population.

\section{Summary \& Future Work}
\label{sec:summary}
The sensitivity and angular resolution of JWST/NIRCam allow not only the morphological selection of AGN candidates but also enable direct studies of the underlying host galaxies. 
    Morphological analysis with NIRCam can identify bright  central emission, often coming from luminous AGN activity. { The star-forming properties of host galaxies of} NIRCam-selected AGN candidates { differ from those of the hosts of X-ray- or radio-selected AGN. The hosts of NIRCam-selected AGN themselves show evidence for two statistically different but partially overlapping components, best understood by their AGN fractions and specific star-formation rates.}
\begin{itemize}
    \item { The host galaxies of NIRCam-selected AGN have  $\rm \Delta SFMS$ between the values for radio-selected AGN hosts ($\rm \Delta SFMS\simeq0$) and X-ray-selected hosts ($\rm \Delta SFMS\simeq-1$). The  $\log_{10} \rm sSFR$ and $f_{\rm AGN}$ distributions also differ for the three selection groups.   The NIRCam-selected stellar mass distribution is consistent with being the same as for the X-ray-selected hosts but differs from that of the radio-selected AGN hosts. These differences persist in the joint parameter spaces $(\log_{10} M_\star, \Delta \rm SFMS)$ and ($f_{\rm AGN}, \log_{10} \rm sSFR$), suggesting that the star-forming processes occurring in NIRCam-selected AGN are different than those  in X-ray and radio AGN.}

    \item Host galaxies of NIRCam-selected AGN { form two groups in $(f_{\rm AGN}, \log_{10} \rm sSFR)$ space. ``Component~I'' has $\rm sSFR\simeq10^{-9}\rm~yr^{-1}$ but a wide range of $f_{\rm AGN}$. ``Component~II'' has $f_{\rm AGN}\simeq0.35$ but a wide range of sSFR down to ${\ll}10^{-12}$~yr$^{-1}$}.   
    
    \item The Component II galaxies mostly lack X-ray and radio counterparts. Although they have compact, bright cores that show point-source features, { these cores may not always indicate an AGN}\null. These objects are either quenched or in transition  between star-forming and quenched. They {  represent} either a non-AGN population  with bright, compact stellar bulges or a low-luminosity or obscured AGN population.
    
    \item The Component I galaxies mostly host radio or X-ray sources. They have $\rm\Delta SFRs$ between $\pm$0.3~{dex} and $\pm$0.6~{dex}. These { galaxies} may have current or recent bursts or quenching of star formation. The AGN are likely impacting the star-forming activity of their hosts and are suggestive of a phase where AGN coexist with galaxies nearby but not on the SFMS\null. This { group} is morphologically diverse { but includes galaxies with the brightest central point-sources, especially in the long-wavelength NIRCam filters}.
\end{itemize}

The star-forming behaviors of the { Component I and Component II galaxies} have no correlation with redshift or stellar mass. { Component I} galaxies have both brighter and more dominant point sources between the two groups and are { hosts of strong} AGN { candidates} by the detections of X-ray and/or radio counterparts. { Component II} galaxies may or may not be AGN hosts as they do not { usually} have X-ray or radio counterparts { and lack the point-source signature typical of AGN in NIRCam bands}.

Future studies similar to this one will offer further insights into whether the host galaxies of NIRCam-selected AGN { do indeed have distinct star-forming properties} in the sense of \S\ref{sec:canonical-SF} { and whether or not the components identified in \S\ref{sec:substructure} arise from different stages of evolution}. This sample contains 36 objects from a single catalog { and therefore has limited statistical power}, but the methods are applicable to any number of objects, and large-scale analyses of NIRCam-selected host galaxies are possible. 

The results of this study suggest rich coevolution between AGN and their host galaxies. However, as this work is concerned expressly with host galaxies' star-forming properties, it is unable to draw conclusions regarding, e.g., SMBH properties or gas kinematics. Results of these kinds are obtainable from studies using codes such as \textsc{AGNfitter} \citep[][]{CalistroRivera2016, MartinezRamirez2024} or \textsc{GELATO} \citep[][]{Hviding2022} that are designed to distinguish AGN and host-galaxy emission. Spectroscopic studies of NIRCam-selected objects will be particularly illuminating on this front and will be useful in identifying the presence and type(s) of feedback mechanisms suggested by this study's results (e.g., quenching).

Most critically, high-resolution observations of these objects and other similar sources, especially in the mid-infrared, will offer additional constraints on host-galaxy properties not well-determined by UV--visible--NIR photometry \citep[][]{Lyu2024} alone. With just these wavelengths, objects' SFHs could not be resolved in any meaningful way, and so ages of the most recent starbursts, etc.\ cannot be identified from SED fitting. For this sample, conclusions can be drawn because X-ray and radio detections provided insight into the timescales of AGN activity, but the timescales of change in the hosts' SFR could not  be quantified directly. Mid-infrared observations at high resolution with, e.g, JWST's Mid-Infrared Instrument (MIRI) would both provide invaluable constraints on the SFH and allow dust attenuation and other host properties to be inferred from SEDs with less degeneracy. In addition, it is ambiguous whether or not the { Component II} galaxies are genuinely AGN hosts: it is possible that they are hosts to highly obscured AGN that are actively quenching star formation, but the NIR does not offer insight here. Mid-infrared observations will be pivotal in determining whether { Component II}-type galaxies do host this type of AGN or if morphological AGN selection is contaminated by centrally bright, non-AGN objects. Given its resolution in its wavelength range, JWST/MIRI is the only suitable tool to advance host-galaxy studies like this one, and its observations will address what NIRCam cannot. Nonetheless, this work presents NIRCam imaging as a prime tool for identifying and analyzing AGN populations.

\begin{acknowledgments}
{ The authors thank the anonymous referee for comments that greatly improved the manuscript.} This work is based on observations associated with programs JWST-GTO-2738 and 1176 made with the NASA/ESA/CSA James Webb Space Telescope (JWST) and observations associated with programs HST-GO-15278, 16252 and 16793 made with the NASA/ESA Hubble Space Telescope (HST). 
The JWST and HST data were obtained from the Mikulski Archive for Space Telescopes at the Space Telescope Science Institute (STScI), which is operated by the Association of Universities for Research in Astronomy, Inc. (AURA), under NASA contracts NAS\,5-03127 (JWST) and NAS\,5-26555 (HST).  This work uses data obtained at the MMT Observatory, a facility jointly operated by the University of Arizona and the Smithsonian Institution.
RAW, SHC, and RAJ acknowledge support from NASA JWST Interdisciplinary Scientist grants NAG5-12460, NNX14AN10G and 80NSSC18K0200 from GSFC.
RAJ, RO, RAW, SHC, AMK, BLF and CNAW acknowledge support from HST grants HST-GO-15278.*, 16252.* and 16793.* from STScI, which is operated by AURA under contract NAS\,5-26555 from NASA.
CNAW acknowledges funding from the JWST/NIRCam contract NASS-0215 to the University of Arizona.
Work by CJC acknowledges support from the European Research Council (ERC) Advanced Investigator Grant EPOCHS (788113). BLF thanks the Berkeley Center for Theoretical Physics for their hospitality during the writing of this paper. This work is sponsored (in part) by the Chinese Academy of Sciences (CAS) through a grant to the CAS South America Center for Astronomy. C.C. acknowledges NSFC grant No. 11803044 and 12173045. This work is supported by the China Manned Space Program with grant no. CMS-CSST-2025-A07. C.C. is supported by Chinese Academy of Sciences South America Center for Astronomy (CASSACA) Key Research Project E52H540301. { HBH and SNM acknowledge support from NASA JWST Interdisciplinary Scientist grant 21-SMDSS21-0013.}

We also acknowledge the indigenous peoples of Arizona, including the Akimel O'odham (Pima) and Pee Posh (Maricopa) Indian Communities, whose care and keeping of the land has enabled us to be at ASU's Tempe campus in the Salt River Valley, where much of our work was conducted.
\end{acknowledgments}

All the JWST and HST data used in this paper can be found in MAST: \dataset[10.17909/eag6-3964]{https://doi.org/10.17909/eag6-3964}.

\software{\textsc{Astropy} \citep{astropy1, astropy2, astropy3}}
\software{\textsc{IDL} Astronomy Library \citep{IDL}}
\software{\textsc{SourceExtractor} \citep{Bertin1996}}
\software{\textsc{Galfit} \citep{Peng2002, Peng2010}}
\software{\textsc{CIGALE} \citep{Boquien2019, Yang2020, Yang2022, Burgarella2025}}
\software{\textsc{STPSF} \citep[formerly \textsc{WebbPSF},][]{Perrin2014}}
\software{\textsc{Trilogy} \citep{Coe2012}}

\facilities{James Webb Space Telescope, Hubble Space Telescope, MMT Observatory, Chandra X-Ray Observatory, Karl G. Jansky Very Large Array, {Mikulski Archive for Space Telescopes} \url{https://archive.stsci.edu}}

\bibliography{bib}{}

@ARTICLE{Ortiz2024,
       author = {{Ortiz}, Rafael and {Windhorst}, Rogier A. and {Cohen}, Seth H. and {Willner}, Steven P. and {Jansen}, Rolf A. and {Carleton}, Timothy and {Kamieneski}, Patrick S. and {Rutkowski}, Michael J. and {Smith}, Brent M. and {Summers}, Jake and {Cheng}, Cheng and {Coe}, Dan and {Conselice}, Christopher J. and {Diego}, Jose M. and {Driver}, Simon P. and {D'Silva}, Jordan C.~J. and {Frye}, Brenda L. and {Gim}, Hansung B. and {Grogin}, Norman A. and {Hammel}, Heidi B. and {Hathi}, Nimish P. and {Holwerda}, Benne W. and {Hyun}, Minhee and {Im}, Myungshin and {Keel}, William C. and {Koekemoer}, Anton M. and {Li}, Juno and {Marshall}, Madeline A. and {McCabe}, Tyler J. and {McLeod}, Noah J. and {Milam}, Stefanie N. and {O'Brien}, Rosalia and {Pirzkal}, Nor and {Robotham}, Aaron S.~G. and {Ryan}, Russell E. and {Willmer}, Christopher N.~A. and {Yan}, Haojing and {Yun}, Min S. and {Zitrin}, Adi},
        title = "{PEARLS: Discovery of Point-source Features within Galaxies in the North Ecliptic Pole Time Domain Field}",
      journal = {\apj},
         year = 2024,
        month = oct,
       volume = {974},
       number = {2},
          eid = {258},
        pages = {258},
          doi = {10.3847/1538-4357/ad6d5e},
archivePrefix = {arXiv},
       eprint = {2404.10709},
 primaryClass = {astro-ph.GA},
       adsurl = {https://ui.adsabs.harvard.edu/abs/2024ApJ...974..258O}
}

@ARTICLE{astropy1,
       author = {{Astropy Collaboration} and {Robitaille}, Thomas P. and {Tollerud}, Erik J. and {Greenfield}, Perry and {Droettboom}, Michael and {Bray}, Erik and {Aldcroft}, Tom and {Davis}, Matt and {Ginsburg}, Adam and {Price-Whelan}, Adrian M. and {Kerzendorf}, Wolfgang E. and {Conley}, Alexander and {Crighton}, Neil and {Barbary}, Kyle and {Muna}, Demitri and {Ferguson}, Henry and {Grollier}, Fr{\'e}d{\'e}ric and {Parikh}, Madhura M. and {Nair}, Prasanth H. and {Unther}, Hans M. and {Deil}, Christoph and {Woillez}, Julien and {Conseil}, Simon and {Kramer}, Roban and {Turner}, James E.~H. and {Singer}, Leo and {Fox}, Ryan and {Weaver}, Benjamin A. and {Zabalza}, Victor and {Edwards}, Zachary I. and {Azalee Bostroem}, K. and {Burke}, D.~J. and {Casey}, Andrew R. and {Crawford}, Steven M. and {Dencheva}, Nadia and {Ely}, Justin and {Jenness}, Tim and {Labrie}, Kathleen and {Lim}, Pey Lian and {Pierfederici}, Francesco and {Pontzen}, Andrew and {Ptak}, Andy and {Refsdal}, Brian and {Servillat}, Mathieu and {Streicher}, Ole},
        title = "{Astropy: A community Python package for astronomy}",
      journal = {\aap},
         year = 2013,
        month = oct,
       volume = {558},
          eid = {A33},
        pages = {A33},
          doi = {10.1051/0004-6361/201322068},
archivePrefix = {arXiv},
       eprint = {1307.6212},
 primaryClass = {astro-ph.IM},
       adsurl = {https://ui.adsabs.harvard.edu/abs/2013A&A...558A..33A}
}

@ARTICLE{astropy2,
       author = {{Astropy Collaboration} and {Price-Whelan}, A.~M. and {Sip{\H{o}}cz}, B.~M. and {G{\"u}nther}, H.~M. and {Lim}, P.~L. and {Crawford}, S.~M. and {Conseil}, S. and {Shupe}, D.~L. and {Craig}, M.~W. and {Dencheva}, N. and {Ginsburg}, A. and {VanderPlas}, J.~T. and {Bradley}, L.~D. and {P{\'e}rez-Su{\'a}rez}, D. and {de Val-Borro}, M. and {Aldcroft}, T.~L. and {Cruz}, K.~L. and {Robitaille}, T.~P. and {Tollerud}, E.~J. and {Ardelean}, C. and {Babej}, T. and {Bach}, Y.~P. and {Bachetti}, M. and {Bakanov}, A.~V. and {Bamford}, S.~P. and {Barentsen}, G. and {Barmby}, P. and {Baumbach}, A. and {Berry}, K.~L. and {Biscani}, F. and {Boquien}, M. and {Bostroem}, K.~A. and {Bouma}, L.~G. and {Brammer}, G.~B. and {Bray}, E.~M. and {Breytenbach}, H. and {Buddelmeijer}, H. and {Burke}, D.~J. and {Calderone}, G. and {Cano Rodr{\'\i}guez}, J.~L. and {Cara}, M. and {Cardoso}, J.~V.~M. and {Cheedella}, S. and {Copin}, Y. and {Corrales}, L. and {Crichton}, D. and {D'Avella}, D. and {Deil}, C. and {Depagne}, {\'E}. and {Dietrich}, J.~P. and {Donath}, A. and {Droettboom}, M. and {Earl}, N. and {Erben}, T. and {Fabbro}, S. and {Ferreira}, L.~A. and {Finethy}, T. and {Fox}, R.~T. and {Garrison}, L.~H. and {Gibbons}, S.~L.~J. and {Goldstein}, D.~A. and {Gommers}, R. and {Greco}, J.~P. and {Greenfield}, P. and {Groener}, A.~M. and {Grollier}, F. and {Hagen}, A. and {Hirst}, P. and {Homeier}, D. and {Horton}, A.~J. and {Hosseinzadeh}, G. and {Hu}, L. and {Hunkeler}, J.~S. and {Ivezi{\'c}}, {\v{Z}}. and {Jain}, A. and {Jenness}, T. and {Kanarek}, G. and {Kendrew}, S. and {Kern}, N.~S. and {Kerzendorf}, W.~E. and {Khvalko}, A. and {King}, J. and {Kirkby}, D. and {Kulkarni}, A.~M. and {Kumar}, A. and {Lee}, A. and {Lenz}, D. and {Littlefair}, S.~P. and {Ma}, Z. and {Macleod}, D.~M. and {Mastropietro}, M. and {McCully}, C. and {Montagnac}, S. and {Morris}, B.~M. and {Mueller}, M. and {Mumford}, S.~J. and {Muna}, D. and {Murphy}, N.~A. and {Nelson}, S. and {Nguyen}, G.~H. and {Ninan}, J.~P. and {N{\"o}the}, M. and {Ogaz}, S. and {Oh}, S. and {Parejko}, J.~K. and {Parley}, N. and {Pascual}, S. and {Patil}, R. and {Patil}, A.~A. and {Plunkett}, A.~L. and {Prochaska}, J.~X. and {Rastogi}, T. and {Reddy Janga}, V. and {Sabater}, J. and {Sakurikar}, P. and {Seifert}, M. and {Sherbert}, L.~E. and {Sherwood-Taylor}, H. and {Shih}, A.~Y. and {Sick}, J. and {Silbiger}, M.~T. and {Singanamalla}, S. and {Singer}, L.~P. and {Sladen}, P.~H. and {Sooley}, K.~A. and {Sornarajah}, S. and {Streicher}, O. and {Teuben}, P. and {Thomas}, S.~W. and {Tremblay}, G.~R. and {Turner}, J.~E.~H. and {Terr{\'o}n}, V. and {van Kerkwijk}, M.~H. and {de la Vega}, A. and {Watkins}, L.~L. and {Weaver}, B.~A. and {Whitmore}, J.~B. and {Woillez}, J. and {Zabalza}, V. and {Astropy Contributors}},
        title = "{The Astropy Project: Building an Open-science Project and Status of the v2.0 Core Package}",
      journal = {\aj},
         year = 2018,
        month = sep,
       volume = {156},
       number = {3},
          eid = {123},
        pages = {123},
          doi = {10.3847/1538-3881/aabc4f},
archivePrefix = {arXiv},
       eprint = {1801.02634},
 primaryClass = {astro-ph.IM},
       adsurl = {https://ui.adsabs.harvard.edu/abs/2018AJ....156..123A}
}

@ARTICLE{astropy3,
       author = {{Astropy Collaboration} and {Price-Whelan}, Adrian M. and {Lim}, Pey Lian and {Earl}, Nicholas and {Starkman}, Nathaniel and {Bradley}, Larry and {Shupe}, David L. and {Patil}, Aarya A. and {Corrales}, Lia and {Brasseur}, C.~E. and {N{\"o}the}, Maximilian and {Donath}, Axel and {Tollerud}, Erik and {Morris}, Brett M. and {Ginsburg}, Adam and {Vaher}, Eero and {Weaver}, Benjamin A. and {Tocknell}, James and {Jamieson}, William and {van Kerkwijk}, Marten H. and {Robitaille}, Thomas P. and {Merry}, Bruce and {Bachetti}, Matteo and {G{\"u}nther}, H. Moritz and {Aldcroft}, Thomas L. and {Alvarado-Montes}, Jaime A. and {Archibald}, Anne M. and {B{\'o}di}, Attila and {Bapat}, Shreyas and {Barentsen}, Geert and {Baz{\'a}n}, Juanjo and {Biswas}, Manish and {Boquien}, M{\'e}d{\'e}ric and {Burke}, D.~J. and {Cara}, Daria and {Cara}, Mihai and {Conroy}, Kyle E. and {Conseil}, Simon and {Craig}, Matthew W. and {Cross}, Robert M. and {Cruz}, Kelle L. and {D'Eugenio}, Francesco and {Dencheva}, Nadia and {Devillepoix}, Hadrien A.~R. and {Dietrich}, J{\"o}rg P. and {Eigenbrot}, Arthur Davis and {Erben}, Thomas and {Ferreira}, Leonardo and {Foreman-Mackey}, Daniel and {Fox}, Ryan and {Freij}, Nabil and {Garg}, Suyog and {Geda}, Robel and {Glattly}, Lauren and {Gondhalekar}, Yash and {Gordon}, Karl D. and {Grant}, David and {Greenfield}, Perry and {Groener}, Austen M. and {Guest}, Steve and {Gurovich}, Sebastian and {Handberg}, Rasmus and {Hart}, Akeem and {Hatfield-Dodds}, Zac and {Homeier}, Derek and {Hosseinzadeh}, Griffin and {Jenness}, Tim and {Jones}, Craig K. and {Joseph}, Prajwel and {Kalmbach}, J. Bryce and {Karamehmetoglu}, Emir and {Ka{\l}uszy{\'n}ski}, Miko{\l}aj and {Kelley}, Michael S.~P. and {Kern}, Nicholas and {Kerzendorf}, Wolfgang E. and {Koch}, Eric W. and {Kulumani}, Shankar and {Lee}, Antony and {Ly}, Chun and {Ma}, Zhiyuan and {MacBride}, Conor and {Maljaars}, Jakob M. and {Muna}, Demitri and {Murphy}, N.~A. and {Norman}, Henrik and {O'Steen}, Richard and {Oman}, Kyle A. and {Pacifici}, Camilla and {Pascual}, Sergio and {Pascual-Granado}, J. and {Patil}, Rohit R. and {Perren}, Gabriel I. and {Pickering}, Timothy E. and {Rastogi}, Tanuj and {Roulston}, Benjamin R. and {Ryan}, Daniel F. and {Rykoff}, Eli S. and {Sabater}, Jose and {Sakurikar}, Parikshit and {Salgado}, Jes{\'u}s and {Sanghi}, Aniket and {Saunders}, Nicholas and {Savchenko}, Volodymyr and {Schwardt}, Ludwig and {Seifert-Eckert}, Michael and {Shih}, Albert Y. and {Jain}, Anany Shrey and {Shukla}, Gyanendra and {Sick}, Jonathan and {Simpson}, Chris and {Singanamalla}, Sudheesh and {Singer}, Leo P. and {Singhal}, Jaladh and {Sinha}, Manodeep and {Sip{\H{o}}cz}, Brigitta M. and {Spitler}, Lee R. and {Stansby}, David and {Streicher}, Ole and {{\v{S}}umak}, Jani and {Swinbank}, John D. and {Taranu}, Dan S. and {Tewary}, Nikita and {Tremblay}, Grant R. and {de Val-Borro}, Miguel and {Van Kooten}, Samuel J. and {Vasovi{\'c}}, Zlatan and {Verma}, Shresth and {de Miranda Cardoso}, Jos{\'e} Vin{\'\i}cius and {Williams}, Peter K.~G. and {Wilson}, Tom J. and {Winkel}, Benjamin and {Wood-Vasey}, W.~M. and {Xue}, Rui and {Yoachim}, Peter and {Zhang}, Chen and {Zonca}, Andrea and {Astropy Project Contributors}},
        title = "{The Astropy Project: Sustaining and Growing a Community-oriented Open-source Project and the Latest Major Release (v5.0) of the Core Package}",
      journal = {\apj},
         year = 2022,
        month = aug,
       volume = {935},
       number = {2},
          eid = {167},
        pages = {167},
          doi = {10.3847/1538-4357/ac7c74},
archivePrefix = {arXiv},
       eprint = {2206.14220},
 primaryClass = {astro-ph.IM},
       adsurl = {https://ui.adsabs.harvard.edu/abs/2022ApJ...935..167A}
}

@INPROCEEDINGS{IDL,
       author = {{Landsman}, W.~B.},
        title = "{The IDL Astronomy User's Library}",
    booktitle = {Astronomical Data Analysis Software and Systems II},
         year = 1993,
       editor = {{Hanisch}, R.~J. and {Brissenden}, R.~J.~V. and {Barnes}, J.},
       series = {Astronomical Society of the Pacific Conference Series},
       volume = {52},
        month = jan,
        pages = {246},
       adsurl = {https://ui.adsabs.harvard.edu/abs/1993ASPC...52..246L}
}

@ARTICLE{Bertin1996,
       author = {{Bertin}, E. and {Arnouts}, S.},
        title = "{SExtractor: Software for source extraction.}",
      journal = {\aaps},
         year = 1996,
        month = jun,
       volume = {117},
        pages = {393-404},
          doi = {10.1051/aas:1996164},
       adsurl = {https://ui.adsabs.harvard.edu/abs/1996A&AS..117..393B}
}

@ARTICLE{Peng2002,
       author = {{Peng}, Chien Y. and {Ho}, Luis C. and {Impey}, Chris D. and {Rix}, Hans-Walter},
        title = "{Detailed Structural Decomposition of Galaxy Images}",
      journal = {\aj},
         year = 2002,
        month = jul,
       volume = {124},
       number = {1},
        pages = {266-293},
          doi = {10.1086/340952},
archivePrefix = {arXiv},
       eprint = {astro-ph/0204182},
 primaryClass = {astro-ph},
       adsurl = {https://ui.adsabs.harvard.edu/abs/2002AJ....124..266P}
}

@ARTICLE{Peng2010,
       author = {{Peng}, Chien Y. and {Ho}, Luis C. and {Impey}, Chris D. and {Rix}, Hans-Walter},
        title = "{Detailed Decomposition of Galaxy Images. II. Beyond Axisymmetric Models}",
      journal = {\aj},
         year = 2010,
        month = jun,
       volume = {139},
       number = {6},
        pages = {2097-2129},
          doi = {10.1088/0004-6256/139/6/2097},
archivePrefix = {arXiv},
       eprint = {0912.0731},
 primaryClass = {astro-ph.CO},
       adsurl = {https://ui.adsabs.harvard.edu/abs/2010AJ....139.2097P}
}

@ARTICLE{Boquien2019,
       author = {{Boquien}, M. and {Burgarella}, D. and {Roehlly}, Y. and {Buat}, V. and {Ciesla}, L. and {Corre}, D. and {Inoue}, A.~K. and {Salas}, H.},
        title = "{CIGALE: a python Code Investigating GALaxy Emission}",
      journal = {\aap},
         year = 2019,
        month = feb,
       volume = {622},
          eid = {A103},
        pages = {A103},
          doi = {10.1051/0004-6361/201834156},
archivePrefix = {arXiv},
       eprint = {1811.03094},
 primaryClass = {astro-ph.GA},
       adsurl = {https://ui.adsabs.harvard.edu/abs/2019A&A...622A.103B}
}

@ARTICLE{Yang2020,
       author = {{Yang}, G. and {Boquien}, M. and {Buat}, V. and {Burgarella}, D. and {Ciesla}, L. and {Duras}, F. and {Stalevski}, M. and {Brandt}, W.~N. and {Papovich}, C.},
        title = "{X-CIGALE: Fitting AGN/galaxy SEDs from X-ray to infrared}",
      journal = {\mnras},
         year = 2020,
        month = jan,
       volume = {491},
       number = {1},
        pages = {740-757},
          doi = {10.1093/mnras/stz3001},
archivePrefix = {arXiv},
       eprint = {2001.08263},
 primaryClass = {astro-ph.GA},
       adsurl = {https://ui.adsabs.harvard.edu/abs/2020MNRAS.491..740Y}
}

@ARTICLE{Yang2022,
       author = {{Yang}, Guang and {Boquien}, M{\'e}d{\'e}ric and {Brandt}, W.~N. and {Buat}, V{\'e}ronique and {Burgarella}, Denis and {Ciesla}, Laure and {Lehmer}, Bret D. and {Ma{\l}ek}, Katarzyna and {Mountrichas}, George and {Papovich}, Casey and {Pons}, Estelle and {Stalevski}, Marko and {Theul{\'e}}, Patrice and {Zhu}, Shifu},
        title = "{Fitting AGN/Galaxy X-Ray-to-radio SEDs with CIGALE and Improvement of the Code}",
      journal = {\apj},
         year = 2022,
        month = mar,
       volume = {927},
       number = {2},
          eid = {192},
        pages = {192},
          doi = {10.3847/1538-4357/ac4971},
archivePrefix = {arXiv},
       eprint = {2201.03718},
 primaryClass = {astro-ph.GA},
       adsurl = {https://ui.adsabs.harvard.edu/abs/2022ApJ...927..192Y}
}

@ARTICLE{Burgarella2025,
       author = {{Burgarella}, Denis and {Buat}, V{\'e}ronique and {Theul{\'e}}, Patrice and {Zavala}, Jorge and {Dickinson}, Mark and {Arrabal Haro}, Pablo and {Bagley}, Micaela B. and {Boquien}, M{\'e}d{\'e}ric and {Cleri}, Nikko and {Dewachter}, Tim and {Ferguson}, Henry C. and {Fern{\`a}ndez}, Vital and {Finkelstein}, Steven L. and {Gawiser}, Eric and {Grazian}, Andrea and {Grogin}, Norman and {Holwerda}, Benne W. and {Kartaltepe}, Jeyhan S. and {Kewley}, Lisa and {Kirkpatrick}, Allison and {Kocevski}, Dale and {Koekemoer}, Anton M. and {Long}, Arianna and {Lotz}, Jennifer and {Lucas}, Ray A. and {Mobasher}, Bahram and {Papovich}, Casey and {P{\'e}rez-Gonz{\`a}lez}, Pablo G. and {Pirzkal}, Nor and {Ravindranath}, Swara and {Rodighiero}, Giulia and {Roehlly}, Yannick and {Rose}, Caitlin and {Seill{\'e}}, Lise-Marie and {Somerville}, Rachel and {Wilkins}, Steve and {Yang}, Guang and {Yung}, L.~Y. Aaron},
        title = "{CEERS: Possibly forging the first dust grains in the universe: A population of galaxies with spectroscopically derived extremely low dust attenuation (GELDA) at 4.0 < z {\ensuremath{\lesssim}} 11.4}",
      journal = {\aap},
         year = 2025,
        month = jul,
       volume = {699},
          eid = {A336},
        pages = {A336},
          doi = {10.1051/0004-6361/202554231},
archivePrefix = {arXiv},
       eprint = {2504.13118},
 primaryClass = {astro-ph.GA},
       adsurl = {https://ui.adsabs.harvard.edu/abs/2025A&A...699A.336B}
}

@ARTICLE{Coe2012,
       author = {{Coe}, Dan and {Umetsu}, Keiichi and {Zitrin}, Adi and {Donahue}, Megan and {Medezinski}, Elinor and {Postman}, Marc and {Carrasco}, Mauricio and {Anguita}, Timo and {Geller}, Margaret J. and {Rines}, Kenneth J. and {Diaferio}, Antonaldo and {Kurtz}, Michael J. and {Bradley}, Larry and {Koekemoer}, Anton and {Zheng}, Wei and {Nonino}, Mario and {Molino}, Alberto and {Mahdavi}, Andisheh and {Lemze}, Doron and {Infante}, Leopoldo and {Ogaz}, Sara and {Melchior}, Peter and {Host}, Ole and {Ford}, Holland and {Grillo}, Claudio and {Rosati}, Piero and {Jim{\'e}nez-Teja}, Yolanda and {Moustakas}, John and {Broadhurst}, Tom and {Ascaso}, Bego{\~n}a and {Lahav}, Ofer and {Bartelmann}, Matthias and {Ben{\'\i}tez}, Narciso and {Bouwens}, Rychard and {Graur}, Or and {Graves}, Genevieve and {Jha}, Saurabh and {Jouvel}, Stephanie and {Kelson}, Daniel and {Moustakas}, Leonidas and {Maoz}, Dan and {Meneghetti}, Massimo and {Merten}, Julian and {Riess}, Adam and {Rodney}, Steve and {Seitz}, Stella},
        title = "{CLASH: Precise New Constraints on the Mass Profile of the Galaxy Cluster A2261}",
      journal = {\apj},
         year = 2012,
        month = sep,
       volume = {757},
       number = {1},
          eid = {22},
        pages = {22},
          doi = {10.1088/0004-637X/757/1/22},
archivePrefix = {arXiv},
       eprint = {1201.1616},
 primaryClass = {astro-ph.CO},
       adsurl = {https://ui.adsabs.harvard.edu/abs/2012ApJ...757...22C}
}

@INPROCEEDINGS{Perrin2014,
       author = {{Perrin}, Marshall D. and {Sivaramakrishnan}, Anand and {Lajoie}, Charles-Philippe and {Elliott}, Erin and {Pueyo}, Laurent and {Ravindranath}, Swara and {Albert}, Lo{\"\i}c.},
        title = "{Updated point spread function simulations for JWST with WebbPSF}",
    booktitle = {Space Telescopes and Instrumentation 2014: Optical, Infrared, and Millimeter Wave},
         year = 2014,
       editor = {{Oschmann}, Jr., Jacobus M. and {Clampin}, Mark and {Fazio}, Giovanni G. and {MacEwen}, Howard A.},
       series = {Society of Photo-Optical Instrumentation Engineers (SPIE) Conference Series},
       volume = {9143},
        month = aug,
          eid = {91433X},
        pages = {91433X},
          doi = {10.1117/12.2056689},
       adsurl = {https://ui.adsabs.harvard.edu/abs/2014SPIE.9143E..3XP}
}

@ARTICLE{York2000,
       author = {{York}, Donald G. and {Adelman}, J. and {Anderson}, Jr., John E. and {Anderson}, Scott F. and {Annis}, James and {Bahcall}, Neta A. and {Bakken}, J.~A. and {Barkhouser}, Robert and {Bastian}, Steven and {Berman}, Eileen and {Boroski}, William N. and {Bracker}, Steve and {Briegel}, Charlie and {Briggs}, John W. and {Brinkmann}, J. and {Brunner}, Robert and {Burles}, Scott and {Carey}, Larry and {Carr}, Michael A. and {Castander}, Francisco J. and {Chen}, Bing and {Colestock}, Patrick L. and {Connolly}, A.~J. and {Crocker}, J.~H. and {Csabai}, Istv{\'a}n and {Czarapata}, Paul C. and {Davis}, John Eric and {Doi}, Mamoru and {Dombeck}, Tom and {Eisenstein}, Daniel and {Ellman}, Nancy and {Elms}, Brian R. and {Evans}, Michael L. and {Fan}, Xiaohui and {Federwitz}, Glenn R. and {Fiscelli}, Larry and {Friedman}, Scott and {Frieman}, Joshua A. and {Fukugita}, Masataka and {Gillespie}, Bruce and {Gunn}, James E. and {Gurbani}, Vijay K. and {de Haas}, Ernst and {Haldeman}, Merle and {Harris}, Frederick H. and {Hayes}, J. and {Heckman}, Timothy M. and {Hennessy}, G.~S. and {Hindsley}, Robert B. and {Holm}, Scott and {Holmgren}, Donald J. and {Huang}, Chi-hao and {Hull}, Charles and {Husby}, Don and {Ichikawa}, Shin-Ichi and {Ichikawa}, Takashi and {Ivezi{\'c}}, {\v{Z}}eljko and {Kent}, Stephen and {Kim}, Rita S.~J. and {Kinney}, E. and {Klaene}, Mark and {Kleinman}, A.~N. and {Kleinman}, S. and {Knapp}, G.~R. and {Korienek}, John and {Kron}, Richard G. and {Kunszt}, Peter Z. and {Lamb}, D.~Q. and {Lee}, B. and {Leger}, R. French and {Limmongkol}, Siriluk and {Lindenmeyer}, Carl and {Long}, Daniel C. and {Loomis}, Craig and {Loveday}, Jon and {Lucinio}, Rich and {Lupton}, Robert H. and {MacKinnon}, Bryan and {Mannery}, Edward J. and {Mantsch}, P.~M. and {Margon}, Bruce and {McGehee}, Peregrine and {McKay}, Timothy A. and {Meiksin}, Avery and {Merelli}, Aronne and {Monet}, David G. and {Munn}, Jeffrey A. and {Narayanan}, Vijay K. and {Nash}, Thomas and {Neilsen}, Eric and {Neswold}, Rich and {Newberg}, Heidi Jo and {Nichol}, R.~C. and {Nicinski}, Tom and {Nonino}, Mario and {Okada}, Norio and {Okamura}, Sadanori and {Ostriker}, Jeremiah P. and {Owen}, Russell and {Pauls}, A. George and {Peoples}, John and {Peterson}, R.~L. and {Petravick}, Donald and {Pier}, Jeffrey R. and {Pope}, Adrian and {Pordes}, Ruth and {Prosapio}, Angela and {Rechenmacher}, Ron and {Quinn}, Thomas R. and {Richards}, Gordon T. and {Richmond}, Michael W. and {Rivetta}, Claudio H. and {Rockosi}, Constance M. and {Ruthmansdorfer}, Kurt and {Sandford}, Dale and {Schlegel}, David J. and {Schneider}, Donald P. and {Sekiguchi}, Maki and {Sergey}, Gary and {Shimasaku}, Kazuhiro and {Siegmund}, Walter A. and {Smee}, Stephen and {Smith}, J. Allyn and {Snedden}, S. and {Stone}, R. and {Stoughton}, Chris and {Strauss}, Michael A. and {Stubbs}, Christopher and {SubbaRao}, Mark and {Szalay}, Alexander S. and {Szapudi}, Istvan and {Szokoly}, Gyula P. and {Thakar}, Anirudda R. and {Tremonti}, Christy and {Tucker}, Douglas L. and {Uomoto}, Alan and {Vanden Berk}, Dan and {Vogeley}, Michael S. and {Waddell}, Patrick and {Wang}, Shu-i. and {Watanabe}, Masaru and {Weinberg}, David H. and {Yanny}, Brian and {Yasuda}, Naoki and {SDSS Collaboration}},
        title = "{The Sloan Digital Sky Survey: Technical Summary}",
      journal = {\aj},
         year = 2000,
        month = sep,
       volume = {120},
       number = {3},
        pages = {1579-1587},
          doi = {10.1086/301513},
archivePrefix = {arXiv},
       eprint = {astro-ph/0006396},
 primaryClass = {astro-ph},
       adsurl = {https://ui.adsabs.harvard.edu/abs/2000AJ....120.1579Y}
}

@ARTICLE{Ciesla2015,
       author = {{Ciesla}, L. and {Charmandaris}, V. and {Georgakakis}, A. and {Bernhard}, E. and {Mitchell}, P.~D. and {Buat}, V. and {Elbaz}, D. and {LeFloc'h}, E. and {Lacey}, C.~G. and {Magdis}, G.~E. and {Xilouris}, M.},
        title = "{Constraining the properties of AGN host galaxies with spectral energy distribution modelling}",
      journal = {\aap},
         year = 2015,
        month = apr,
       volume = {576},
          eid = {A10},
        pages = {A10},
          doi = {10.1051/0004-6361/201425252},
archivePrefix = {arXiv},
       eprint = {1501.03672},
 primaryClass = {astro-ph.GA},
       adsurl = {https://ui.adsabs.harvard.edu/abs/2015A&A...576A..10C}
}

@ARTICLE{Ramirez2014,
       author = {{Ram{\'\i}rez}, E.~A. and {Tadhunter}, C.~N. and {Dicken}, D. and {Rose}, M. and {Axon}, D. and {Sparks}, W. and {Packham}, C.},
        title = "{Hubble Space Telescope and Spitzer point-source detection and optical extinction in powerful narrow-line radio galaxies}",
      journal = {\mnras},
         year = 2014,
        month = apr,
       volume = {439},
       number = {2},
        pages = {1270-1285},
          doi = {10.1093/mnras/stt2444},
archivePrefix = {arXiv},
       eprint = {1401.1223},
 primaryClass = {astro-ph.GA},
       adsurl = {https://ui.adsabs.harvard.edu/abs/2014MNRAS.439.1270R}
}

@ARTICLE{Kotilainen1992a,
       author = {{Kotilainen}, J.~K. and {Ward}, M.~J. and {Boisson}, C. and {Depoy}, D.~L. and {Bryant}, L.~R. and {Smith}, M.~G.},
        title = "{Near-infrared imaging of hard X-ray selected active galaxies - I. Decomposition into nuclear and stellar components.}",
      journal = {\mnras},
         year = 1992,
        month = may,
       volume = {256},
        pages = {125-148},
          doi = {10.1093/mnras/256.1.125},
       adsurl = {https://ui.adsabs.harvard.edu/abs/1992MNRAS.256..125K}
}

@ARTICLE{Kotilainen1992b,
       author = {{Kotilainen}, J.~K. and {Ward}, M.~J. and {Boisson}, C. and {Depoy}, D.~L. and {Smith}, M.~G.},
        title = "{Near-infrared imaging of hard X-ray selected active galaxies - II. The non-stellar continuum.}",
      journal = {\mnras},
         year = 1992,
        month = may,
       volume = {256},
        pages = {149-165},
          doi = {10.1093/mnras/256.1.149},
       adsurl = {https://ui.adsabs.harvard.edu/abs/1992MNRAS.256..149K}
}

@ARTICLE{Stern2012,
       author = {{Stern}, Daniel and {Assef}, Roberto J. and {Benford}, Dominic J. and {Blain}, Andrew and {Cutri}, Roc and {Dey}, Arjun and {Eisenhardt}, Peter and {Griffith}, Roger L. and {Jarrett}, T.~H. and {Lake}, Sean and {Masci}, Frank and {Petty}, Sara and {Stanford}, S.~A. and {Tsai}, Chao-Wei and {Wright}, E.~L. and {Yan}, Lin and {Harrison}, Fiona and {Madsen}, Kristin},
        title = "{Mid-infrared Selection of Active Galactic Nuclei with the Wide-Field Infrared Survey Explorer. I. Characterizing WISE-selected Active Galactic Nuclei in COSMOS}",
      journal = {\apj},
         year = 2012,
        month = jul,
       volume = {753},
       number = {1},
          eid = {30},
        pages = {30},
          doi = {10.1088/0004-637X/753/1/30},
archivePrefix = {arXiv},
       eprint = {1205.0811},
 primaryClass = {astro-ph.CO},
       adsurl = {https://ui.adsabs.harvard.edu/abs/2012ApJ...753...30S}
}

@ARTICLE{Assef2013,
       author = {{Assef}, R.~J. and {Stern}, D. and {Kochanek}, C.~S. and {Blain}, A.~W. and {Brodwin}, M. and {Brown}, M.~J.~I. and {Donoso}, E. and {Eisenhardt}, P.~R.~M. and {Jannuzi}, B.~T. and {Jarrett}, T.~H. and {Stanford}, S.~A. and {Tsai}, C.-W. and {Wu}, J. and {Yan}, L.},
        title = "{Mid-infrared Selection of Active Galactic Nuclei with the Wide-field Infrared Survey Explorer. II. Properties of WISE-selected Active Galactic Nuclei in the NDWFS Bo{\"o}tes Field}",
      journal = {\apj},
         year = 2013,
        month = jul,
       volume = {772},
       number = {1},
          eid = {26},
        pages = {26},
          doi = {10.1088/0004-637X/772/1/26},
archivePrefix = {arXiv},
       eprint = {1209.6055},
 primaryClass = {astro-ph.CO},
       adsurl = {https://ui.adsabs.harvard.edu/abs/2013ApJ...772...26A}
}

@ARTICLE{Jansen2018,
       author = {{Jansen}, Rolf A. and {Windhorst}, Rogier A.},
        title = "{The James Webb Space Telescope North Ecliptic Pole Time-domain Field. I. Field Selection of a JWST Community Field for Time-domain Studies}",
      journal = {\pasp},
         year = 2018,
        month = dec,
       volume = {130},
       number = {994},
        pages = {124001},
          doi = {10.1088/1538-3873/aae476},
archivePrefix = {arXiv},
       eprint = {1807.05278},
 primaryClass = {astro-ph.GA},
       adsurl = {https://ui.adsabs.harvard.edu/abs/2018PASP..130l4001J}
}

@ARTICLE{Windhorst2023,
       author = {{Windhorst}, Rogier A. and {Cohen}, Seth H. and {Jansen}, Rolf A. and {Summers}, Jake and {Tompkins}, Scott and {Conselice}, Christopher J. and {Driver}, Simon P. and {Yan}, Haojing and {Coe}, Dan and {Frye}, Brenda and {Grogin}, Norman and {Koekemoer}, Anton and {Marshall}, Madeline A. and {O'Brien}, Rosalia and {Pirzkal}, Nor and {Robotham}, Aaron and {Ryan}, Russell E. and {Willmer}, Christopher N.~A. and {Carleton}, Timothy and {Diego}, Jose M. and {Keel}, William C. and {Porto}, Paolo and {Redshaw}, Caleb and {Scheller}, Sydney and {Wilkins}, Stephen M. and {Willner}, S.~P. and {Zitrin}, Adi and {Adams}, Nathan J. and {Austin}, Duncan and {Arendt}, Richard G. and {Beacom}, John F. and {Bhatawdekar}, Rachana A. and {Bradley}, Larry D. and {Broadhurst}, Tom and {Cheng}, Cheng and {Civano}, Francesca and {Dai}, Liang and {Dole}, Herv{\'e} and {D'Silva}, Jordan C.~J. and {Duncan}, Kenneth J. and {Fazio}, Giovanni G. and {Ferrami}, Giovanni and {Ferreira}, Leonardo and {Finkelstein}, Steven L. and {Furtak}, Lukas J. and {Gim}, Hansung B. and {Griffiths}, Alex and {Hammel}, Heidi B. and {Harrington}, Kevin C. and {Hathi}, Nimish P. and {Holwerda}, Benne W. and {Honor}, Rachel and {Huang}, Jia-Sheng and {Hyun}, Minhee and {Im}, Myungshin and {Joshi}, Bhavin A. and {Kamieneski}, Patrick S. and {Kelly}, Patrick and {Larson}, Rebecca L. and {Li}, Juno and {Lim}, Jeremy and {Ma}, Zhiyuan and {Maksym}, Peter and {Manzoni}, Giorgio and {Meena}, Ashish Kumar and {Milam}, Stefanie N. and {Nonino}, Mario and {Pascale}, Massimo and {Petric}, Andreea and {Pierel}, Justin D.~R. and {Polletta}, Maria del Carmen and {R{\"o}ttgering}, Huub J.~A. and {Rutkowski}, Michael J. and {Smail}, Ian and {Straughn}, Amber N. and {Strolger}, Louis-Gregory and {Swirbul}, Andi and {Trussler}, James A.~A. and {Wang}, Lifan and {Welch}, Brian and {B. Wyithe}, J. Stuart and {Yun}, Min and {Zackrisson}, Erik and {Zhang}, Jiashuo and {Zhao}, Xiurui},
        title = "{JWST PEARLS. Prime Extragalactic Areas for Reionization and Lensing Science: Project Overview and First Results}",
      journal = {\aj},
         year = 2023,
        month = jan,
       volume = {165},
       number = {1},
          eid = {13},
        pages = {13},
          doi = {10.3847/1538-3881/aca163},
archivePrefix = {arXiv},
       eprint = {2209.04119},
 primaryClass = {astro-ph.CO},
       adsurl = {https://ui.adsabs.harvard.edu/abs/2023AJ....165...13W}
}

@ARTICLE{Obrien2024,
       author = {{O'Brien}, Rosalia and {Jansen}, Rolf A. and {Grogin}, Norman A. and {Cohen}, Seth H. and {Smith}, Brent M. and {Silver}, Ross M. and {Maksym}, W.~P. and {Windhorst}, Rogier A. and {Carleton}, Timothy and {Koekemoer}, Anton M. and {Hathi}, Nimish P. and {Willmer}, Christopher N.~A. and {Frye}, Brenda L. and {Alpaslan}, M. and {Ashby}, M.~L.~N. and {Ashcraft}, T.~A. and {Bonoli}, S. and {Brisken}, W. and {Cappelluti}, N. and {Civano}, F. and {Conselice}, C.~J. and {Dhillon}, V.~S. and {Driver}, S.~P. and {Duncan}, K.~J. and {Dupke}, R. and {Elvis}, M. and {Fazio}, G.~G. and {Finkelstein}, S.~L. and {Gim}, H.~B. and {Griffiths}, A. and {Hammel}, H.~B. and {Hyun}, M. and {Im}, M. and {Jones}, V.~R. and {Kim}, D. and {Ladjelate}, B. and {Larson}, R.~L. and {Malhotra}, S. and {Marshall}, M.~A. and {Milam}, S.~N. and {Pierel}, J.~D.~R. and {Rhoads}, J.~E. and {Rodney}, S.~A. and {R{\"o}ttgering}, H.~J.~A. and {Rutkowski}, M.~J. and {Ryan}, R.~E. and {Ward}, M.~J. and {White}, C.~W. and {van Weeren}, R.~J. and {Zhao}, X. and {Summers}, J. and {D'Silva}, J.~C.~J. and {Ortiz}, R. and {Robotham}, A.~S.~G. and {Coe}, D. and {Nonino}, M. and {Pirzkal}, N. and {Yan}, H. and {Acharya}, T.},
        title = "{TREASUREHUNT: Transients and Variability Discovered with HST in the JWST North Ecliptic Pole Time-domain Field}",
      journal = {\apjs},
         year = 2024,
        month = may,
       volume = {272},
       number = {1},
          eid = {19},
        pages = {19},
          doi = {10.3847/1538-4365/ad3948},
archivePrefix = {arXiv},
       eprint = {2401.04944},
 primaryClass = {astro-ph.GA},
       adsurl = {https://ui.adsabs.harvard.edu/abs/2024ApJS..272...19O}
}

@ARTICLE{Zhao2024,
       author = {{Zhao}, Xiurui and {Civano}, Francesca and {Willmer}, Christopher N.~A. and {Bonoli}, Silvia and {Chen}, Chien-Ting and {Creech}, Samantha and {Dupke}, Renato and {Fornasini}, Francesca M. and {Jansen}, Rolf A. and {Kikuta}, Satoshi and {Koekemoer}, Anton M. and {Laha}, Sibasish and {Marchesi}, Stefano and {O'Brien}, Rosalia and {Silver}, Ross and {Willner}, S.~P. and {Windhorst}, Rogier A. and {Yan}, Haojing and {Alcaniz}, Jailson and {Benitez}, Narciso and {Carneiro}, Saulo and {Cenarro}, Javier and {Crist{\'o}bal-Hornillos}, David and {Ederoclite}, Alessandro and {Hern{\'a}n-Caballero}, Antonio and {L{\'o}pez-Sanjuan}, Carlos and {Mar{\'\i}n-Franch}, Antonio and {de Oliveira}, Claudia Mendes and {Moles}, Mariano and {Sodr{\'e}}, Jr., Laerte and {Taylor}, Keith and {Varela}, Jes{\'u}s and {Rami{\'o}}, H{\'e}ctor V{\'a}zquez},
        title = "{PEARLS: NuSTAR and XMM-Newton Extragalactic Survey of the JWST North Ecliptic Pole Time-domain Field II}",
      journal = {\apj},
         year = 2024,
        month = apr,
       volume = {965},
       number = {2},
          eid = {188},
        pages = {188},
          doi = {10.3847/1538-4357/ad2b61},
archivePrefix = {arXiv},
       eprint = {2402.13508},
 primaryClass = {astro-ph.HE},
       adsurl = {https://ui.adsabs.harvard.edu/abs/2024ApJ...965..188Z}
}

@ARTICLE{Hyun2023,
       author = {{Hyun}, Minhee and {Im}, Myungshin and {Smail}, Ian R. and {Cotton}, William D. and {Birkin}, Jack E. and {Kikuta}, Satoshi and {Shim}, Hyunjin and {Willmer}, Christopher N.~A. and {Condon}, James J. and {Windhorst}, Rogier A. and {Cohen}, Seth H. and {Jansen}, Rolf A. and {Ly}, Chun and {Matsuda}, Yuichi and {Fazio}, Giovanni G. and {Swinbank}, A.~M. and {Yan}, Haojing},
        title = "{The JCMT SCUBA-2 Survey of the James Webb Space Telescope North Ecliptic Pole Time-Domain Field}",
      journal = {\apjs},
         year = 2023,
        month = jan,
       volume = {264},
       number = {1},
          eid = {19},
        pages = {19},
          doi = {10.3847/1538-4365/ac9bf4},
archivePrefix = {arXiv},
       eprint = {2301.02786},
 primaryClass = {astro-ph.GA},
       adsurl = {https://ui.adsabs.harvard.edu/abs/2023ApJS..264...19H}
}

@ARTICLE{Willner2023,
       author = {{Willner}, S.~P. and {Gim}, Hansung B. and {Polletta}, Maria del Carmen and {Cohen}, Seth H. and {Willmer}, Christopher N.~A. and {Zhao}, Xiurui and {D'Silva}, Jordan C.~J. and {Jansen}, Rolf A. and {Koekemoer}, Anton M. and {Summers}, Jake and {Windhorst}, Rogier A. and {Coe}, Dan and {Conselice}, Christopher J. and {Driver}, Simon P. and {Frye}, Brenda and {Grogin}, Norman A. and {Marshall}, Madeline A. and {Nonino}, Mario and {Ortiz}, Rafael and {Pirzkal}, Nor and {Robotham}, Aaron and {Rutkowski}, Michael J. and {Ryan}, Russell E. and {Tompkins}, Scott and {Yan}, Haojing and {Hammel}, Heidi B. and {Milam}, Stefanie N. and {Adams}, Nathan J. and {Beacom}, John F. and {Bhatawdekar}, Rachana and {Cheng}, Cheng and {Civano}, F. and {Cotton}, W. and {Hyun}, Minhee and {Kikuta}, Satoshi and {Nyland}, K.~E. and {Peters}, W.~M. and {Petric}, Andreea and {R{\"o}ttgering}, Huub J.~A. and {Shimwell}, T. and {Yun}, Min S.},
        title = "{PEARLS: JWST Counterparts of Microjansky Radio Sources in the Time Domain Field}",
      journal = {\apj},
         year = 2023,
        month = dec,
       volume = {958},
       number = {2},
          eid = {176},
        pages = {176},
          doi = {10.3847/1538-4357/acfdfb},
archivePrefix = {arXiv},
       eprint = {2309.13008},
 primaryClass = {astro-ph.GA},
       adsurl = {https://ui.adsabs.harvard.edu/abs/2023ApJ...958..176W}
}

@ARTICLE{Willner2026,
       author = {{Willner}, S.~P. and {Gim}, Hansung B. and {Polletta}, Maria del Carmen and {Bowling}, Gibson B. and {Cohen}, Seth H. and {Estrada-Carpenter}, Vicente and {Koekemoer}, Anton M. and {O'Brien}, Rosalia and {Pigarelli}, Alex and {Willmer}, Christopher N.~A. and {Windhorst}, Rogier A. and {Jansen}, Rolf A. and {Ortiz}, III, Rafael and {Summers}, Jake and {Cotton}, William and {Civano}, Francesca and {Grogin}, Norman A. and {Maksym}, W.~P. and {Saikia}, Payaswini and {Silver}, Ross M. and {Zhao}, Xiurui},
        title = "{PEARLS: JWST Counterparts of Microjansky Radio Sources in the North Ecliptic Pole Time Domain Field. II. All Four Spokes}",
      journal = {\apj},
         year = 2026,
        month = aug,
       volume = {1006},
       number = {2},
          eid = {219},
        pages = {219},
          doi = {10.3847/1538-4357/ae796c},
archivePrefix = {arXiv},
       eprint = {2605.17040},
 primaryClass = {astro-ph.GA},
       adsurl = {https://ui.adsabs.harvard.edu/abs/2026ApJ..1006..219W}
}

@ARTICLE{Casertano2000,
       author = {{Casertano}, Stefano and {de Mello}, Du{\'\i}lia and {Dickinson}, Mark and {Ferguson}, Henry C. and {Fruchter}, Andrew S. and {Gonzalez-Lopezlira}, Rosa A. and {Heyer}, Inge and {Hook}, Richard N. and {Levay}, Zolt and {Lucas}, Ray A. and {Mack}, Jennifer and {Makidon}, Russell B. and {Mutchler}, Max and {Smith}, T. Ed and {Stiavelli}, Massimo and {Wiggs}, Michael S. and {Williams}, Robert E.},
        title = "{WFPC2 Observations of the Hubble Deep Field South}",
      journal = {\aj},
         year = 2000,
        month = dec,
       volume = {120},
       number = {6},
        pages = {2747-2824},
          doi = {10.1086/316851},
archivePrefix = {arXiv},
       eprint = {astro-ph/0010245},
 primaryClass = {astro-ph},
       adsurl = {https://ui.adsabs.harvard.edu/abs/2000AJ....120.2747C}
}

@INPROCEEDINGS{Krist2011,
       author = {{Krist}, John E. and {Hook}, Richard N. and {Stoehr}, Felix},
        title = "{20 years of Hubble Space Telescope optical modeling using Tiny Tim}",
    booktitle = {Optical Modeling and Performance Predictions V},
         year = 2011,
       editor = {{Kahan}, Mark A.},
       series = {Society of Photo-Optical Instrumentation Engineers (SPIE) Conference Series},
       volume = {8127},
        month = oct,
          eid = {81270J},
        pages = {81270J},
          doi = {10.1117/12.892762},
       adsurl = {https://ui.adsabs.harvard.edu/abs/2011SPIE.8127E..0JK}
}

@INPROCEEDINGS{Krist1993,
       author = {{Krist}, J.},
        title = "{Tiny Tim : an HST PSF Simulator}",
    booktitle = {Astronomical Data Analysis Software and Systems II},
         year = 1993,
       editor = {{Hanisch}, R.~J. and {Brissenden}, R.~J.~V. and {Barnes}, J.},
       series = {Astronomical Society of the Pacific Conference Series},
       volume = {52},
        month = jan,
        pages = {536},
       adsurl = {https://ui.adsabs.harvard.edu/abs/1993ASPC...52..536K}
}

@ARTICLE{Windhorst2011,
       author = {{Windhorst}, Rogier A. and {Cohen}, Seth H. and {Hathi}, Nimish P. and {McCarthy}, Patrick J. and {Ryan}, Jr., Russell E. and {Yan}, Haojing and {Baldry}, Ivan K. and {Driver}, Simon P. and {Frogel}, Jay A. and {Hill}, David T. and {Kelvin}, Lee S. and {Koekemoer}, Anton M. and {Mechtley}, Matt and {O'Connell}, Robert W. and {Robotham}, Aaron S.~G. and {Rutkowski}, Michael J. and {Seibert}, Mark and {Straughn}, Amber N. and {Tuffs}, Richard J. and {Balick}, Bruce and {Bond}, Howard E. and {Bushouse}, Howard and {Calzetti}, Daniela and {Crockett}, Mark and {Disney}, Michael J. and {Dopita}, Michael A. and {Hall}, Donald N.~B. and {Holtzman}, Jon A. and {Kaviraj}, Sugata and {Kimble}, Randy A. and {MacKenty}, John W. and {Mutchler}, Max and {Paresce}, Francesco and {Saha}, Abihit and {Silk}, Joseph I. and {Trauger}, John T. and {Walker}, Alistair R. and {Whitmore}, Bradley C. and {Young}, Erick T.},
        title = "{The Hubble Space Telescope Wide Field Camera 3 Early Release Science Data: Panchromatic Faint Object Counts for 0.2-2 {\ensuremath{\mu}}m Wavelength}",
      journal = {\apjs},
         year = 2011,
        month = apr,
       volume = {193},
       number = {2},
          eid = {27},
        pages = {27},
          doi = {10.1088/0067-0049/193/2/27},
archivePrefix = {arXiv},
       eprint = {1005.2776},
 primaryClass = {astro-ph.CO},
       adsurl = {https://ui.adsabs.harvard.edu/abs/2011ApJS..193...27W}
}

@ARTICLE{vanderWel2012,
       author = {{van der Wel}, A. and {Bell}, E.~F. and {H{\"a}ussler}, B. and {McGrath}, E.~J. and {Chang}, Yu-Yen and {Guo}, Yicheng and {McIntosh}, D.~H. and {Rix}, H.-W. and {Barden}, M. and {Cheung}, E. and {Faber}, S.~M. and {Ferguson}, H.~C. and {Galametz}, A. and {Grogin}, N.~A. and {Hartley}, W. and {Kartaltepe}, J.~S. and {Kocevski}, D.~D. and {Koekemoer}, A.~M. and {Lotz}, J. and {Mozena}, M. and {Peth}, M.~A. and {Peng}, Chien Y.},
        title = "{Structural Parameters of Galaxies in CANDELS}",
      journal = {\apjs},
         year = 2012,
        month = dec,
       volume = {203},
       number = {2},
          eid = {24},
        pages = {24},
          doi = {10.1088/0067-0049/203/2/24},
archivePrefix = {arXiv},
       eprint = {1211.6954},
 primaryClass = {astro-ph.CO},
       adsurl = {https://ui.adsabs.harvard.edu/abs/2012ApJS..203...24V}
}

@ARTICLE{Sersic1963,
       author = {{S{\'e}rsic}, J.~L.},
        title = "{Influence of the atmospheric and instrumental dispersion on the brightness distribution in a galaxy}",
      journal = {Boletin de la Asociacion Argentina de Astronomia La Plata Argentina},
         year = 1963,
        month = feb,
       volume = {6},
        pages = {41-43},
       adsurl = {https://ui.adsabs.harvard.edu/abs/1963BAAA....6...41S}
}

@ARTICLE{deVaucouleurs1948,
       author = {{de Vaucouleurs}, Gerard},
        title = "{Recherches sur les Nebuleuses Extragalactiques}",
      journal = {Annales d'Astrophysique},
         year = 1948,
        month = jan,
       volume = {11},
        pages = {247},
       adsurl = {https://ui.adsabs.harvard.edu/abs/1948AnAp...11..247D}
}

@BOOK{Sersic1968,
       author = {{S{\'e}rsic}, J.~L.},
        title = "{Atlas de Galaxias Australes}",
         year = 1968,
       adsurl = {https://ui.adsabs.harvard.edu/abs/1968adga.book.....S}
}

@ARTICLE{Blanton2003,
       author = {{Blanton}, Michael R. and {Hogg}, David W. and {Bahcall}, Neta A. and {Baldry}, Ivan K. and {Brinkmann}, J. and {Csabai}, Istv{\'a}n and {Eisenstein}, Daniel and {Fukugita}, Masataka and {Gunn}, James E. and {Ivezi{\'c}}, {\v{Z}}eljko and {Lamb}, D.~Q. and {Lupton}, Robert H. and {Loveday}, Jon and {Munn}, Jeffrey A. and {Nichol}, R.~C. and {Okamura}, Sadanori and {Schlegel}, David J. and {Shimasaku}, Kazuhiro and {Strauss}, Michael A. and {Vogeley}, Michael S. and {Weinberg}, David H.},
        title = "{The Broadband Optical Properties of Galaxies with Redshifts 0.02<z<0.22}",
      journal = {\apj},
         year = 2003,
        month = sep,
       volume = {594},
       number = {1},
        pages = {186-207},
          doi = {10.1086/375528},
archivePrefix = {arXiv},
       eprint = {astro-ph/0209479},
 primaryClass = {astro-ph},
       adsurl = {https://ui.adsabs.harvard.edu/abs/2003ApJ...594..186B}
}

@ARTICLE{Driver2011,
       author = {{Driver}, S.~P. and {Hill}, D.~T. and {Kelvin}, L.~S. and {Robotham}, A.~S.~G. and {Liske}, J. and {Norberg}, P. and {Baldry}, I.~K. and {Bamford}, S.~P. and {Hopkins}, A.~M. and {Loveday}, J. and {Peacock}, J.~A. and {Andrae}, E. and {Bland-Hawthorn}, J. and {Brough}, S. and {Brown}, M.~J.~I. and {Cameron}, E. and {Ching}, J.~H.~Y. and {Colless}, M. and {Conselice}, C.~J. and {Croom}, S.~M. and {Cross}, N.~J.~G. and {de Propris}, R. and {Dye}, S. and {Drinkwater}, M.~J. and {Ellis}, S. and {Graham}, Alister W. and {Grootes}, M.~W. and {Gunawardhana}, M. and {Jones}, D.~H. and {van Kampen}, E. and {Maraston}, C. and {Nichol}, R.~C. and {Parkinson}, H.~R. and {Phillipps}, S. and {Pimbblet}, K. and {Popescu}, C.~C. and {Prescott}, M. and {Roseboom}, I.~G. and {Sadler}, E.~M. and {Sansom}, A.~E. and {Sharp}, R.~G. and {Smith}, D.~J.~B. and {Taylor}, E. and {Thomas}, D. and {Tuffs}, R.~J. and {Wijesinghe}, D. and {Dunne}, L. and {Frenk}, C.~S. and {Jarvis}, M.~J. and {Madore}, B.~F. and {Meyer}, M.~J. and {Seibert}, M. and {Staveley-Smith}, L. and {Sutherland}, W.~J. and {Warren}, S.~J.},
        title = "{Galaxy and Mass Assembly (GAMA): survey diagnostics and core data release}",
      journal = {\mnras},
         year = 2011,
        month = may,
       volume = {413},
       number = {2},
        pages = {971-995},
          doi = {10.1111/j.1365-2966.2010.18188.x},
archivePrefix = {arXiv},
       eprint = {1009.0614},
 primaryClass = {astro-ph.CO},
       adsurl = {https://ui.adsabs.harvard.edu/abs/2011MNRAS.413..971D}
}

@ARTICLE{Luc2011,
       author = {{Simard}, Luc and {Mendel}, J. Trevor and {Patton}, David R. and {Ellison}, Sara L. and {McConnachie}, Alan W.},
        title = "{A Catalog of Bulge+disk Decompositions and Updated Photometry for 1.12 Million Galaxies in the Sloan Digital Sky Survey}",
      journal = {\apjs},
         year = 2011,
        month = sep,
       volume = {196},
       number = {1},
          eid = {11},
        pages = {11},
          doi = {10.1088/0067-0049/196/1/11},
archivePrefix = {arXiv},
       eprint = {1107.1518},
 primaryClass = {astro-ph.CO},
       adsurl = {https://ui.adsabs.harvard.edu/abs/2011ApJS..196...11S}
}

@ARTICLE{Ciotti1999,
       author = {{Ciotti}, L. and {Bertin}, G.},
        title = "{Analytical properties of the R$^{1/m}$ law}",
      journal = {\aap},
         year = 1999,
        month = dec,
       volume = {352},
        pages = {447-451},
          doi = {10.48550/arXiv.astro-ph/9911078},
archivePrefix = {arXiv},
       eprint = {astro-ph/9911078},
 primaryClass = {astro-ph},
       adsurl = {https://ui.adsabs.harvard.edu/abs/1999A&A...352..447C}
}

@ARTICLE{Gabor2009,
       author = {{Gabor}, J.~M. and {Impey}, C.~D. and {Jahnke}, K. and {Simmons}, B.~D. and {Trump}, J.~R. and {Koekemoer}, A.~M. and {Brusa}, M. and {Cappelluti}, N. and {Schinnerer}, E. and {Smol{\v{c}}i{\'c}}, V. and {Salvato}, M. and {Rhodes}, J.~D. and {Mobasher}, B. and {Capak}, P. and {Massey}, R. and {Leauthaud}, A. and {Scoville}, N.},
        title = "{Active Galactic Nucleus Host Galaxy Morphologies in COSMOS}",
      journal = {\apj},
         year = 2009,
        month = jan,
       volume = {691},
       number = {1},
        pages = {705-722},
          doi = {10.1088/0004-637X/691/1/705},
archivePrefix = {arXiv},
       eprint = {0809.0309},
 primaryClass = {astro-ph},
       adsurl = {https://ui.adsabs.harvard.edu/abs/2009ApJ...691..705G}
}

@ARTICLE{Haussler2007,
       author = {{H{\"a}ussler}, Boris and {McIntosh}, Daniel H. and {Barden}, Marco and {Bell}, Eric F. and {Rix}, Hans-Walter and {Borch}, Andrea and {Beckwith}, Steven V.~W. and {Caldwell}, John A.~R. and {Heymans}, Catherine and {Jahnke}, Knud and {Jogee}, Shardha and {Koposov}, Sergey E. and {Meisenheimer}, Klaus and {S{\'a}nchez}, Sebastian F. and {Somerville}, Rachel S. and {Wisotzki}, Lutz and {Wolf}, Christian},
        title = "{GEMS: Galaxy Fitting Catalogs and Testing Parametric Galaxy Fitting Codes: GALFIT and GIM2D}",
      journal = {\apjs},
         year = 2007,
        month = oct,
       volume = {172},
       number = {2},
        pages = {615-633},
          doi = {10.1086/518836},
archivePrefix = {arXiv},
       eprint = {0704.2601},
 primaryClass = {astro-ph},
       adsurl = {https://ui.adsabs.harvard.edu/abs/2007ApJS..172..615H}
}

@ARTICLE{Caon1993,
       author = {{Caon}, N. and {Capaccioli}, M. and {D'Onofrio}, M.},
        title = "{On the shape of the light profiles of early-type galaxies.}",
      journal = {\mnras},
         year = 1993,
        month = dec,
       volume = {265},
        pages = {1013-1021},
          doi = {10.1093/mnras/265.4.1013},
archivePrefix = {arXiv},
       eprint = {astro-ph/9309013},
 primaryClass = {astro-ph},
       adsurl = {https://ui.adsabs.harvard.edu/abs/1993MNRAS.265.1013C}
}

@ARTICLE{Dewsnap2025,
       author = {{Dewsnap}, Callum and {Barmby}, Pauline and {Gallagher}, Sarah C.},
        title = "{AGN-Host Galaxy Image Decomposition with JWST: Limitations of S{\'e}rsic Profile Models}",
      journal = {\pasp},
         year = 2025,
        month = nov,
       volume = {137},
       number = {11},
          eid = {114101},
        pages = {114101},
          doi = {10.1088/1538-3873/ae1662},
archivePrefix = {arXiv},
       eprint = {2510.27214},
 primaryClass = {astro-ph.GA},
       adsurl = {https://ui.adsabs.harvard.edu/abs/2025PASP..137k4101D}
}

@ARTICLE{Papovich2004,
       author = {{Papovich}, Casey and {Dickinson}, Mark and {Ferguson}, Henry C. and {Giavalisco}, Mauro and {Lotz}, Jennifer and {Madau}, Piero and {Idzi}, Rafal and {Kretchmer}, Claudia and {Moustakas}, Leonidas A. and {de Mello}, Duilia F. and {Gardner}, Jonathan P. and {Rieke}, Marcia J. and {Somerville}, Rachel S. and {Stern}, Daniel},
        title = "{Evolution in the Colors of Lyman Break Galaxies from z\raisebox{-0.5ex}\textasciitilde4 to z\raisebox{-0.5ex}\textasciitilde3}",
      journal = {\apjl},
         year = 2004,
        month = jan,
       volume = {600},
       number = {2},
        pages = {L111-L114},
          doi = {10.1086/381075},
archivePrefix = {arXiv},
       eprint = {astro-ph/0310888},
 primaryClass = {astro-ph},
       adsurl = {https://ui.adsabs.harvard.edu/abs/2004ApJ...600L.111P}
}

@ARTICLE{CarvajalBohorquez2025,
       author = {{Carvajal-Bohorquez}, C. and {Ciesla}, L. and {Laporte}, N. and {Boquien}, M. and {Buat}, V. and {Ilbert}, O. and {Aufort}, G. and {Shuntov}, M. and {Witten}, C. and {Oesch}, P.~A. and {Covelo-Paz}, A.},
        title = "{Stochastic star formation activity of galaxies within the first billion years probed by JWST}",
      journal = {\aap},
         year = 2025,
        month = dec,
       volume = {704},
          eid = {A290},
        pages = {A290},
          doi = {10.1051/0004-6361/202556471},
archivePrefix = {arXiv},
       eprint = {2507.13160},
 primaryClass = {astro-ph.GA},
       adsurl = {https://ui.adsabs.harvard.edu/abs/2025A&A...704A.290C}
}

@ARTICLE{Bruzual2003,
       author = {{Bruzual}, G. and {Charlot}, S.},
        title = "{Stellar population synthesis at the resolution of 2003}",
      journal = {\mnras},
         year = 2003,
        month = oct,
       volume = {344},
       number = {4},
        pages = {1000-1028},
          doi = {10.1046/j.1365-8711.2003.06897.x},
archivePrefix = {arXiv},
       eprint = {astro-ph/0309134},
 primaryClass = {astro-ph},
       adsurl = {https://ui.adsabs.harvard.edu/abs/2003MNRAS.344.1000B}
}

@ARTICLE{Chabrier2003,
       author = {{Chabrier}, Gilles},
        title = "{Galactic Stellar and Substellar Initial Mass Function}",
      journal = {\pasp},
         year = 2003,
        month = jul,
       volume = {115},
       number = {809},
        pages = {763-795},
          doi = {10.1086/376392},
archivePrefix = {arXiv},
       eprint = {astro-ph/0304382},
 primaryClass = {astro-ph},
       adsurl = {https://ui.adsabs.harvard.edu/abs/2003PASP..115..763C}
}

@ARTICLE{Calzetti2000,
       author = {{Calzetti}, Daniela and {Armus}, Lee and {Bohlin}, Ralph C. and {Kinney}, Anne L. and {Koornneef}, Jan and {Storchi-Bergmann}, Thaisa},
        title = "{The Dust Content and Opacity of Actively Star-forming Galaxies}",
      journal = {\apj},
         year = 2000,
        month = apr,
       volume = {533},
       number = {2},
        pages = {682-695},
          doi = {10.1086/308692},
archivePrefix = {arXiv},
       eprint = {astro-ph/9911459},
 primaryClass = {astro-ph},
       adsurl = {https://ui.adsabs.harvard.edu/abs/2000ApJ...533..682C}
}

@ARTICLE{Dale2014,
       author = {{Dale}, Daniel A. and {Helou}, George and {Magdis}, Georgios E. and {Armus}, Lee and {D{\'\i}az-Santos}, Tanio and {Shi}, Yong},
        title = "{A Two-parameter Model for the Infrared/Submillimeter/Radio Spectral Energy Distributions of Galaxies and Active Galactic Nuclei}",
      journal = {\apj},
         year = 2014,
        month = mar,
       volume = {784},
       number = {1},
          eid = {83},
        pages = {83},
          doi = {10.1088/0004-637X/784/1/83},
archivePrefix = {arXiv},
       eprint = {1402.1495},
 primaryClass = {astro-ph.GA},
       adsurl = {https://ui.adsabs.harvard.edu/abs/2014ApJ...784...83D}
}

@ARTICLE{Meiksin2006,
       author = {{Meiksin}, Avery},
        title = "{Colour corrections for high-redshift objects due to intergalactic attenuation}",
      journal = {\mnras},
         year = 2006,
        month = jan,
       volume = {365},
       number = {3},
        pages = {807-812},
          doi = {10.1111/j.1365-2966.2005.09756.x},
archivePrefix = {arXiv},
       eprint = {astro-ph/0512435},
 primaryClass = {astro-ph},
       adsurl = {https://ui.adsabs.harvard.edu/abs/2006MNRAS.365..807M}
}

@ARTICLE{Kim2019,
       author = {{Kim}, Duho and {Jansen}, Rolf A. and {Windhorst}, Rogier A. and {Cohen}, Seth H. and {McCabe}, Tyler J.},
        title = "{Analysis of the Spatially Resolved V-3.6 {\ensuremath{\mu}}m Colors and Dust Extinction in 257 Nearby NGC and IC Galaxies}",
      journal = {\apj},
         year = 2019,
        month = oct,
       volume = {884},
       number = {1},
          eid = {21},
        pages = {21},
          doi = {10.3847/1538-4357/ab385c},
archivePrefix = {arXiv},
       eprint = {1901.00565},
 primaryClass = {astro-ph.GA},
       adsurl = {https://ui.adsabs.harvard.edu/abs/2019ApJ...884...21K}
}

@INPROCEEDINGS{Blank2012,
       author = {{Blank}, M. and {Duschl}, W.~J.},
        title = "{A fresh look at the starburst-AGN connection}",
    booktitle = {Journal of Physics Conference Series},
         year = 2012,
       series = {Journal of Physics Conference Series},
       volume = {372},
        month = jul,
    publisher = {IOP},
          eid = {012053},
        pages = {012053},
          doi = {10.1088/1742-6596/372/1/012053},
       adsurl = {https://ui.adsabs.harvard.edu/abs/2012JPhCS.372a2053B}
}

@ARTICLE{Farrah2003,
       author = {{Farrah}, D. and {Afonso}, J. and {Efstathiou}, A. and {Rowan-Robinson}, M. and {Fox}, M. and {Clements}, D.},
        title = "{Starburst and AGN activity in ultraluminous infrared galaxies}",
      journal = {\mnras},
         year = 2003,
        month = aug,
       volume = {343},
       number = {2},
        pages = {585-607},
          doi = {10.1046/j.1365-8711.2003.06696.x},
archivePrefix = {arXiv},
       eprint = {astro-ph/0304154},
 primaryClass = {astro-ph},
       adsurl = {https://ui.adsabs.harvard.edu/abs/2003MNRAS.343..585F}
}

@ARTICLE{Stalevski2012,
       author = {{Stalevski}, Marko},
        title = "{SKIRTOR - database of modelled AGN dusty torus SEDs}",
      journal = {Bulgarian Astronomical Journal},
         year = 2012,
        month = sep,
       volume = {18},
       number = {3},
        pages = {3},
       adsurl = {https://ui.adsabs.harvard.edu/abs/2012BlgAJ..18c...3S}
}

@ARTICLE{Stalevski2016,
       author = {{Stalevski}, Marko and {Ricci}, Claudio and {Ueda}, Yoshihiro and {Lira}, Paulina and {Fritz}, Jacopo and {Baes}, Maarten},
        title = "{The dust covering factor in active galactic nuclei}",
      journal = {\mnras},
         year = 2016,
        month = may,
       volume = {458},
       number = {3},
        pages = {2288-2302},
          doi = {10.1093/mnras/stw444},
archivePrefix = {arXiv},
       eprint = {1602.06954},
 primaryClass = {astro-ph.GA},
       adsurl = {https://ui.adsabs.harvard.edu/abs/2016MNRAS.458.2288S}
}

@ARTICLE{Hatziminaoglou2009,
       author = {{Hatziminaoglou}, E. and {Fritz}, J. and {Jarrett}, T.~H.},
        title = "{Properties of dusty tori in active galactic nuclei - II. Type 2 AGN}",
      journal = {\mnras},
         year = 2009,
        month = nov,
       volume = {399},
       number = {3},
        pages = {1206-1222},
          doi = {10.1111/j.1365-2966.2009.15390.x},
archivePrefix = {arXiv},
       eprint = {0907.2389},
 primaryClass = {astro-ph.CO},
       adsurl = {https://ui.adsabs.harvard.edu/abs/2009MNRAS.399.1206H}
}

@ARTICLE{Pozzi2010,
       author = {{Pozzi}, F. and {Vignali}, C. and {Comastri}, A. and {Bellocchi}, E. and {Fritz}, J. and {Gruppioni}, C. and {Mignoli}, M. and {Maiolino}, R. and {Pozzetti}, L. and {Brusa}, M. and {Fiore}, F. and {Zamorani}, G.},
        title = "{The HELLAS2XMM survey. XIII. Multi-component analysis of the spectral energy distribution of obscured AGN}",
      journal = {\aap},
         year = 2010,
        month = jul,
       volume = {517},
          eid = {A11},
        pages = {A11},
          doi = {10.1051/0004-6361/200913043},
archivePrefix = {arXiv},
       eprint = {1003.5563},
 primaryClass = {astro-ph.CO},
       adsurl = {https://ui.adsabs.harvard.edu/abs/2010A&A...517A..11P}
}

@ARTICLE{Treister2004,
       author = {{Treister}, Ezequiel and {Urry}, C. Megan and {Chatzichristou}, Eleni and {Bauer}, Franz and {Alexander}, David M. and {Koekemoer}, Anton and {Van Duyne}, Jeffrey and {Brandt}, William N. and {Bergeron}, Jacqueline and {Stern}, Daniel and {Moustakas}, Leonidas A. and {Chary}, Ranga-Ram and {Conselice}, Christopher and {Cristiani}, Stefano and {Grogin}, Norman},
        title = "{Obscured Active Galactic Nuclei and the X-Ray, Optical, and Far-Infrared Number Counts of Active Galactic Nuclei in the GOODS Fields}",
      journal = {\apj},
         year = 2004,
        month = nov,
       volume = {616},
       number = {1},
        pages = {123-135},
          doi = {10.1086/424891},
archivePrefix = {arXiv},
       eprint = {astro-ph/0408099},
 primaryClass = {astro-ph},
       adsurl = {https://ui.adsabs.harvard.edu/abs/2004ApJ...616..123T}
}

@ARTICLE{Yang2023,
       author = {{Yang}, G. and {Caputi}, K.~I. and {Papovich}, C. and {Arrabal Haro}, P. and {Bagley}, M.~B. and {Behroozi}, P. and {Bell}, E.~F. and {Bisigello}, L. and {Buat}, V. and {Burgarella}, D. and {Cheng}, Y. and {Cleri}, N.~J. and {Dav{\'e}}, R. and {Dickinson}, M. and {Elbaz}, D. and {Ferguson}, H.~C. and {Finkelstein}, S.~L. and {Grogin}, N.~A. and {Hathi}, N.~P. and {Hirschmann}, M. and {Holwerda}, B.~W. and {Huertas-Company}, M. and {Hutchison}, T.~A. and {Iani}, E. and {Kartaltepe}, J.~S. and {Kirkpatrick}, A. and {Kocevski}, D.~D. and {Koekemoer}, A.~M. and {Kokorev}, V. and {Larson}, R.~L. and {Lucas}, R.~A. and {P{\'e}rez-Gonz{\'a}lez}, P.~G. and {Rinaldi}, P. and {Shen}, L. and {Trump}, J.~R. and {de la Vega}, A. and {Yung}, L.~Y.~A. and {Zavala}, J.~A.},
        title = "{CEERS Key Paper. VI. JWST/MIRI Uncovers a Large Population of Obscured AGN at High Redshifts}",
      journal = {\apjl},
         year = 2023,
        month = jun,
       volume = {950},
       number = {1},
          eid = {L5},
        pages = {L5},
          doi = {10.3847/2041-8213/acd639},
archivePrefix = {arXiv},
       eprint = {2303.11736},
 primaryClass = {astro-ph.GA},
       adsurl = {https://ui.adsabs.harvard.edu/abs/2023ApJ...950L...5Y}
}

@ARTICLE{Oke1983,
       author = {{Oke}, J.~B. and {Gunn}, J.~E.},
        title = "{Secondary standard stars for absolute spectrophotometry.}",
      journal = {\apj},
         year = 1983,
        month = mar,
       volume = {266},
        pages = {713-717},
          doi = {10.1086/160817},
       adsurl = {https://ui.adsabs.harvard.edu/abs/1983ApJ...266..713O}
}

@ARTICLE{Planck2016,
       author = {{Planck Collaboration} and {Ade}, P.~A.~R. and {Aghanim}, N. and {Arnaud}, M. and {Ashdown}, M. and {Aumont}, J. and {Baccigalupi}, C. and {Banday}, A.~J. and {Barreiro}, R.~B. and {Bartlett}, J.~G. and {Bartolo}, N. and {Battaner}, E. and {Battye}, R. and {Benabed}, K. and {Beno{\^\i}t}, A. and {Benoit-L{\'e}vy}, A. and {Bernard}, J.-P. and {Bersanelli}, M. and {Bielewicz}, P. and {Bock}, J.~J. and {Bonaldi}, A. and {Bonavera}, L. and {Bond}, J.~R. and {Borrill}, J. and {Bouchet}, F.~R. and {Boulanger}, F. and {Bucher}, M. and {Burigana}, C. and {Butler}, R.~C. and {Calabrese}, E. and {Cardoso}, J.-F. and {Catalano}, A. and {Challinor}, A. and {Chamballu}, A. and {Chary}, R.-R. and {Chiang}, H.~C. and {Chluba}, J. and {Christensen}, P.~R. and {Church}, S. and {Clements}, D.~L. and {Colombi}, S. and {Colombo}, L.~P.~L. and {Combet}, C. and {Coulais}, A. and {Crill}, B.~P. and {Curto}, A. and {Cuttaia}, F. and {Danese}, L. and {Davies}, R.~D. and {Davis}, R.~J. and {de Bernardis}, P. and {de Rosa}, A. and {de Zotti}, G. and {Delabrouille}, J. and {D{\'e}sert}, F.-X. and {Di Valentino}, E. and {Dickinson}, C. and {Diego}, J.~M. and {Dolag}, K. and {Dole}, H. and {Donzelli}, S. and {Dor{\'e}}, O. and {Douspis}, M. and {Ducout}, A. and {Dunkley}, J. and {Dupac}, X. and {Efstathiou}, G. and {Elsner}, F. and {En{\ss}lin}, T.~A. and {Eriksen}, H.~K. and {Farhang}, M. and {Fergusson}, J. and {Finelli}, F. and {Forni}, O. and {Frailis}, M. and {Fraisse}, A.~A. and {Franceschi}, E. and {Frejsel}, A. and {Galeotta}, S. and {Galli}, S. and {Ganga}, K. and {Gauthier}, C. and {Gerbino}, M. and {Ghosh}, T. and {Giard}, M. and {Giraud-H{\'e}raud}, Y. and {Giusarma}, E. and {Gjerl{\o}w}, E. and {Gonz{\'a}lez-Nuevo}, J. and {G{\'o}rski}, K.~M. and {Gratton}, S. and {Gregorio}, A. and {Gruppuso}, A. and {Gudmundsson}, J.~E. and {Hamann}, J. and {Hansen}, F.~K. and {Hanson}, D. and {Harrison}, D.~L. and {Helou}, G. and {Henrot-Versill{\'e}}, S. and {Hern{\'a}ndez-Monteagudo}, C. and {Herranz}, D. and {Hildebrandt}, S.~R. and {Hivon}, E. and {Hobson}, M. and {Holmes}, W.~A. and {Hornstrup}, A. and {Hovest}, W. and {Huang}, Z. and {Huffenberger}, K.~M. and {Hurier}, G. and {Jaffe}, A.~H. and {Jaffe}, T.~R. and {Jones}, W.~C. and {Juvela}, M. and {Keih{\"a}nen}, E. and {Keskitalo}, R. and {Kisner}, T.~S. and {Kneissl}, R. and {Knoche}, J. and {Knox}, L. and {Kunz}, M. and {Kurki-Suonio}, H. and {Lagache}, G. and {L{\"a}hteenm{\"a}ki}, A. and {Lamarre}, J.-M. and {Lasenby}, A. and {Lattanzi}, M. and {Lawrence}, C.~R. and {Leahy}, J.~P. and {Leonardi}, R. and {Lesgourgues}, J. and {Levrier}, F. and {Lewis}, A. and {Liguori}, M. and {Lilje}, P.~B. and {Linden-V{\o}rnle}, M. and {L{\'o}pez-Caniego}, M. and {Lubin}, P.~M. and {Mac{\'\i}as-P{\'e}rez}, J.~F. and {Maggio}, G. and {Maino}, D. and {Mandolesi}, N. and {Mangilli}, A. and {Marchini}, A. and {Maris}, M. and {Martin}, P.~G. and {Martinelli}, M. and {Mart{\'\i}nez-Gonz{\'a}lez}, E. and {Masi}, S. and {Matarrese}, S. and {McGehee}, P. and {Meinhold}, P.~R. and {Melchiorri}, A. and {Melin}, J.-B. and {Mendes}, L. and {Mennella}, A. and {Migliaccio}, M. and {Millea}, M. and {Mitra}, S. and {Miville-Desch{\^e}nes}, M.-A. and {Moneti}, A. and {Montier}, L. and {Morgante}, G. and {Mortlock}, D. and {Moss}, A. and {Munshi}, D. and {Murphy}, J.~A. and {Naselsky}, P. and {Nati}, F. and {Natoli}, P. and {Netterfield}, C.~B. and {N{\o}rgaard-Nielsen}, H.~U. and {Noviello}, F. and {Novikov}, D. and {Novikov}, I. and {Oxborrow}, C.~A. and {Paci}, F. and {Pagano}, L. and {Pajot}, F. and {Paladini}, R. and {Paoletti}, D. and {Partridge}, B. and {Pasian}, F. and {Patanchon}, G. and {Pearson}, T.~J. and {Perdereau}, O. and {Perotto}, L. and {Perrotta}, F. and {Pettorino}, V. and {Piacentini}, F. and {Piat}, M. and {Pierpaoli}, E. and {Pietrobon}, D. and {Plaszczynski}, S. and {Pointecouteau}, E. and {Polenta}, G. and {Popa}, L. and {Pratt}, G.~W. and {Pr{\'e}zeau}, G.},
        title = "{Planck 2015 results. XIII. Cosmological parameters}",
      journal = {\aap},
         year = 2016,
        month = sep,
       volume = {594},
          eid = {A13},
        pages = {A13},
          doi = {10.1051/0004-6361/201525830},
archivePrefix = {arXiv},
       eprint = {1502.01589},
 primaryClass = {astro-ph.CO},
       adsurl = {https://ui.adsabs.harvard.edu/abs/2016A&A...594A..13P}
}

@ARTICLE{Planck2020,
       author = {{Planck Collaboration} and {Aghanim}, N. and {Akrami}, Y. and {Ashdown}, M. and {Aumont}, J. and {Baccigalupi}, C. and {Ballardini}, M. and {Banday}, A.~J. and {Barreiro}, R.~B. and {Bartolo}, N. and {Basak}, S. and {Battye}, R. and {Benabed}, K. and {Bernard}, J.-P. and {Bersanelli}, M. and {Bielewicz}, P. and {Bock}, J.~J. and {Bond}, J.~R. and {Borrill}, J. and {Bouchet}, F.~R. and {Boulanger}, F. and {Bucher}, M. and {Burigana}, C. and {Butler}, R.~C. and {Calabrese}, E. and {Cardoso}, J.-F. and {Carron}, J. and {Challinor}, A. and {Chiang}, H.~C. and {Chluba}, J. and {Colombo}, L.~P.~L. and {Combet}, C. and {Contreras}, D. and {Crill}, B.~P. and {Cuttaia}, F. and {de Bernardis}, P. and {de Zotti}, G. and {Delabrouille}, J. and {Delouis}, J.-M. and {Di Valentino}, E. and {Diego}, J.~M. and {Dor{\'e}}, O. and {Douspis}, M. and {Ducout}, A. and {Dupac}, X. and {Dusini}, S. and {Efstathiou}, G. and {Elsner}, F. and {En{\ss}lin}, T.~A. and {Eriksen}, H.~K. and {Fantaye}, Y. and {Farhang}, M. and {Fergusson}, J. and {Fernandez-Cobos}, R. and {Finelli}, F. and {Forastieri}, F. and {Frailis}, M. and {Fraisse}, A.~A. and {Franceschi}, E. and {Frolov}, A. and {Galeotta}, S. and {Galli}, S. and {Ganga}, K. and {G{\'e}nova-Santos}, R.~T. and {Gerbino}, M. and {Ghosh}, T. and {Gonz{\'a}lez-Nuevo}, J. and {G{\'o}rski}, K.~M. and {Gratton}, S. and {Gruppuso}, A. and {Gudmundsson}, J.~E. and {Hamann}, J. and {Handley}, W. and {Hansen}, F.~K. and {Herranz}, D. and {Hildebrandt}, S.~R. and {Hivon}, E. and {Huang}, Z. and {Jaffe}, A.~H. and {Jones}, W.~C. and {Karakci}, A. and {Keih{\"a}nen}, E. and {Keskitalo}, R. and {Kiiveri}, K. and {Kim}, J. and {Kisner}, T.~S. and {Knox}, L. and {Krachmalnicoff}, N. and {Kunz}, M. and {Kurki-Suonio}, H. and {Lagache}, G. and {Lamarre}, J.-M. and {Lasenby}, A. and {Lattanzi}, M. and {Lawrence}, C.~R. and {Le Jeune}, M. and {Lemos}, P. and {Lesgourgues}, J. and {Levrier}, F. and {Lewis}, A. and {Liguori}, M. and {Lilje}, P.~B. and {Lilley}, M. and {Lindholm}, V. and {L{\'o}pez-Caniego}, M. and {Lubin}, P.~M. and {Ma}, Y.-Z. and {Mac{\'\i}as-P{\'e}rez}, J.~F. and {Maggio}, G. and {Maino}, D. and {Mandolesi}, N. and {Mangilli}, A. and {Marcos-Caballero}, A. and {Maris}, M. and {Martin}, P.~G. and {Martinelli}, M. and {Mart{\'\i}nez-Gonz{\'a}lez}, E. and {Matarrese}, S. and {Mauri}, N. and {McEwen}, J.~D. and {Meinhold}, P.~R. and {Melchiorri}, A. and {Mennella}, A. and {Migliaccio}, M. and {Millea}, M. and {Mitra}, S. and {Miville-Desch{\^e}nes}, M.-A. and {Molinari}, D. and {Montier}, L. and {Morgante}, G. and {Moss}, A. and {Natoli}, P. and {N{\o}rgaard-Nielsen}, H.~U. and {Pagano}, L. and {Paoletti}, D. and {Partridge}, B. and {Patanchon}, G. and {Peiris}, H.~V. and {Perrotta}, F. and {Pettorino}, V. and {Piacentini}, F. and {Polastri}, L. and {Polenta}, G. and {Puget}, J.-L. and {Rachen}, J.~P. and {Reinecke}, M. and {Remazeilles}, M. and {Renzi}, A. and {Rocha}, G. and {Rosset}, C. and {Roudier}, G. and {Rubi{\~n}o-Mart{\'\i}n}, J.~A. and {Ruiz-Granados}, B. and {Salvati}, L. and {Sandri}, M. and {Savelainen}, M. and {Scott}, D. and {Shellard}, E.~P.~S. and {Sirignano}, C. and {Sirri}, G. and {Spencer}, L.~D. and {Sunyaev}, R. and {Suur-Uski}, A.-S. and {Tauber}, J.~A. and {Tavagnacco}, D. and {Tenti}, M. and {Toffolatti}, L. and {Tomasi}, M. and {Trombetti}, T. and {Valenziano}, L. and {Valiviita}, J. and {Van Tent}, B. and {Vibert}, L. and {Vielva}, P. and {Villa}, F. and {Vittorio}, N. and {Wandelt}, B.~D. and {Wehus}, I.~K. and {White}, M. and {White}, S.~D.~M. and {Zacchei}, A. and {Zonca}, A.},
        title = "{Planck 2018 results. VI. Cosmological parameters}",
      journal = {\aap},
         year = 2020,
        month = sep,
       volume = {641},
          eid = {A6},
        pages = {A6},
          doi = {10.1051/0004-6361/201833910},
archivePrefix = {arXiv},
       eprint = {1807.06209},
 primaryClass = {astro-ph.CO},
       adsurl = {https://ui.adsabs.harvard.edu/abs/2020A&A...641A...6P}
}

@ARTICLE{Cardelli1989,
       author = {{Cardelli}, Jason A. and {Clayton}, Geoffrey C. and {Mathis}, John S.},
        title = "{The Relationship between Infrared, Optical, and Ultraviolet Extinction}",
      journal = {\apj},
         year = 1989,
        month = oct,
       volume = {345},
        pages = {245},
          doi = {10.1086/167900},
       adsurl = {https://ui.adsabs.harvard.edu/abs/1989ApJ...345..245C}
}

@ARTICLE{Carnall2019,
       author = {{Carnall}, Adam C. and {Leja}, Joel and {Johnson}, Benjamin D. and {McLure}, Ross J. and {Dunlop}, James S. and {Conroy}, Charlie},
        title = "{How to Measure Galaxy Star Formation Histories. I. Parametric Models}",
      journal = {\apj},
         year = 2019,
        month = mar,
       volume = {873},
       number = {1},
          eid = {44},
        pages = {44},
          doi = {10.3847/1538-4357/ab04a2},
archivePrefix = {arXiv},
       eprint = {1811.03635},
 primaryClass = {astro-ph.GA},
       adsurl = {https://ui.adsabs.harvard.edu/abs/2019ApJ...873...44C}
}

@ARTICLE{Leja2019,
       author = {{Leja}, Joel and {Carnall}, Adam C. and {Johnson}, Benjamin D. and {Conroy}, Charlie and {Speagle}, Joshua S.},
        title = "{How to Measure Galaxy Star Formation Histories. II. Nonparametric Models}",
      journal = {\apj},
         year = 2019,
        month = may,
       volume = {876},
       number = {1},
          eid = {3},
        pages = {3},
          doi = {10.3847/1538-4357/ab133c},
archivePrefix = {arXiv},
       eprint = {1811.03637},
 primaryClass = {astro-ph.GA},
       adsurl = {https://ui.adsabs.harvard.edu/abs/2019ApJ...876....3L}
}

@ARTICLE{Popesso2023,
       author = {{Popesso}, P. and {Concas}, A. and {Cresci}, G. and {Belli}, S. and {Rodighiero}, G. and {Inami}, H. and {Dickinson}, M. and {Ilbert}, O. and {Pannella}, M. and {Elbaz}, D.},
        title = "{The main sequence of star-forming galaxies across cosmic times}",
      journal = {\mnras},
         year = 2023,
        month = feb,
       volume = {519},
       number = {1},
        pages = {1526-1544},
          doi = {10.1093/mnras/stac3214},
archivePrefix = {arXiv},
       eprint = {2203.10487},
 primaryClass = {astro-ph.GA},
       adsurl = {https://ui.adsabs.harvard.edu/abs/2023MNRAS.519.1526P}
}

@ARTICLE{Rodighiero2011,
       author = {{Rodighiero}, G. and {Daddi}, E. and {Baronchelli}, I. and {Cimatti}, A. and {Renzini}, A. and {Aussel}, H. and {Popesso}, P. and {Lutz}, D. and {Andreani}, P. and {Berta}, S. and {Cava}, A. and {Elbaz}, D. and {Feltre}, A. and {Fontana}, A. and {F{\"o}rster Schreiber}, N.~M. and {Franceschini}, A. and {Genzel}, R. and {Grazian}, A. and {Gruppioni}, C. and {Ilbert}, O. and {Le Floch}, E. and {Magdis}, G. and {Magliocchetti}, M. and {Magnelli}, B. and {Maiolino}, R. and {McCracken}, H. and {Nordon}, R. and {Poglitsch}, A. and {Santini}, P. and {Pozzi}, F. and {Riguccini}, L. and {Tacconi}, L.~J. and {Wuyts}, S. and {Zamorani}, G.},
        title = "{The Lesser Role of Starbursts in Star Formation at z = 2}",
      journal = {\apjl},
         year = 2011,
        month = oct,
       volume = {739},
       number = {2},
          eid = {L40},
        pages = {L40},
          doi = {10.1088/2041-8205/739/2/L40},
archivePrefix = {arXiv},
       eprint = {1108.0933},
 primaryClass = {astro-ph.CO},
       adsurl = {https://ui.adsabs.harvard.edu/abs/2011ApJ...739L..40R}
}

@ARTICLE{Renzini2015,
       author = {{Renzini}, Alvio and {Peng}, Ying-jie},
        title = "{An Objective Definition for the Main Sequence of Star-forming Galaxies}",
      journal = {\apjl},
         year = 2015,
        month = mar,
       volume = {801},
       number = {2},
          eid = {L29},
        pages = {L29},
          doi = {10.1088/2041-8205/801/2/L29},
archivePrefix = {arXiv},
       eprint = {1502.01027},
 primaryClass = {astro-ph.GA},
       adsurl = {https://ui.adsabs.harvard.edu/abs/2015ApJ...801L..29R}
}

@ARTICLE{Inoue2011,
       author = {{Inoue}, Akio K.},
        title = "{Rest-frame ultraviolet-to-optical spectral characteristics of extremely metal-poor and metal-free galaxies}",
      journal = {\mnras},
         year = 2011,
        month = aug,
       volume = {415},
       number = {3},
        pages = {2920-2931},
          doi = {10.1111/j.1365-2966.2011.18906.x},
archivePrefix = {arXiv},
       eprint = {1102.5150},
 primaryClass = {astro-ph.CO},
       adsurl = {https://ui.adsabs.harvard.edu/abs/2011MNRAS.415.2920I}
}

@ARTICLE{MadauDickinson2014,
       author = {{Madau}, Piero and {Dickinson}, Mark},
        title = "{Cosmic Star-Formation History}",
      journal = {\araa},
         year = 2014,
        month = aug,
       volume = {52},
        pages = {415-486},
          doi = {10.1146/annurev-astro-081811-125615},
archivePrefix = {arXiv},
       eprint = {1403.0007},
 primaryClass = {astro-ph.CO},
       adsurl = {https://ui.adsabs.harvard.edu/abs/2014ARA&A..52..415M}
}

@ARTICLE{Speagle2014,
       author = {{Speagle}, J.~S. and {Steinhardt}, C.~L. and {Capak}, P.~L. and {Silverman}, J.~D.},
        title = "{A Highly Consistent Framework for the Evolution of the Star-Forming ``Main Sequence'' from z \raisebox{-0.5ex}\textasciitilde 0-6}",
      journal = {\apjs},
         year = 2014,
        month = oct,
       volume = {214},
       number = {2},
          eid = {15},
        pages = {15},
          doi = {10.1088/0067-0049/214/2/15},
archivePrefix = {arXiv},
       eprint = {1405.2041},
 primaryClass = {astro-ph.GA},
       adsurl = {https://ui.adsabs.harvard.edu/abs/2014ApJS..214...15S}
}

@INPROCEEDINGS{Iyer2026,
       author = {{Iyer}, Kartheik G. and {Pacifici}, Camilla and {Calistro-Rivera}, Gabriela and {Lovell}, Christopher C.},
        title = "{The spectral energy distributions of galaxies}",
    booktitle = {Encyclopedia of Astrophysics},
         year = 2026,
       volume = {4},
        month = jan,
        pages = {236-281},
          doi = {10.1016/B978-0-443-21439-4.00127-9},
       adsurl = {https://ui.adsabs.harvard.edu/abs/2026enap....4..236I}
}

@ARTICLE{Noeske2007,
       author = {{Noeske}, K.~G. and {Weiner}, B.~J. and {Faber}, S.~M. and {Papovich}, C. and {Koo}, D.~C. and {Somerville}, R.~S. and {Bundy}, K. and {Conselice}, C.~J. and {Newman}, J.~A. and {Schiminovich}, D. and {Le Floc'h}, E. and {Coil}, A.~L. and {Rieke}, G.~H. and {Lotz}, J.~M. and {Primack}, J.~R. and {Barmby}, P. and {Cooper}, M.~C. and {Davis}, M. and {Ellis}, R.~S. and {Fazio}, G.~G. and {Guhathakurta}, P. and {Huang}, J. and {Kassin}, S.~A. and {Martin}, D.~C. and {Phillips}, A.~C. and {Rich}, R.~M. and {Small}, T.~A. and {Willmer}, C.~N.~A. and {Wilson}, G.},
        title = "{Star Formation in AEGIS Field Galaxies since z=1.1: The Dominance of Gradually Declining Star Formation, and the Main Sequence of Star-forming Galaxies}",
      journal = {\apjl},
         year = 2007,
        month = may,
       volume = {660},
       number = {1},
        pages = {L43-L46},
          doi = {10.1086/517926},
archivePrefix = {arXiv},
       eprint = {astro-ph/0701924},
 primaryClass = {astro-ph},
       adsurl = {https://ui.adsabs.harvard.edu/abs/2007ApJ...660L..43N}
}

@ARTICLE{Bisigello2018,
       author = {{Bisigello}, L. and {Caputi}, K.~I. and {Grogin}, N. and {Koekemoer}, A.},
        title = "{Analysis of the SFR-M$^{{\ensuremath{*}}}$ plane at z < 3: single fitting versus multi-Gaussian decomposition}",
      journal = {\aap},
         year = 2018,
        month = jan,
       volume = {609},
          eid = {A82},
        pages = {A82},
          doi = {10.1051/0004-6361/201731399},
archivePrefix = {arXiv},
       eprint = {1706.06154},
 primaryClass = {astro-ph.GA},
       adsurl = {https://ui.adsabs.harvard.edu/abs/2018A&A...609A..82B}
}

@ARTICLE{Daddi2007,
       author = {{Daddi}, E. and {Dickinson}, M. and {Morrison}, G. and {Chary}, R. and {Cimatti}, A. and {Elbaz}, D. and {Frayer}, D. and {Renzini}, A. and {Pope}, A. and {Alexander}, D.~M. and {Bauer}, F.~E. and {Giavalisco}, M. and {Huynh}, M. and {Kurk}, J. and {Mignoli}, M.},
        title = "{Multiwavelength Study of Massive Galaxies at z\raisebox{-0.5ex}\textasciitilde2. I. Star Formation and Galaxy Growth}",
      journal = {\apj},
         year = 2007,
        month = nov,
       volume = {670},
       number = {1},
        pages = {156-172},
          doi = {10.1086/521818},
archivePrefix = {arXiv},
       eprint = {0705.2831},
 primaryClass = {astro-ph},
       adsurl = {https://ui.adsabs.harvard.edu/abs/2007ApJ...670..156D}
}

@ARTICLE{Pannella2009,
       author = {{Pannella}, M. and {Carilli}, C.~L. and {Daddi}, E. and {McCracken}, H.~J. and {Owen}, F.~N. and {Renzini}, A. and {Strazzullo}, V. and {Civano}, F. and {Koekemoer}, A.~M. and {Schinnerer}, E. and {Scoville}, N. and {Smol{\v{c}}i{\'c}}, V. and {Taniguchi}, Y. and {Aussel}, H. and {Kneib}, J.~P. and {Ilbert}, O. and {Mellier}, Y. and {Salvato}, M. and {Thompson}, D. and {Willott}, C.~J.},
        title = "{Star Formation and Dust Obscuration at z {\ensuremath{\approx}} 2: Galaxies at the Dawn of Downsizing}",
      journal = {\apjl},
         year = 2009,
        month = jun,
       volume = {698},
       number = {2},
        pages = {L116-L120},
          doi = {10.1088/0004-637X/698/2/L116},
archivePrefix = {arXiv},
       eprint = {0905.1674},
 primaryClass = {astro-ph.CO},
       adsurl = {https://ui.adsabs.harvard.edu/abs/2009ApJ...698L.116P}
}

@ARTICLE{Windhorst1991,
       author = {{Windhorst}, Rogier A. and {Burstein}, David and {Mathis}, Doug F. and {Neuschaefer}, Lyman W. and {Bertola}, F. and {Buson}, L.~M. and {Koo}, David C. and {Matthews}, Keith and {Barthel}, Peter D. and {Chambers}, K.~C.},
        title = "{The Discovery of a Young Radio Galaxy at Z = 2.390: Probing Initial Star Formation at Z 2pt< -11.5pt-3pt approximately 3.0}",
      journal = {\apj},
         year = 1991,
        month = oct,
       volume = {380},
        pages = {362},
          doi = {10.1086/170596},
       adsurl = {https://ui.adsabs.harvard.edu/abs/1991ApJ...380..362W}
}

@ARTICLE{Windhorst1992,
       author = {{Windhorst}, Rogier A. and {Mathis}, Douglas F. and {Keel}, William C.},
        title = "{Deep Hubble Space Telescope Imaging of a Compact Radio Galaxy at Z = 2.390}",
      journal = {\apjl},
         year = 1992,
        month = nov,
       volume = {400},
        pages = {L1},
          doi = {10.1086/186634},
       adsurl = {https://ui.adsabs.harvard.edu/abs/1992ApJ...400L...1W}
}

@ARTICLE{Kocevski2023,
       author = {{Kocevski}, Dale D. and {Barro}, Guillermo and {McGrath}, Elizabeth J. and {Finkelstein}, Steven L. and {Bagley}, Micaela B. and {Ferguson}, Henry C. and {Jogee}, Shardha and {Yang}, Guang and {Dickinson}, Mark and {Hathi}, Nimish P. and {Backhaus}, Bren E. and {Bell}, Eric F. and {Bisigello}, Laura and {Buat}, V{\'e}ronique and {Burgarella}, Denis and {Casey}, Caitlin M. and {Cleri}, Nikko J. and {Cooper}, M.~C. and {Costantin}, Luca and {Croton}, Darren and {Daddi}, Emanuele and {Fontana}, Adriano and {Fujimoto}, Seiji and {Gardner}, Jonathan P. and {Gawiser}, Eric and {Giavalisco}, Mauro and {Grazian}, Andrea and {Grogin}, Norman A. and {Guo}, Yuchen and {Arrabal Haro}, Pablo and {Hirschmann}, Michaela and {Holwerda}, Benne W. and {Huertas-Company}, Marc and {Hutchison}, Taylor A. and {Iyer}, Kartheik G. and {Jones}, Brenda and {Juneau}, St{\'e}phanie and {Kartaltepe}, Jeyhan S. and {Kewley}, Lisa J. and {Kirkpatrick}, Allison and {Koekemoer}, Anton M. and {Kurczynski}, Peter and {Le Bail}, Aur{\'e}lien and {Long}, Arianna S. and {Lotz}, Jennifer M. and {Lucas}, Ray A. and {Papovich}, Casey and {Pentericci}, Laura and {P{\'e}rez-Gonz{\'a}lez}, Pablo G. and {Pirzkal}, Nor and {Rafelski}, Marc and {Ravindranath}, Swara and {Somerville}, Rachel S. and {Straughn}, Amber N. and {Tacchella}, Sandro and {Trump}, Jonathan R. and {Wilkins}, Stephen M. and {Wuyts}, Stijn and {Yung}, L.~Y. Aaron and {Zavala}, Jorge A.},
        title = "{CEERS Key Paper. II. A First Look at the Resolved Host Properties of AGN at 3 < z < 5 with JWST}",
      journal = {\apjl},
         year = 2023,
        month = mar,
       volume = {946},
       number = {1},
          eid = {L14},
        pages = {L14},
          doi = {10.3847/2041-8213/acad00},
archivePrefix = {arXiv},
       eprint = {2208.14480},
 primaryClass = {astro-ph.GA},
       adsurl = {https://ui.adsabs.harvard.edu/abs/2023ApJ...946L..14K}
}

@ARTICLE{Donley2012,
       author = {{Donley}, J.~L. and {Koekemoer}, A.~M. and {Brusa}, M. and {Capak}, P. and {Cardamone}, C.~N. and {Civano}, F. and {Ilbert}, O. and {Impey}, C.~D. and {Kartaltepe}, J.~S. and {Miyaji}, T. and {Salvato}, M. and {Sanders}, D.~B. and {Trump}, J.~R. and {Zamorani}, G.},
        title = "{Identifying Luminous Active Galactic Nuclei in Deep Surveys: Revised IRAC Selection Criteria}",
      journal = {\apj},
         year = 2012,
        month = apr,
       volume = {748},
       number = {2},
          eid = {142},
        pages = {142},
          doi = {10.1088/0004-637X/748/2/142},
archivePrefix = {arXiv},
       eprint = {1201.3899},
 primaryClass = {astro-ph.CO},
       adsurl = {https://ui.adsabs.harvard.edu/abs/2012ApJ...748..142D}
}

@ARTICLE{DiMatteo2005,
       author = {{Di Matteo}, Tiziana and {Springel}, Volker and {Hernquist}, Lars},
        title = "{Energy input from quasars regulates the growth and activity of black holes and their host galaxies}",
      journal = {\nat},
         year = 2005,
        month = feb,
       volume = {433},
       number = {7026},
        pages = {604-607},
          doi = {10.1038/nature03335},
archivePrefix = {arXiv},
       eprint = {astro-ph/0502199},
 primaryClass = {astro-ph},
       adsurl = {https://ui.adsabs.harvard.edu/abs/2005Natur.433..604D}
}

@ARTICLE{Hopkins2006,
       author = {{Hopkins}, Philip F. and {Hernquist}, Lars and {Cox}, Thomas J. and {Di Matteo}, Tiziana and {Robertson}, Brant and {Springel}, Volker},
        title = "{A Unified, Merger-driven Model of the Origin of Starbursts, Quasars, the Cosmic X-Ray Background, Supermassive Black Holes, and Galaxy Spheroids}",
      journal = {\apjs},
         year = 2006,
        month = mar,
       volume = {163},
       number = {1},
        pages = {1-49},
          doi = {10.1086/499298},
archivePrefix = {arXiv},
       eprint = {astro-ph/0506398},
 primaryClass = {astro-ph},
       adsurl = {https://ui.adsabs.harvard.edu/abs/2006ApJS..163....1H}
}

@ARTICLE{Dekel2014,
       author = {{Dekel}, A. and {Burkert}, A.},
        title = "{Wet disc contraction to galactic blue nuggets and quenching to red nuggets}",
      journal = {\mnras},
         year = 2014,
        month = feb,
       volume = {438},
       number = {2},
        pages = {1870-1879},
          doi = {10.1093/mnras/stt2331},
archivePrefix = {arXiv},
       eprint = {1310.1074},
 primaryClass = {astro-ph.CO},
       adsurl = {https://ui.adsabs.harvard.edu/abs/2014MNRAS.438.1870D}
}

@ARTICLE{Zolotov2015,
       author = {{Zolotov}, Adi and {Dekel}, Avishai and {Mandelker}, Nir and {Tweed}, Dylan and {Inoue}, Shigeki and {DeGraf}, Colin and {Ceverino}, Daniel and {Primack}, Joel R. and {Barro}, Guillermo and {Faber}, Sandra M.},
        title = "{Compaction and quenching of high-z galaxies in cosmological simulations: blue and red nuggets}",
      journal = {\mnras},
         year = 2015,
        month = jul,
       volume = {450},
       number = {3},
        pages = {2327-2353},
          doi = {10.1093/mnras/stv740},
archivePrefix = {arXiv},
       eprint = {1412.4783},
 primaryClass = {astro-ph.GA},
       adsurl = {https://ui.adsabs.harvard.edu/abs/2015MNRAS.450.2327Z}
}

@ARTICLE{Hopkins2009,
       author = {{Hopkins}, Philip F. and {Hernquist}, Lars},
        title = "{Quasars Are Not Light Bulbs: Testing Models of Quasar Lifetimes with the Observed Eddington Ratio Distribution}",
      journal = {\apj},
         year = 2009,
        month = jun,
       volume = {698},
       number = {2},
        pages = {1550-1569},
          doi = {10.1088/0004-637X/698/2/1550},
archivePrefix = {arXiv},
       eprint = {0809.3789},
 primaryClass = {astro-ph},
       adsurl = {https://ui.adsabs.harvard.edu/abs/2009ApJ...698.1550H}
}

@ARTICLE{Sanders1998,
       author = {{Sanders}, D.~B. and {Soifer}, B.~T. and {Elias}, J.~H. and {Madore}, B.~F. and {Matthews}, K. and {Neugebauer}, G. and {Scoville}, N.~Z.},
        title = "{Ultraluminous Infrared Galaxies and the Origin of Quasars}",
      journal = {\apj},
         year = 1988,
        month = feb,
       volume = {325},
        pages = {74},
          doi = {10.1086/165983},
       adsurl = {https://ui.adsabs.harvard.edu/abs/1988ApJ...325...74S}
}

@ARTICLE{Fabian2012,
       author = {{Fabian}, A.~C.},
        title = "{Observational Evidence of Active Galactic Nuclei Feedback}",
      journal = {\araa},
         year = 2012,
        month = sep,
       volume = {50},
        pages = {455-489},
          doi = {10.1146/annurev-astro-081811-125521},
archivePrefix = {arXiv},
       eprint = {1204.4114},
 primaryClass = {astro-ph.CO},
       adsurl = {https://ui.adsabs.harvard.edu/abs/2012ARA&A..50..455F}
}

@ARTICLE{Bird2008,
       author = {{Bird}, Jonathan and {Martini}, Paul and {Kaiser}, Christian},
        title = "{The Lifetime of FR II Sources in Groups and Clusters: Implications for Radio-Mode Feedback}",
      journal = {\apj},
         year = 2008,
        month = mar,
       volume = {676},
       number = {1},
        pages = {147-162},
          doi = {10.1086/527534},
archivePrefix = {arXiv},
       eprint = {0709.2167},
 primaryClass = {astro-ph},
       adsurl = {https://ui.adsabs.harvard.edu/abs/2008ApJ...676..147B}
}

@ARTICLE{Schawinski2015,
       author = {{Schawinski}, Kevin and {Koss}, Michael and {Berney}, Simon and {Sartori}, Lia F.},
        title = "{Active galactic nuclei flicker: an observational estimate of the duration of black hole growth phases of {\ensuremath{\sim}}{}10$^{5}$ yr}",
      journal = {\mnras},
         year = 2015,
        month = aug,
       volume = {451},
       number = {3},
        pages = {2517-2523},
          doi = {10.1093/mnras/stv1136},
archivePrefix = {arXiv},
       eprint = {1505.06733},
 primaryClass = {astro-ph.GA},
       adsurl = {https://ui.adsabs.harvard.edu/abs/2015MNRAS.451.2517S}
}

@ARTICLE{ErrozFerrer2013,
       author = {{Erroz-Ferrer}, Santiago and {Knapen}, Johan H. and {Mohd Noh Velast{\'\i}n}, Elena A.~N. and {Ryon}, Jenna E. and {Hagen}, Lea M.~Z.},
        title = "{Massive star formation in galaxies with excess ultraviolet emission}",
      journal = {\mnras},
         year = 2013,
        month = dec,
       volume = {436},
       number = {4},
        pages = {3135-3146},
          doi = {10.1093/mnras/stt1797},
archivePrefix = {arXiv},
       eprint = {1309.5908},
 primaryClass = {astro-ph.CO},
       adsurl = {https://ui.adsabs.harvard.edu/abs/2013MNRAS.436.3135E}
}

@ARTICLE{CalistroRivera2016,
       author = {{Calistro Rivera}, Gabriela and {Lusso}, Elisabeta and {Hennawi}, Joseph F. and {Hogg}, David W.},
        title = "{AGNfitter: A Bayesian MCMC Approach to Fitting Spectral Energy Distributions of AGNs}",
      journal = {\apj},
         year = 2016,
        month = dec,
       volume = {833},
       number = {1},
          eid = {98},
        pages = {98},
          doi = {10.3847/1538-4357/833/1/98},
archivePrefix = {arXiv},
       eprint = {1606.05648},
 primaryClass = {astro-ph.GA},
       adsurl = {https://ui.adsabs.harvard.edu/abs/2016ApJ...833...98C}
}

@ARTICLE{MartinezRamirez2024,
       author = {{Mart{\'\i}nez-Ram{\'\i}rez}, L.~N. and {Calistro Rivera}, G. and {Lusso}, E. and {Bauer}, F.~E. and {Nardini}, E. and {Buchner}, J. and {Brown}, M.~J.~I. and {Pineda}, J.~C.~B. and {Temple}, M.~J. and {Banerji}, M. and {Stalevski}, M. and {Hennawi}, J.~F.},
        title = "{AGNFITTER-RX: Modeling the radio-to-X-ray spectral energy distributions of AGNs}",
      journal = {\aap},
         year = 2024,
        month = aug,
       volume = {688},
          eid = {A46},
        pages = {A46},
          doi = {10.1051/0004-6361/202449329},
archivePrefix = {arXiv},
       eprint = {2405.12111},
 primaryClass = {astro-ph.GA},
       adsurl = {https://ui.adsabs.harvard.edu/abs/2024A&A...688A..46M}
}

@ARTICLE{Hviding2022,
       author = {{Hviding}, Raphael E. and {Hainline}, Kevin N. and {Rieke}, Marcia and {Juneau}, St{\'e}phanie and {Lyu}, Jianwei and {Pucha}, Ragadeepika},
        title = "{A New Infrared Criterion for Selecting Active Galactic Nuclei to Lower Luminosities}",
      journal = {\aj},
         year = 2022,
        month = may,
       volume = {163},
       number = {5},
          eid = {224},
        pages = {224},
          doi = {10.3847/1538-3881/ac5e33},
archivePrefix = {arXiv},
       eprint = {2203.11217},
 primaryClass = {astro-ph.GA},
       adsurl = {https://ui.adsabs.harvard.edu/abs/2022AJ....163..224H}
}

@ARTICLE{Kormendy1995,
       author = {{Kormendy}, John and {Richstone}, Douglas},
        title = "{Inward Bound---The Search For Supermassive Black Holes In Galactic Nuclei}",
      journal = {\araa},
         year = 1995,
        month = jan,
       volume = {33},
        pages = {581},
          doi = {10.1146/annurev.aa.33.090195.003053},
       adsurl = {https://ui.adsabs.harvard.edu/abs/1995ARA&A..33..581K}
}

@ARTICLE{Magorrian1998,
       author = {{Magorrian}, John and {Tremaine}, Scott and {Richstone}, Douglas and {Bender}, Ralf and {Bower}, Gary and {Dressler}, Alan and {Faber}, S.~M. and {Gebhardt}, Karl and {Green}, Richard and {Grillmair}, Carl and {Kormendy}, John and {Lauer}, Tod},
        title = "{The Demography of Massive Dark Objects in Galaxy Centers}",
      journal = {\aj},
         year = 1998,
        month = jun,
       volume = {115},
       number = {6},
        pages = {2285-2305},
          doi = {10.1086/300353},
archivePrefix = {arXiv},
       eprint = {astro-ph/9708072},
 primaryClass = {astro-ph},
       adsurl = {https://ui.adsabs.harvard.edu/abs/1998AJ....115.2285M}
}

@ARTICLE{Richstone1998,
       author = {{Richstone}, D. and {Ajhar}, E.~A. and {Bender}, R. and {Bower}, G. and {Dressler}, A. and {Faber}, S.~M. and {Filippenko}, A.~V. and {Gebhardt}, K. and {Green}, R. and {Ho}, L.~C. and {Kormendy}, J. and {Lauer}, T.~R. and {Magorrian}, J. and {Tremaine}, S.},
        title = "{Supermassive black holes and the evolution of galaxies.}",
      journal = {\nat},
         year = 1998,
        month = oct,
       volume = {385},
       number = {6701},
        pages = {A14},
          doi = {10.48550/arXiv.astro-ph/9810378},
archivePrefix = {arXiv},
       eprint = {astro-ph/9810378},
 primaryClass = {astro-ph},
       adsurl = {https://ui.adsabs.harvard.edu/abs/1998Natur.395A..14R}
}

@ARTICLE{Seyfert1943,
       author = {{Seyfert}, Carl K.},
        title = "{Nuclear Emission in Spiral Nebulae.}",
      journal = {\apj},
         year = 1943,
        month = jan,
       volume = {97},
        pages = {28},
          doi = {10.1086/144488},
       adsurl = {https://ui.adsabs.harvard.edu/abs/1943ApJ....97...28S}
}

@ARTICLE{Tran1995a,
       author = {{Tran}, Hien D.},
        title = "{The Nature of Seyfert 2 Galaxies with Obscured Broad-Line Regions. I. Observations}",
      journal = {\apj},
         year = 1995,
        month = feb,
       volume = {440},
        pages = {565},
          doi = {10.1086/175296},
       adsurl = {https://ui.adsabs.harvard.edu/abs/1995ApJ...440..565T}
}

@ARTICLE{Tran1995b,
       author = {{Tran}, Hien D.},
        title = "{The Nature of Seyfert 2 Galaxies with Obscured Broad-Line Regions. II. Individual Objects}",
      journal = {\apj},
         year = 1995,
        month = feb,
       volume = {440},
        pages = {578},
          doi = {10.1086/175297},
       adsurl = {https://ui.adsabs.harvard.edu/abs/1995ApJ...440..578T}
}

@ARTICLE{Tran1995c,
       author = {{Tran}, Hien D.},
        title = "{The Nature of Seyfert 2 Galaxies with Obscured Broad-Line Regions. III. Interpretation}",
      journal = {\apj},
         year = 1995,
        month = feb,
       volume = {440},
        pages = {597},
          doi = {10.1086/175298},
       adsurl = {https://ui.adsabs.harvard.edu/abs/1995ApJ...440..597T}
}

@ARTICLE{Burbidge1970,
       author = {{Burbidge}, G.},
        title = "{The Nature of Galactic Nuclei}",
      journal = {Comments on Astrophysics and Space Physics},
         year = 1970,
        month = jul,
       volume = {2},
        pages = {144},
       adsurl = {https://ui.adsabs.harvard.edu/abs/1970CoASP...2..144B}
}

@ARTICLE{Khachikian1974,
       author = {{Khachikian}, E.~Y. and {Weedman}, D.~W.},
        title = "{An atlas of Seyfert galaxies.}",
      journal = {\apj},
         year = 1974,
        month = sep,
       volume = {192},
        pages = {581-589},
          doi = {10.1086/153093},
       adsurl = {https://ui.adsabs.harvard.edu/abs/1974ApJ...192..581K}
}

@ARTICLE{Heckman1980,
       author = {{Heckman}, T.~M.},
        title = "{An Optical and Radio Survey of the Nuclei of Bright Galaxies - Activity in the Normal Galactic Nuclei}",
      journal = {\aap},
         year = 1980,
        month = jul,
       volume = {87},
        pages = {152},
       adsurl = {https://ui.adsabs.harvard.edu/abs/1980A&A....87..152H}
}

@ARTICLE{Elvis1978,
       author = {{Elvis}, M. and {Maccacaro}, T. and {Wilson}, A.~S. and {Ward}, M.~J. and {Penston}, M.~V. and {Fosbury}, R.~A.~E. and {Perola}, G.~C.},
        title = "{Seyfert galaxies as X-ray sources.}",
      journal = {\mnras},
         year = 1978,
        month = apr,
       volume = {183},
        pages = {129-157},
          doi = {10.1093/mnras/183.2.129},
       adsurl = {https://ui.adsabs.harvard.edu/abs/1978MNRAS.183..129E}
}

@ARTICLE{Stern2005,
       author = {{Stern}, Daniel and {Eisenhardt}, Peter and {Gorjian}, Varoujan and {Kochanek}, Christopher S. and {Caldwell}, Nelson and {Eisenstein}, Daniel and {Brodwin}, Mark and {Brown}, Michael J.~I. and {Cool}, Richard and {Dey}, Arjun and {Green}, Paul and {Jannuzi}, Buell T. and {Murray}, Stephen S. and {Pahre}, Michael A. and {Willner}, S.~P.},
        title = "{Mid-Infrared Selection of Active Galaxies}",
      journal = {\apj},
         year = 2005,
        month = sep,
       volume = {631},
       number = {1},
        pages = {163-168},
          doi = {10.1086/432523},
archivePrefix = {arXiv},
       eprint = {astro-ph/0410523},
 primaryClass = {astro-ph},
       adsurl = {https://ui.adsabs.harvard.edu/abs/2005ApJ...631..163S}
}

@ARTICLE{FanaroffRiley1974,
       author = {{Fanaroff}, B.~L. and {Riley}, J.~M.},
        title = "{The morphology of extragalactic radio sources of high and low luminosity}",
      journal = {\mnras},
         year = 1974,
        month = may,
       volume = {167},
        pages = {31P-36P},
          doi = {10.1093/mnras/167.1.31P},
       adsurl = {https://ui.adsabs.harvard.edu/abs/1974MNRAS.167P..31F}
}

@ARTICLE{Schmidt1963,
       author = {{Schmidt}, M.},
        title = "{3C 273 : A Star-Like Object with Large Red-Shift}",
      journal = {\nat},
         year = 1963,
        month = mar,
       volume = {197},
       number = {4872},
        pages = {1040},
          doi = {10.1038/1971040a0},
       adsurl = {https://ui.adsabs.harvard.edu/abs/1963Natur.197.1040S}
}

@ARTICLE{Giacconi2001,
       author = {{Giacconi}, R. and {Rosati}, P. and {Tozzi}, P. and {Nonino}, M. and {Hasinger}, G. and {Norman}, C. and {Bergeron}, J. and {Borgani}, S. and {Gilli}, R. and {Gilmozzi}, R. and {Zheng}, W.},
        title = "{First Results from the X-Ray and Optical Survey of the Chandra Deep Field South}",
      journal = {\apj},
         year = 2001,
        month = apr,
       volume = {551},
       number = {2},
        pages = {624-634},
          doi = {10.1086/320222},
archivePrefix = {arXiv},
       eprint = {astro-ph/0007240},
 primaryClass = {astro-ph},
       adsurl = {https://ui.adsabs.harvard.edu/abs/2001ApJ...551..624G}
}

@ARTICLE{Dunlop2003,
       author = {{Dunlop}, J.~S. and {McLure}, R.~J. and {Kukula}, M.~J. and {Baum}, S.~A. and {O'Dea}, C.~P. and {Hughes}, D.~H.},
        title = "{Quasars, their host galaxies and their central black holes}",
      journal = {\mnras},
         year = 2003,
        month = apr,
       volume = {340},
       number = {4},
        pages = {1095-1135},
          doi = {10.1046/j.1365-8711.2003.06333.x},
archivePrefix = {arXiv},
       eprint = {astro-ph/0108397},
 primaryClass = {astro-ph},
       adsurl = {https://ui.adsabs.harvard.edu/abs/2003MNRAS.340.1095D}
}

@ARTICLE{Alexander2003,
       author = {{Alexander}, D.~M. and {Bauer}, F.~E. and {Brandt}, W.~N. and {Schneider}, D.~P. and {Hornschemeier}, A.~E. and {Vignali}, C. and {Barger}, A.~J. and {Broos}, P.~S. and {Cowie}, L.~L. and {Garmire}, G.~P. and {Townsley}, L.~K. and {Bautz}, M.~W. and {Chartas}, G. and {Sargent}, W.~L.~W.},
        title = "{The Chandra Deep Field North Survey. XIII. 2 Ms Point-Source Catalogs}",
      journal = {\aj},
         year = 2003,
        month = aug,
       volume = {126},
       number = {2},
        pages = {539-574},
          doi = {10.1086/376473},
archivePrefix = {arXiv},
       eprint = {astro-ph/0304392},
 primaryClass = {astro-ph},
       adsurl = {https://ui.adsabs.harvard.edu/abs/2003AJ....126..539A}
}
\bibliographystyle{aasjournalv7}

\appendix

\startlongtable
\begin{deluxetable}{rccccccccccc}
\label{tab:objects}
\tablecaption{NIRCam-Selected AGN-Host Candidates}
\tablehead{\colhead{ID} & \colhead{R.A.} & \colhead{Dec.} & \colhead{Spoke} & \colhead{$z$} & \colhead{Core Type} & \colhead{$f_\text{AGN}$} & CI/CII & $P(\rm CI)$\tablenotemark{$\dagger$} & $P(\rm CII)$\tablenotemark{$\dagger$} & \colhead{X-Ray ID} & \colhead{Radio ID} \\
 (1) & (2) & (3) & (4) & (5) & (6) & (7) & (8) & (9) & (10) & (11) & (12)}
\startdata
2 & 260.756207 & 65.713177 & 1b & 1.37 & P & 0.50 & I & $>$0.999 & $<$0.001 & \no & 314 \\
6 & 260.701250 & 65.776180 & 1ab & 1.00 & P & 0.83 & I & $>$0.999 & $<$0.001 & 451 & 232 \\
7 & 260.697887 & 65.777933 & 1ab & 1.66 & P & 0.33 & II & 0.074 & 0.926  & 452 & \no \\
9 & 260.695583 & 65.784352 & 1b & 0.95 & P & 0.57 & I & $>$0.999 & $<$0.001  & 453 & 223 \\
10 & 260.746297 & 65.784221 & 1b & 2.04 & P & 0.51 & I & $>$0.999 & $<$0.001  & 102 & 300 \\
11 & 260.719022 & 65.786260 & 1b & 1.17 & U & 0.35 & II & 0.004 & 0.996  & \no & \no \\
13 & 260.641043 & 65.787557 & 4 & 1.16 & U & 0.34 & II & $<$0.001 & $>$0.999  & \no & \no \\
14 & 260.536970 & 65.795297 & 4 & 2.2508 & P & 0.84 & I & $>$0.999 & $<$0.001  & 119 & \no \\
23 & 260.456228 & 65.807676 & 4 & 1.12 & B & 0.34 & II & 0.001 & 0.999  & \no & \no \\
24 & 260.524706 & 65.809982 & 4 & 1.20 & B & 0.33 & II & $<$0.001 & $>$0.999  & \no & \no \\
25 & 260.483404 & 65.811818 & 4 & 1.20 & P & 0.31 & II & 0.360 & 0.640  & 463 & \no \\
27 & 260.531751 & 65.815627 & 4 & 1.08 & B & 0.32 & II & 0.073 & 0.927  & \no & 63 \\
28 & 260.896667 & 65.817463 & 2 & 0.7791 & P & 0.71 & I & $>$0.999 & $<$0.001 & 106 & 497 \\
30 & 260.814880 & 65.821168 & 2 & 1.58 & P & 0.32 & II & $<$0.001 & $>$0.999  & \no & \no \\
32 & 260.898138 & 65.821819 & 2 & 1.47 & U & 0.36 & II & 0.005 & 0.995  & \no & \no \\
33 & 260.423208 & 65.823911 & 4 & 1.07 & B & 0.37 & II & 0.031 & 0.969  & \no & \no \\
34 & 260.768573 & 65.826200 & 2 & 1.51 & U & 0.38 & II & 0.003 & 0.997  & \no & \no \\
35 & 260.538451 & 65.827557 & 4 & 0.9442 & P & 0.29 & II & 0.357 & 0.643  & 155 & \no \\
38 & 260.919468 & 65.831341 & 2 & 0.33 & P & 0.25 & I & 0.970 & 0.030 & \no & 528 \\
39 & 260.965050 & 65.834642 & 2 & 0.6726 & P & 0.23 & I & $>$0.999 & $<$0.001  & 126 & \no \\
40 & 260.661331 & 65.832686 & 3 & 1.53 & U & 0.39 & II & $<$0.001 & $>$0.999  & \no & \no \\
42 & 260.498867 & 65.833758 & 4 & 1.8016 & P & 0.44 & I & 0.993 & 0.007  & 162 & \no \\
43 & 260.661990 & 65.841400 & 3 & 1.42 & B & 0.36 & II & 0.002 & 0.998  & \no & \no \\
48 & 260.639560 & 65.844960 & 3 & 1.3369 & B & 0.80 & I & $>$0.999 & $<$0.001  & 109 & 142 \\
52 & 260.624274 & 65.867878 & 3 & 0.84 & B & 0.33 & II & 0.191 & 0.809  & 165 & 121 \\
53 & 260.681038 & 65.869282 & 3 & 1.88 & B & 0.72 & I & $>$0.999 & $<$0.001  & \no & 207 \\
55 & 260.738030 & 65.902876 & 3 & 1.36 & P & 0.66 & I & $>$0.999 & $<$0.001  & \no & \no \\
56 & 260.708784 & 65.876954 & 3 & 1.79 & U & 0.36 & II & 0.003 & 0.997 & \no & \no \\
57 & 260.651265 & 65.893590 & 3 & 1.05 & U & 0.34 & II & 0.003 & 0.997  & \no & \no \\
58 & 260.668029 & 65.877818 & 3 & 1.43 & B & 0.40 & I & 0.504 & 0.496  & \no & 187 \\
59 & 260.665718 & 65.877257 & 3 & 1.91 & B & 0.38 & II & $<$0.001 & $>$0.999  & \no & \no \\
60 & 260.690274 & 65.899640 & 3 & 1.37 & U & 0.30 & II & 0.246 & 0.754  & \no & \no \\
62 & 260.709682 & 65.880570 & 3 & 1.50 & U & 0.33 & II & 0.170 & 0.830  & \no & \no \\
63 & 260.629030 & 65.874101 & 3 & 1.77 & U & 0.32 & II & 0.166 & 0.834   & \no & \no \\
64 & 260.649498 & 65.870951 & 3 & 1.20 & P & 0.44 & I & 0.992 & 0.008  & 132 & \no \\
66 & 260.672783 & 65.911116 & 3 & 1.70 & P & 0.37 & II & 0.040 & 0.960  & \no & \no
\enddata
\tablecomments{Column descriptions are as follows: (1) ID from \citetalias{Ortiz2024}; (2,3) ICRS positions in decimal degrees from \citetalias{Ortiz2024}; (4) spoke of the JWST NEP-TDF containing the object; (5) redshift, where spectroscopically-confirmed redshifts are listed to 4 decimal places and \citetalias{Ortiz2024} photometric redshifts are listed to 2 decimal places; (6) JWST/F444W core type from \citetalias{Ortiz2024} (B = bulge, P = point source, U = undetermined); (7) AGN fraction from 0.1--30.0~$\mu\rm m$ from \citetalias{Ortiz2024}; { (8) GMM component assignment; (9) posterior probability of membership in Component I; (10) posterior probability of membership in Component II;} (11) X-ray counterpart ID from W. P. Maksym et al. (in prep.); (12) radio counterpart ID from \citet{Hyun2023}.}
\tablenotemark{$\dagger$}{The posterior membership probabilities for the two components sum to unity for each galaxy.}
\end{deluxetable}

\startlongtable
\begin{deluxetable}{ccccccccccc}
\label{tab:galfit-best-fit-params}
\tablecaption{Table of \textsc{Galfit} Output Parameters for JWST Filters}
\tablehead{\colhead{ID} & \colhead{Filter} & \colhead{Disk mag} & \colhead{Disk $r_e$} & \colhead{Disk $n$} & \colhead{Bulge mag} & \colhead{Bulge $r_e$} & \colhead{Bulge $n$} & \colhead{$m_{\rm PS}$} & \colhead{$\delta m_{\rm PS}$} & \colhead{$\chi^2_\nu$} \\
 & & mag & pix & & mag & pix & & mag & mag \\
 (1) & (2) & (3) & (4) & (5) & (6) & (7) & (8) & (9) & (10) & (11)}
\startdata
\textbf{2} & F444W & 20.46 & 4.00 & 1.92 & 23.19 & 0.79 & 3.00 & 23.35 & 0.20 & 1.87 \\
 & F410M & 20.96 & 4.00 & 1.31 & 21.52 & 3.84 & 4.29 & 23.46 & 0.32 & 1.41 \\
 & F356W & 20.72 & 4.00 & 1.93 & 23.72 & 0.00 & 3.00 & 23.72 & 0.15 & 2.32 \\
 & F277W & 21.12 & 4.00 & 1.36 & 22.38 & 13.19 & 3.00 & 23.92 & 0.31 & 2.34 \\
 & F200W & 21.53 & 5.72 & 1.20 & 24.44 & 25.50 & 3.00 & 24.44 & 0.31 & 0.78 \\
 & F150W & 22.26 & 6.26 & 0.78 & 22.52 & 6.58 & 3.00 & 24.61 & 0.31 & 0.73 \\
 & F115W & 22.97 & 7.52 & 0.58 & 22.84 & 8.27 & 3.00 & 25.16 & 1.14 & 0.69 \\
 & F090W & 23.26 & 6.31 & 0.99 & 26.11 & 0.00 & 3.00 & 26.11 & 0.31 & 0.66 \\
\textbf{6} & F444W & 19.97 & 4.00 & 1.55 & 22.91 & 10.84 & 3.00 & 22.91 & 0.30 & 4.89 \\
 & F410M & 20.29 & 4.99 & 1.63 & 21.77 & 1.08 & 3.00 & 22.44 & 0.40 & 2.99 \\
 & F356W & 20.35 & 5.29 & 1.44 & 19.47 & 37664.35 & 9.09 & 22.47 & 0.27 & 6.59 \\
 & F277W & 20.63 & 6.09 & 1.23 & 20.52 & 7648.14 & 5.61 & 23.52 & 0.30 & 5.14 \\
 & F200W & 21.36 & 7.66 & 0.97 & 23.26 & 2.71 & 3.00 & 24.14 & 0.89 & 1.34 \\
 & F150W & 22.30 & 9.19 & 0.48 & 22.24 & 7.09 & 3.00 & 24.52 & 0.90 & 1.41 \\
 & F115W & 22.09 & 7.75 & 1.01 & 25.02 & 0.00 & 3.00 & 25.02 & 0.29 & 1.35 \\
 & F090W & 23.20 & 7.94 & 0.51 & 23.00 & 66.10 & 3.00 & 25.79 & 0.28 & 1.00 \\
$\cdots$ & $\cdots$ & $\cdots$ & $\cdots$ & $\cdots$ & $\cdots$ & $\cdots$ & $\cdots$ & $\cdots$ & $\cdots$ & $\cdots$
\enddata
\tablecomments{The full Table~\ref{tab:galfit-best-fit-params} is available as a machine-readable table (MRT). Column descriptions are as follows: (1) ID from \citetalias{Ortiz2024}; (2) JWST filter; (3) magnitude of disk S\'ersic; (4) half-light radius of the disk S\'ersic; (5) S\'ersic index of the disk S\'ersic; (6) magnitude of the bulge S\'ersic; (7) half-light radius of the bulge S\'ersic; (8) S\'ersic index of the bulge S\'ersic; (9) magnitude of the point-source; (10) uncertainty in the magnitude of the point-source; (11) $\chi^2_\nu$ of the \textsc{Galfit} model. Each ID (bolded) has 8 associated rows, one for each JWST filter. Pixel $r_e$ values listed as 0.00 denote fits for which the $r_e$ was much less than 1 pixel --- see the end of Appendix \ref{app:galfit-configuration} for discussion of S\'ersic component radii. The measurement of the uncertainty in the point-source magnitude is not done by \textsc{Galfit} and is detailed in \S\ref{sec:PS-mag-uncertainties}.}
\end{deluxetable}

\startlongtable
\begin{deluxetable}{lllc}
    \label{tab:cigale-parameters}
    \tablecaption{\textsc{CIGALE} Configuration for SED Fitting}
    \tablehead{\colhead{Model Component} & \colhead{SED Module} & \colhead{Parameter} & \colhead{Details}}
    \startdata
    Star-formation history & \texttt{sfhstochastic\_carvajal2025} & $\tau_\text{main}$$^a$ &  0.1, 0.5, 1.0, 2.0 Gyr \\
     &  & age$_\text{oldest}$$^b$ & 0.1, 0.5, 1, 2, 2.5, 3.0 Gyr \\
     & & $\alpha_\text{PSD}$$^c$ & 1.5, 2.0, 2.5 \\
     & & $\tau_\text{break}$$^d$ & 0.1, 0.15, 0.2 Gyr \\
     & & $\sigma_\text{SFH}$$^e$ & 0.2, 0.4, 0.6 \\
     & & Number of SFH seeds$^f$ & 5 \\
    \hline
    Stellar populations & \texttt{bc03} & IMF & \citet{Chabrier2003} \\
     & & $Z_\star$$^g$ & 0.02 \\
     & & age$_\text{separation}$$^h$ & 0.01, 0.05, 0.1 Gyr \\
    \hline
    Nebular emission & \texttt{nebular} & $\log_{10} U$$^i$ & $-2.0$ \\
     & & $Z_\text{gas}$$^j$ & 0.02 \\
     & & $n_{\mathrm{e}^-}$$^k$ & 100 cm$^{-3}$ \\
     & & $f_\text{escape}$$^l$ & 0.0 \\
     & & $f_\text{dust}$$^m$ & 0.0 \\
     & & \texttt{lines\_width} & 300 km s$^{-1}$ \\
     \hline
    Dust attenuation & \texttt{dustatt\_modified\_starburst} & $E(B-V)_\text{lines}$$^n$ & 0.1, 0.3, 0.5, 0.7, 1.0, 2.0 \\
     & & $E(B-V)_\text{factor}$$^o$ & 0.44 \\
     & & \texttt{uv\_bump\_wavelength}$^p$ & 217.5 nm \\
     & & \texttt{uv\_bump\_width}$^q$ & 35.0 nm \\
     & & \texttt{uv\_bump\_amplitude}$^r$ & 3.0 (Milky Way) \\
     & & $\delta_\text{attenuation}$$^s$ & $-0.5$, $-0.25$, 0.0 \\
     & & \texttt{Ext\_law\_emission\_lines}$^t$ & Milky Way \\
     & & $R_V = A_V/E(B-V)$$^u$ & 3.1 (Milky Way) \\
     \hline
    Dust emission & \texttt{dale2014} & AGN fraction$^v$ & 0.0 \\
     & & $\alpha_\text{dust}$$^w$ & 1.5, 2.0, 2.5 \\
     \hline
    Redshift \& IGM & \texttt{redshifting} & Source redshift $z$ & spec-$z$, else \citetalias{Ortiz2024} photo-$z$$^x$
    \enddata
    \tablecomments{\textsc{CIGALE} fitting was done using 17,321,040 models, with 524,880 per redshift. The precision on ages and timescales is $\pm 1$ Myr. \tablenotemark{a}{$e$-folding time of the main stellar population}, \tablenotemark{b}{age of the oldest stars in the galaxy}, \tablenotemark{c}{slope of the power spectrum density (damped random walk)}, \tablenotemark{d}{decorrelation timescale}, \tablenotemark{e}{dex amplitude of long-term SFH variability}, \tablenotemark{f}{number of seeds for each SFH model computed, each of which corresponding to a unique SFH \citep[see][]{Burgarella2025}}, \tablenotemark{g}{stellar populations' metallicity}, \tablenotemark{h}{age of the separation between young and old stellar populations}, \tablenotemark{i}{logarithm of the ionization parameter}, \tablenotemark{j}{gas metallicity}, \tablenotemark{k}{electron (e$^-$) number density}, \tablenotemark{l}{fraction of Lyman continuum photons escaping the galaxy}, \tablenotemark{m}{fraction of Lyman continuum photons absorbed by dust}, \tablenotemark{n}{color excess of nebular lines}, \tablenotemark{o}{reduction factor on $E(B-V)_\text{lines}$ when computing stellar continuum attenuation}, \tablenotemark{p}{central wavelength of UV bump}, \tablenotemark{q}{FWHM of UV bump}, \tablenotemark{r}{amplitude of UV bump}, \tablenotemark{s}{power-law slope modifying the \citet{Calzetti2000} attenuation curve}, \tablenotemark{t}{extinction law for computing emission line fluxes; the Milky Way extinction law was modeled as done in \citet{Cardelli1989}}, \tablenotemark{u}{ratio of V-band dust extinction to $E(B-V)$, also from \citet{Cardelli1989}}, \tablenotemark{v}{fractional AGN contribution to AGN emission, set to zero for this study; note that this is not the SED fractional AGN contribution $f_\text{AGN}$ in \citetalias{Ortiz2024}}, \tablenotemark{w}{power-law slope of dust mass distribution with respect to radiation field heating intensities \citep[see Equation (1) of][]{Dale2014}}, \tablenotemark{x}{see \S\ref{sec:data}; we used spectroscopic redshifts if available and \citetalias{Ortiz2024} photometric redshifts otherwise}.}
\end{deluxetable}

\section{Technical \textsc{Galfit} Configuration}
\label{app:galfit-configuration}
\textsc{Galfit} was provided with $266\times266$ pixel (${\sim} 8''\times8''$) galaxy images and masks of nearby objects derived from \textsc{SourceExtractor} \citep[][]{Bertin1996} segmentation maps. For fitting in the JWST filters, \textsc{Galfit} was given stamps from pipeline-generated error (ERR) images as sigma images. For fitting in the HST filters, \textsc{Galfit} was allowed to compute internal sigma images because no global error images were  produced in the HST data reduction. Weight (WHT) images were produced in the HST data reduction, but due to image pixel noise correlation \citep[see, e.g., the Appendix of][]{Casertano2000} and the non-uniform depth of the science images \citep[see][their Figure 1]{Obrien2024}, the naive $\text{ERR} = 1/\sqrt{\text{WHT}}$ prescription fails to produce acceptable error images. Preliminary fitting attempts using ERR images computed this way almost never converged. All fitting of HST data was done using models with an additional sky component to ensure robust background RMS estimation, where initial sky-background estimates were taken from the global background RMS computed by \textsc{SourceExtractor} before generating a catalog. The sensitivity of \textsc{Galfit}'s best-fit parameters to the sigma image is weak, but  the sigma image must be of acceptable quality for \textsc{Galfit} to compute accurate $\chi^2_\nu$. \textsc{Galfit} was provided with PSF models, which are discussed in Appendix~\ref{app:psfs}.

All models included S\'ersic components. The S\'ersic profile is treated by \textsc{Galfit} as a surface-brightness profile parametrized by radial coordinate $r$ and S\'ersic index $n>0$ as
\begin{equation}
    \label{eqn:sersic-profile}
    \Sigma(r) = \Sigma_e \exp \left\{ -\kappa_n \left[ \left( \frac{r}{r_e} \right)^{1/n} - 1 \right] \right\}
\end{equation}
for half-light radius $r_e$ and $\Sigma_e \equiv \Sigma(r=r_e)$. The number $\kappa_n$ is computed so that, for a given $n$, $r_e$ is indeed the half-light radius.\footnote{See \citet{Ciotti1999} for further discussion. Equation (3) of \citet[][]{Caon1993} is the equivalent of  Equation~\eqref{eqn:sersic-profile} in terms of intensities.}  The \citet{deVaucouleurs1948} and exponential ``disk'' profiles are recovered by Equation \eqref{eqn:sersic-profile} for $n=4$ and $n=1$, respectively, and the formula's versatility for virtually any $n$ has set a precedent of fitting galaxies with S\'ersic profiles \citep[see \eg][and the references within]{Sersic1968, Blanton2003, Driver2011, Luc2011}. While Equation~\eqref{eqn:sersic-profile} describes a radially symmetric distribution, S\'ersic components in the light-profile models were permitted to be elliptical,  handled internally by \textsc{Galfit}. 

\begin{deluxetable}{ccccc}[h]
\tablecaption{\textsc{Galfit} Constraints for Light Profile Modeling}
    \label{tab:galfit-constraints}
    \tablehead{
    \multicolumn{2}{c}{Singular Constraints} &
    \multicolumn{3}{c}{Relational Constraints} \\
    \cline{1-2} \cline{3-5}
    Component & Constraint & Component 1 & Component 2 & Constraint
    }
    \startdata
    Disk S\'ersic & $\pm3$ \textsc{SourceExtractor} mag & Disk S\'ersic & Bulge S\'ersic & Center, \texttt{ratio} \\
    Bulge S\'ersic & $\pm3$ \textsc{SourceExtractor} mag & Disk S\'ersic & point source & Center, \texttt{ratio} \\
    point source & $\pm3$ \textsc{SourceExtractor} mag & Disk S\'ersic & Bulge S\'ersic & Position angle, \texttt{ratio} \\
    Disk S\'ersic & $0.2 \leq n \leq 2.0$ & Disk S\'ersic & point source & $\pm3$ mag \\
    Bulge S\'ersic & $3.0 \leq n \leq 10.0$ & Bulge S\'ersic & point source & $\pm3$ mag \\
    Disk S\'ersic & $r_e \geq 4~\text{pix} = 0\farcs12$ & --- & --- & --- 
    \enddata
    \tablecomments{Constraints with $\pm3$ mag enforce that components are within 3 magnitudes, not necessarily that components are separated by 3 magnitudes. The \texttt{ratio} constraint is a hard constraint that fixes components' parameters by the ratios of their initial values.}
\end{deluxetable}

This study used one S\'ersic component (the ``disk S\'ersic'') to model the extended component of a galaxy and a point source to model unresolved emission from AGN\null. A second S\'ersic component (the ``bulge S\'ersic'') was used to model any compact, central bulge-like feature whose light may contaminate measurements of the point source.
\textsc{Galfit} constraints, listed in Table~\ref{tab:galfit-constraints}, were chosen to encourage convergence on physically sensible solutions; e.g.,  components were required to share a common center to avoid ``latching onto'' small-scale (but resolved), non-axisymmetric features (e.g., clumps), and all model components were restricted to be within $\pm$3 magnitudes of galaxies' measured magnitude to avoid too-bright or too-faint components ``dragging'' the whole fit into non-convergence or an absurd result. Relational constraints on the S\'ersic components were imposed so that they remain distinct (i.e., not two disk-like components overlapping each other or two bulge-like components with no disk).

Initial fitting parameters were derived filter-wise for each object from \textsc{SourceExtractor} catalogs run in dual-image mode. For JWST filters, the photometry from  \citetalias{Ortiz2024} with JWST/F444W as the detection image was used. New photometry was measured in the HST filters following \S2.2 of \citetalias{Ortiz2024} but with HST/F606W as the detection image to avoid apertures ``latching onto'' the JWST/F444W PSF's ``wings'' from a point source that are not present in HST images. For the S\'ersic components, the initial position angle and axis ratio were set to the \textsc{SourceExtractor} \texttt{THETA\_IMAGE} and the ratio of the \textsc{SourceExtractor} \texttt{B\_IMAGE} to \texttt{A\_IMAGE}. The initial half-light radius was estimated with $r_e = \text{FWHM/3}$. The initial S\'ersic indices of the ``disk'' and ``bulge'' components profiles were set to $n=1$ and $n=5$, respectively. All initial point-source magnitudes were set to the \textsc{SourceExtractor} catalog's \texttt{MAG\_AUTO} for that filter. \textsc{Galfit} was allowed to iterate up to 100 times before either converging or being forced to stop. 

The S\'ersic components' $r_e$ were left unconstrained in fitting except that they were required to be non-zero. This was done so that \textsc{Galfit} was allowed to fit a larger or smaller extended galaxy with the ``bulge'' S\'ersic component if the data preferred an $n$ in the ``bulge'' range of Table \ref{tab:galfit-constraints} and vice versa. The result of this was that the ``bulge'' S\'ersic was allowed  to converge on a larger $r_e$ than the ``disk component''. The bulge S\'ersics were occasionally found to be very small or very large. This is not problematic \textit{for the purposes of measuring a point-source magnitude} because the integrated magnitudes in these cases are not extreme, in the sense that they are not excessively bright or faint. For these reasons, the \textsc{Galfit} results in Table~\ref{tab:galfit-best-fit-params} are not true bulge-disk decomposition results but are suitable for subtracting the galaxy's central flux and measuring a point-source magnitude.

\section{Point-Spread-Function Models}
\label{app:psfs}
\begin{figure}[H]
    \centering
    \includegraphics[width=\linewidth]{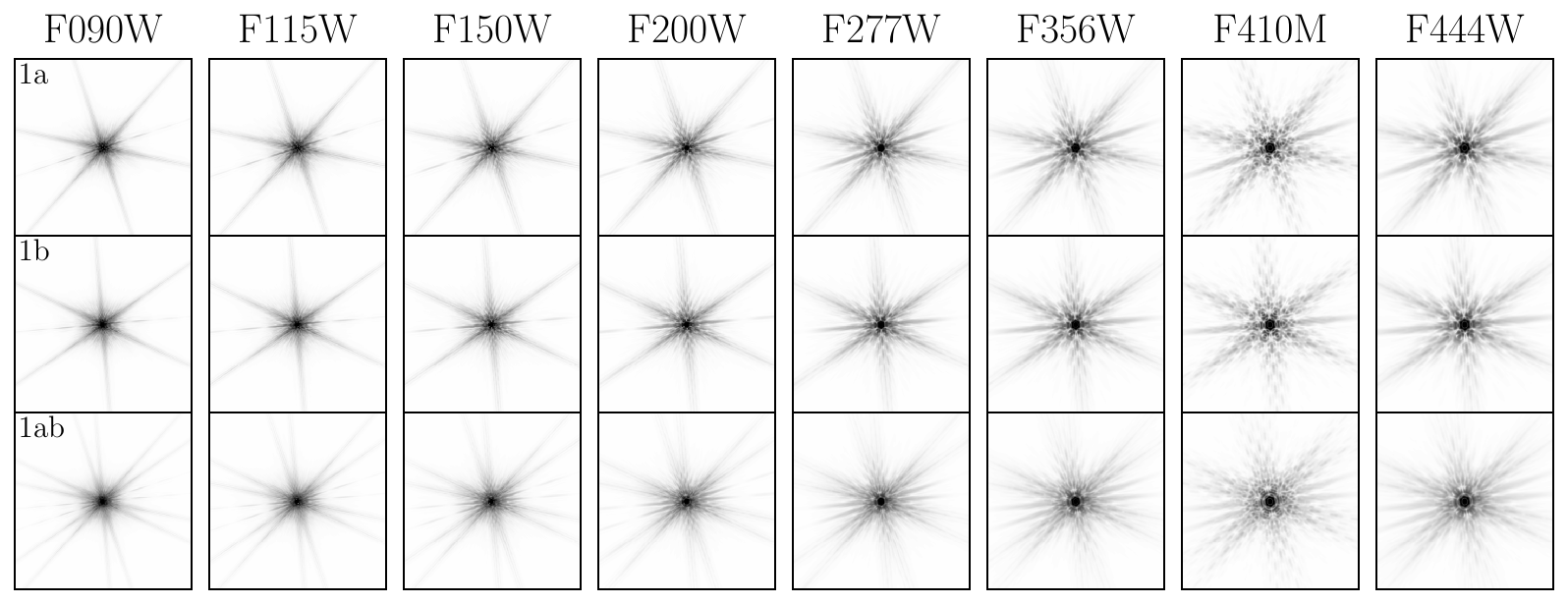}
    \caption{Renders of the PSF models given to \textsc{Galfit} for spoke 1's various position angles{{, which are $\sim$10$''$ on a side as shown here}}. The position angle indicators a, b, and ab are as in Table~\ref{tab:objects}. Each row shows a unique position angle, and each column shows a specific filter. These are the \texttt{OVERSAMP} extensions of the \textsc{STPSF} outputs, matched to the data's pixel scale so that \textsc{Galfit}'s PSF fine sampling is 1.}
    \label{fig:hybrid-psfs}
\end{figure}

High-quality PSF models are the most important part of modeling any point source in \textsc{Galfit}\null. For JWST fitting, simulated PSF models from \textsc{STPSF} \citep[formerly \textsc{WebbPSF},][]{Perrin2014} were used for each object's position angle in the JWST NEP-TDF, as determined by its spoke. Two objects in the sample have multiple position angles due to being observed in multiple overlapping ``spokes'' of the field. In these cases, the PSF models at the appropriate position angles were median-combined, and the results were used as the PSF models given to \textsc{Galfit}\null. Images of the hybrid PSF models are shown in Figure~\ref{fig:hybrid-psfs}.

Unlike the JWST mosaics, the HST mosaics have a large number of position angles drizzled onto the same pixel grid \citep[see Figure 1 of][]{Obrien2024}, which prevents simulation or construction (in the sense of co-adding bright, isolated, high-S/N stars) of PSF images with ``wings'' aligned with the sample galaxies' position angles. Galactic stars in the NEP-TDF have low areal density, and the multiplicity of overlapping position angles means that PSF models from stacked high S/N stars are not ``well-behaved'' in the sense of, e.g., \citet{Zhuang2024} or \citet{Dewsnap2025}, and so they cannot be used to construct PSF models for AGN-host decomposition. This was tested explicitly using both \textsc{PSFEx} \citep[][]{Bertin2013} and \textsc{Photutils}'s \texttt{EPSFBuilder} \citep[][]{photutils}, which failed to produce results of acceptable quality. Simulation of HST PSFs to median-combine as done for JWST was not feasible either, because the standard HST PSF simulation software \textsc{TinyTim} \citep[][]{Krist1993, Krist2011} is no longer officially supported by STScI, as HST's instruments have aged and legacy models no longer reproduce their performance.

Because the majority of the HST PSF's light is concentrated in its center, not the wings, a Gaussian HST PSF model was assumed with  
\begin{equation}
    \label{eqn:hst-psf-sigma}
    \sigma_\text{PSF} = \frac{\text{FWHM}}{2\sqrt{2\ln2}}
\end{equation}
for the PSF full width at half maximum (FWHM). We used the PSF FWHMs listed in Table 2 of \citet{Windhorst2011}, which are corrected to account for instrument-specific pixel sampling and telescope optics corrections. The HST PSF models provided to \textsc{Galfit} were drawn directly onto a pixel grid with standard deviations $\sigma_\text{PSF}$ following Equation \eqref{eqn:hst-psf-sigma} after converting to units of pixels via the $0\farcs030~\text{pixel}^{-1}$  scale.

\end{document}